\documentclass[12pt,preprint]{aastex} 
\usepackage{emulateapj5}
\usepackage [nolists,noheads,nomarkers]{endfloat}
\usepackage{graphicx}
\usepackage{epsfig}
\usepackage{verbatim}
\usepackage{appendix}

\newcommand{\etal}{{et al.~}}

\newcommand{\bq}{\begin{equation}}
\newcommand{\eq}{\end{equation}}

\def\gtsim{\lower.5ex\hbox{$\buSildrel > \over\sim$}}
\def\ltsim{\lower.5ex\hbox{$\buildrel < \over\sim$}}
\def\arcsec{^{\prime\prime}}

\def\farcs{\hbox{$.\!\!^{\prime\prime}$}}

\def\sun{\mbox{$_\odot$}}
\def\apjl{ApJL}
\def\apj{ApJ}
\def\apjs{ApJS}
\def\mnras{MNRAS}
\def\araa{ARAA}
\def\aj{AJ}
\def\aap{A\&A}

\def\nat{Nature}

\begin{document}

\title{Insights on the Formation, Evolution, and Activity of Massive
Galaxies From Ultra-Compact and Disky Galaxies at $z=2-3$}

\author{Tim Weinzirl\altaffilmark{1}, Shardha Jogee\altaffilmark{1}, Christopher J. Conselice\altaffilmark{2}, Casey Papovich\altaffilmark{3}, Ranga-Ram Chary\altaffilmark{4}, Asa Bluck\altaffilmark{5}, Ruth Gr{\"u}tzbauch\altaffilmark{2}, Fernando Buitrago\altaffilmark{2}, Bahram Mobasher\altaffilmark{6}, Ray A. Lucas\altaffilmark{7}, Mark Dickinson\altaffilmark{8}, Amanda E. Bauer\altaffilmark{9}}
\altaffiltext{1}{Department of Astronomy, University of Texas at
Austin, Austin, TX }
\altaffiltext{2}{University of Nottingham, School of Physics \& Astronomy, Nottingham NG7 2RD, U.K.}
\altaffiltext{3}{George P. and Cynthia Woods Mitchell Institute for Fundamental Physics and Astronomy, Department of Physics and Astronomy, Texas A\&M University, 4242 TAMU, College Station, TX 77843}
\altaffiltext{4}{U.S. Planck Data Center, MS220-6 Caltech, Pasadena, CA 91125}
\altaffiltext{5}{Gemini Observatory, Hilo, HI 96720, USA}
\altaffiltext{6}{Department of Physics and Astronomy, University of California, Riverside, CA 92521}
\altaffiltext{7}{Space Telescope Science Institute, 3700 San Martin Drive, Baltimore, MD, 21218}
\altaffiltext{8}{NOAO, 950 N. Cherry Avenue, Tucson, AZ 85719, USA}
\altaffiltext{9}{Australian Astronomical Observatory, P.O. Box 296, Epping, NSW 1710, Australia}

\begin{abstract}
We present our results on  the structure and activity of massive galaxies
at $z=1-3$ using one of the largest (166 with $M_\star\geq5\times10^{10}$
$M_\odot$) and most
diverse samples of massive galaxies derived from the GOODS-NICMOS survey:
(1)~S\'ersic fits to deep NIC3/F160W images
indicate that the rest-frame optical structures of  massive galaxies
are very different at $z=2-3$ compared to $z\sim0$.
Approximately 40\% of massive galaxies are ultra-compact
($r_e\le2$ kpc), compared to less than 1\% at $z\sim0$.
Furthermore, most ($\sim65$\%)  systems at $z=2-3$ have a low S\'ersic index
$n\le2$,  compared to $\sim13\%$ at $z\sim0$.
We present evidence that the $n\le2$ systems  at $z=2-3$  likely contain 
prominent disks, unlike most  massive $z\sim0$ systems.
(2)~There is a correlation between structure and star formation rates (SFR).
The majority ($\sim85$\%)  of non-AGN massive
galaxies at $z=2-3$,  with SFR high enough to yield
a 5$\sigma$ (30$\mu$Jy)
24 $\mu$m $Spitzer$ detection have low $n\leq 2$.
Such $n\leq 2$ systems host the highest SFR.
(3)~The frequency of AGN is $\sim40\%$ at $z=2-3$.
Most  ($\sim65\%$) AGN hosts have disky ($n\leq2$) morphologies.
Ultra-compact galaxies  appear quiescent
in terms of both AGN activity and star formation.
(4)~Large stellar surface densities imply massive galaxies
at $z=2-3$ formed via rapid, highly dissipative events at $z>2$. 
The large fraction of $n\le2$ disky systems suggests
cold mode accretion complements gas-rich major mergers at $z>2$. 
In order for massive galaxies at $z=2-3$ to evolve into present-day 
massive E/S0s, they need to significantly increase ($n$, $r_e$). 
Dry minor and major mergers may play an important role in this 
process.
\end{abstract}


\keywords{galaxies: bulges --- galaxies: evolution --- galaxies: formation --- galaxies: fundamental parameters --- galaxies: interactions --- galaxies: structure}

\section{Introduction}\label{intro}
Studies of high-redshift galaxies are essential for testing and constraining
models of galaxy formation.  Conventional wisdom suggests galaxies
are assembled and shaped by a combination of
mergers, smooth accretion, and internal secular evolution.
Galaxies form inside cold dark matter halos that grow hierarchically
through mergers with other halos and gas accretion
(Somerville \& Primack(1999); Cole et al. 2000; Steinmetz \& Navarro 2002; 
Birnboim \& Dekel 2003; Kere{\v s} et al. 2005; Dekel
\& Birnboim 2006; Dekel et al. 2009a; Dekel et al. 2009b; 
Kere{\v s} et al. 2005;  Kere{\v s} et al. 2009; Brooks et al. 2009; 
Ceverino et al. 2010), while  internal secular
evolution (Kormendy \& Kennicutt, 2004; Jogee et al. 2005)  redistributes 
accreted material. Within the paradigm of hierarchical assembly, a number 
of issues remain. It is not known when and how the main baryonic components 
of modern galaxies (bulges, disks, and bars) formed, 
but  the global stellar mass density rose substantially between
$z\sim1-3$, reaching $\sim50\%$ of its present value by $z\sim1$ (Dickinson et al. 2003b; 
Drory et al. 2005; Conselice et al. 2007; Elsner et al. 2008; P{\'e}rez-Gonz{\'a}lez et al. 2008).

It is also not clear how high-redshift galaxies evolve into
present-day galaxies.  Complex baryonic physics such as mergers,
gas dissipation, and feedback are all at work to an extent.
There is also mounting evidence that cold-mode accretion
(Birnboim \& Dekel 2003; Kere{\v s} et al. 2005; Dekel \& Birnboim 2006; 
Dekel et al. 2009a; Dekel et al. 2009b; Kere{\v s} et al. 2005; Kere{\v s} et al. 2009; 
Brooks et al. 2009; Ceverino et al. 2010) is important for building  star-forming 
galaxies. This process is particularly
effective in galaxies with halos of mass below $10^{12}$ $M_\odot$ such that
cold-mode accretion dominates the global growth of galaxies at
high redshifts and the growth of lower mass objects at late times.

High-redshift galaxies are different from local galaxies.
Within the framework of hierarchical assembly,
early, high-redshift galaxies are expected to be smaller, at a given mass,
than their present-day counterparts.
The size difference is predicted to be a factor of a few at
$z=2-3$ (Loeb \& Peebles 2003; Robertson et al. 2006;
Khochfar \& Silk 2006; Naab et al. 2007).  Several recent studies
using rest-frame optical data
provide evidence for size evolution among massive galaxies
(Guzman et al. 1997; Daddi et al. 2005; Trujillo et al. 2006, 2007;
Zirm et al. 2007; Toft et al. 2007; Longhetti et al. 2007;
Cimatti et al. 2008; Buitrago et al. 2008; van Dokkum et al. 2008, 2010;
van der Wel et al. 2011). Aside from size evolution, there is some evidence 
that the nature of red galaxies changes at higher redshift.
At $z\lesssim 1$, the red sequence primarily consists of
old, passively evolving galaxies (Bell et al. 2004). Among extremely
red galaxies (EROs) at $z=1-2$, less than 40\% are morphologically
early types (Yan \& Thompson 2003; Moustakas et al. 2004).
It is well known that star formation rates were more intense
at higher redshift (Daddi et al. 2007; Drory \& Alvarez 2008),
and a link has been found between star formation, size, and morphology
at $z\sim2.5$. Toft et al. (2007) and Zirm et al. (2007) find from
NICMOS rest-frame optical imaging that blue star-forming galaxies
are significantly more extended than red quiescent galaxies.
Additionally, examples of rapidly star-forming galaxies
(SFR $\sim50-200$ $M_\odot$ yr$^{-1}$) at $z\sim2-3$, whose ionized gas
kinematics are consistent with turbulent rotating disks, are found in the 
SINS survey 
(F{\"o}rster Schreiber et al. 2009; Genzel et al. 2008; Shapiro et al. 2008).

Progress on understanding the evolution of massive galaxies at high
redshift has been hindered by significant observational challenges.
The deep optical surveys carried out by $HST$ $ACS$,
such as the Hubble Ultra Deep Field (HUDF, Beckwith et al. 2006) and
the Great Observatories Origins Deep Survey (GOODS, Giavalisco et al. 2004),
trace rest-frame optical galaxy morphology only out to $z\sim1$.
At higher $z$, bandpass shifting effects cause filters to trace progressively
bluer bands, and optical filters trace rest-frame UV at $z\gtrsim2$.
UV light traces massive young stars, but manages to set few constraints about
the overall mass distribution, making it difficult to probe the structure and
mass of galaxy components at early epochs. 

Without high-resolution, deep, rest-frame optical
imaging, it is not possible to robustly compare structural parameters in 
galaxies across redshift.
NIR imaging is required to probe the rest-frame optical at $z\sim1-3$.
Unfortunately, deep NIR imaging with $HST$ has been completed for a limited
number of galaxies over relatively small fields and small volumes at $z>1$,
with most pointings being within the Hubble Deep Fields and the Hubble Ultra
Deep Field due to the inefficiency of the NICMOS camera in covering large areas
(e.g., Dickinson et al. 2004; Thompson et al. 2005; Zirm et al. 2007;
van Dokkum et al. 2008).
While ground-based NIR imaging surveys (e.g., Kajisawa et al. 2006;
Retzlaff et al. 2010) efficiently cover wide fields at
resolutions almost comparable to $HST$ NICMOS, the depths reached are
at least an order of magnitude shallower.

A large area, high-resolution, deep, space-based NIR survey would be 
bountiful for galaxy formation studies.  The GOODS-NICMOS Survey
(GNS; Conselice et al. 2011), covering 44 arcmin$^{2}$ of
the GOODS fields with NIC3, is a strong first effort.
The GOODS-North and GOODS-South are among the best-studied regions in the
sky and are a natural choice for such a survey.
The GOODS fields already have deep data from
$HST$ $ACS$ (Giavalisco et al. 2004),
$Spitzer$ IRAC/MIPS (Dickinson et al. 2003a),
and Chandra (Giacconi et al. 2002; Alexander et al. 2003; Lehmer et al. 2005;
Luo et al. 2008), among others. GNS consists of 60 pointings centered on 
massive ($M_\star>10^{11}$ $M_\odot$) galaxies at $z>2$, observed to a 
depth of $H=26.8$ magnitudes. The value of GNS lies in the fact that the 
target fields were optimized to include massive galaxies selected by multiple 
methods in order to create an unbiased sample (see Conselice et al. 2011).
There are additional massive galaxies in each field beyond the 60 main targets,
so that there are 82 galaxies with $M_\star \geq 10^{11}$ $M_\odot$ at $z=1-3$ 
across all pointings. Thus, the GNS data contain one of the largest
samples of very massive galaxies at high redshift with rest-frame
optical imaging, and they robustly probe massive galaxies when the 
Universe was less than 1/3 of its current age, during the epoch of 
bulge and disk formation.

The goal of this work is to investigate the evolution of massive galaxies 
over $z=1-3$ with this unique sample.
We take advantage of the existing rich ancillary data to derive star formation
rates (SFR) from 24 $\mu$m detections and look for AGN activity
based on X-ray detections and mid-IR SEDs.
We correlate rest-frame optical structural parameters with SFR to gain insight 
into how massive galaxies are expected to evolve.

The plan of this paper is as follows.  We discuss the data and sample
properties in \S\ref{data}.
In \S\ref{sprop} we describe the measurement of structural parameters, and 
in \S\ref{sressp} we make a 
detailed comparison with $z\sim0$ galaxies of similar stellar mass.
A detailed artificial redshifting experiment is conducted in \S\ref{redshift}
to explore the impact of instrumental and redshift-dependent effects on
structural parameters.
In $\S$\ref{sfrate}, we measure
star formation properties based on $Spitzer$ MIPS  24 $\mu$m detections
and discuss how they relate to structural properties.
Estimates of  the mass and fraction of cold gas in massive star-forming
galaxies  at $z=2-3$ are presented in   $\S$\ref{scold}.
In $\S$\ref{agn}, we use a variety of techniques (X-ray properties,
IR power-law, and IR-to-optical excess) to identify AGN and consider how
galaxy activity relates to galaxy structure.
Finally, in \S\ref{discuss} and \S\ref{summary}, we discuss and summarize
our results. All calculations assume a flat $\Lambda$CDM
cosmology with $\Omega_\Lambda = 0.7$ and $H_0 = 70$ km s$^{-1}$ Mpc$^{-1}$.

\section{Data and Sample}\label{data}
\subsection{ Observations and Pointing  Selections for GNS}\label{dataobserv}

Our data comes from the GOODS-NICMOS Survey (GNS; Conselice et al. 2011).
GNS is a deep, 180-orbit survey with the $HST$ NICMOS-3 camera in the F160W
($H$) band that probes optical light from galaxies between $z\sim 1-3$.
The coverage extends over both $ACS$ GOODS fields and is divided
between 60 pointings centered on massive $M_\star \geq 10^{11} M_\odot$
galaxies at $z>2$.
Each pointing covers $51\farcs2$ x $51\farcs2$ and was observed to
a depth of three orbits in nine exposures of $\sim 900$ seconds
($\sim 135$ minutes per pointing). A total of $\sim8300$ sources
were detected across an effective area of $\sim44$ arcmin$^2$.
The $5\sigma$ limiting magnitude for an extended source with a
$0\farcs7$ diameter is $H$=26.8 (AB). The NIC-3 images were drizzled 
with a pixfrac of 0.7 and output platescale of $0\farcs1$.  The NIC3 
camera is currently out of focus,and after detailed investigation 
(see \S\ref{decomppsf}), we find the point spread function (PSF) spans 
a full width half maximum (FWHM) of $0\farcs26-0\farcs36$ with a mean 
value of $0\farcs3$.

The 60 GNS pointings were planned by identifying massive galaxies
having a photometric redshift of $1.5<z<2.9$ and stellar mass 
$M_\star > 10^{11}$ $M_\odot$ via three color selection criteria.
The target galaxies include Distant Red Galaxies (DRGs, Papovich et al. 2006),
Extremely Red Objects (EROs, Yan et al. 2004), and $BzK$-selected galaxies
(Daddi et al. 2004).  All of these methods are designed to select red dusty
or red passively evolving  galaxies.
DRGs have evolved stellar populations that are identified with $J-K>2.3$
(Vega mag). EROs are selected based on $Spitzer$ and NIR data via
$f_{\nu}$(3.6$\mu m$)/$f_{\nu}(z850)>20$.
This selection is sensitive to red populations that are either old or
reddened, so EROs contain a mixture of young and old stellar populations.
$BzK$ galaxies are selected based on the quantity
$BzK \equiv (z-K)_{AB} - (B-z)_{AB}$.  Galaxies with $BzK>-0.2$ at $z>1.4$
are identified as star-forming galaxies.  Redder and possibly more evolved
galaxies are identified with $BzK<-0.2$ and $(z-K)_{AB}> 2.5$.
The final pointings were designed to include at least one red
massive galaxy and to also maximize the total number of additional galaxies
(e.g., Lyman-break galaxies and sub-mm galaxies) within each pointing.

\subsection{ Our Sample  of Massive Galaxies at z=1-3}\label{datamassz}

The sample of massive galaxies that we work with in this paper
is not limited to the original color-selected  massive galaxies  at $z>1.5$
defining the original  60 GNS pointings.  Instead, our sample of massive
galaxies at $z=1-3$ is derived from the set of
all  galaxies mapped with NIC3/F160W across the 60 fields,
and for  which a reliable stellar  mass and photometric redshift
was estimated by Conselice et al. (2011), based on  SED fits
to the NIC3/F160W and optical  imaging.
A detailed description of
how these quantities were estimated is in Conselice et al. (2011), and
we only briefly summarize the methodology here.

The source extraction catalog for the NICMOS images across the 60
pointings of the GNS survey contains $\sim8300$ sources with $H<28$ and $V<30$.
For those galaxies detected in the $ACS$ $BViz$ and NICMOS $H$ bands, we
use the available photometric redshifts and stellar masses from Conselice et al. (2011).
Photometric redshifts were determined by fitting template spectra to
the $BVizH$ data.
Stellar masses were measured by fitting the $BVizH$
magnitudes to a grid of SEDs generated from Bruzual \& Charlot (2003) stellar
population synthesis models, assuming a Salpeter IMF\footnote{
In \S\ref{sfrate} we use a Chabrier IMF for SFR estimates. Using a
Chabrier IMF rather than a Salpeter IMF in estimating the stellar
mass would lower the values by a factor of 0.25 dex or less.}. 
The grid includes different colors, ages of stellar populations,
metallicities, dust content, and
star formation histories as characterized by exponentially
declining models. 
In general,  the  stellar masses derived depend on the SED used and
the assumptions  used in the SED modeling, such as the IMF, the 
metallicity, the extinction law, and star formation history   (e.g., 
Borch et al. 2006;  Marchesini et al. 2009; Conselice et al. 2011). 
The typical uncertainty in stellar mass across the sample is 
a factor of $\sim2-3$.

In order to account for a small number (15) of additional massive 
($M_\star \geq 5\times 10^{10}$ $M_\odot$) red systems, which are
undetected in the GOODS $ACS$ $BV$ and therefore do not have
viable stellar masses from the above techniques, we use
available masses and redshifts (Buitrago et al. 2008; Bluck et al. 2009)
based on deep ground-based $RIJHK$ data along with
ACS $iz$ data, where available.  
Photometric redshifts are determined with a mixture of techniques
(e.g., neural networks and Bayesian techniques) described
more fully in Conselice et al. (2007).
Stellar masses were measured from these data
with uncertainties of a factor of $\sim2-3$ with the multi-color stellar
population fitting techniques from Conselice et al. (2007, 2008).
As with the larger sample described above, a stellar mass is produced by 
fitting model SEDs to the observed SED for each galaxy.
A Salpeter IMF is assumed, and the SED grids are
constructed from Bruzual \& Charlot (2003) stellar population synthesis
models. 

From the sample of galaxies with photometric redshifts and stellar
masses determined as described above, we define the sample of massive
galaxies used in this paper. 
 We restrict our analysis to the redshift interval  $z=1-3$
over which our NIC3/F160W images probe the rest-frame optical light
in order to avoid bandpass shifts into the rest-frame UV.
This ensures that we measure all structural parameters in the
rest-frame  optical across $z=1-3$, thereby reducing bandshift biases
(see \S\ref{decomp} for a quantitative estimate).
Although the mass functions calculated for GNS by Mortlock et al. 
(2011) show that the mass completeness limit is 
$\sim3\times 10^{9}$ $M_\odot$ at $z\sim3$, we apply
a higher mass cut of $5 \times 10^{10}$ $M_\odot$ as 
our interest is specifically with the most massive galaxies.

Our final sample consists of the 166 (82) massive galaxies with
$M_\star \geq 5\times 10^{10}$ $M_\odot$
($M_\star \geq 1\times 10^{11}$ $M_\odot$) and $z=1-3$.
This is the largest $HST$-based dataset with rest-frame optical
imaging of massive galaxies over $z=1-3$.  The galaxies with
$M_\star \geq 10^{11}$ $M_\odot$ from Buitrago et al. (2008) are 
part of the sample. The other previous $HST$ NICMOS studies
(e.g., Toft et al. 2007; Zirm et al. 2007; van Dokkum et al. 2008)
each contain, at most, $10-20$ systems with $M_\star \geq 10^{11}$ $M_\odot$.
The full distributions of apparent $H$ and $V$ magnitude, stellar mass,
and redshift for this sample are shown in Figure \ref{maghist}.

Figure~\ref{massfunc}  shows  a comparison of the galaxy stellar
mass  function (SMF)  of our GNS-based sample to the published SMF of
other NIR-selected samples in the literature, such as the 
 $K$-selected samples of Fontana et al. (2006), Kajisawa et al. 
 (2009)  and   Marchesini et al. (2009), as well as the IRAC-selected 
sample of P\'erez-Gonzalez  et al. (2008). This figure essentially
shows that for the mass range ($M_\star \geq 5\times 10^{10}$ $M_\odot$) 
relevant for the GNS-based sample used in our paper, 
there is good agreement between the SMF of our sample and 
those from  these four studies.
In particular,  at $M_\star \geq 5\times 10^{10}$ $M_\odot$, 
the top panel shows that  there is
very good agreement with our sample, Fontana et al. (2006),  
and  P\'erez-Gonzalez et al. (2008)  for three different redshift  bins 
between $z=1.5$ and  $z=3.0$.
In the lower panel, at $M_\star \geq 5\times 10^{10}$ $M_\odot$,  
the average SMF from Kajisawa et al. (2009) agrees with that 
of our sample within a factor of $\sim2$ over  $1.5<z<2.5$.
The SMF from our GNS-based sample and Marchesini et al.  (2009) 
show good agreement at $z=2-3$, and are slightly offset at 
z=1.3 to 2.0. The small offset may not be statistically significant 
if one includes all the sources of error. The error bars on the GNS
mass functions include Poisson errors only. Marchesini et al. (2009) 
show that the dominant sources of error regarding stellar mass 
functions are cosmic variance and systematics from the assumptions 
used in the SED modeling. 
For a discussion of the SMF for 
lower mass ($M_\star \geq 5\times 10^{10}$ $M_\odot$) galaxies,  
which are not included in the sample used in this paper,  we refer 
the reader to Mortlock et al. (2011).

In our sample of 166, massive galaxies, spectroscopic redshifts are 
available for 44 galaxies  ($26.5 \pm 3.4\%$ of the sample). 
These 44 galaxies  are all bright with $V\leq27$ and $H_{\rm AB}<23$.
Among these 44 galaxies,  the median photometric redshift error is
$\delta z/(1+z)=0.071$ (Gr{\"u}tzbauch et al. 2010), 
7/44 ($15.9 \pm 5.5\%$) have $\delta z/(1+z)>0.2$,  
and none have $\delta z/(1+z)>0.5$.\footnote{
While figure 6 of Conselice et  al. (2011)
shows that $\sim15-20\%$ of bright ($20<H_{\rm AB}<23$) galaxies with spectroscopic 
redshifts are catastrophic outliers in photometric redshift with
$\delta z/(1+z)>0.5$,  it should be noted that there are no 
catastrophic outliers with such large $\delta z/(1+z)>0.5$ 
among the 44 galaxies with spectroscopic redshifts in our sample of
massive  ($M_\star \ge 5\times 10^{10} M_\odot$) galaxies at $z=1-3$. 
The outliers with $\delta z/(1+z)>0.5$ in the GNS survey have 
stellar masses below the cutoff value of our sample or/and
lie outside its redshift range.}
For the remaining 122/166 ($73.5 \pm 3.4\%$) of our sample galaxies without 
spectroscopic redshifts, photometric redshifts are used.
Among these 122 galaxies, 60 ($49.2 \pm 4.5\%$) are fainter than $V>27$, 
and the uncertainties in photometric redshifts may be larger than the 
median value of 0.071 cited above.

\subsection{ Properties and Selection Biases in the Sample} \label{ssbias}

We estimate the  number density of massive 
($M_\star \geq 5\times 10^{10}$ $M_\odot$) galaxies over $z=2-3$ to be
$\sim5 \times 10^{-4}$ Mpc$^{-3}$
(see Conselice et al. 2011 for a detailed discussion of the number 
density of massive galaxies in the GNS sample). 
The corresponding stellar mass density is
$\sim 6 \times 10^7$~$M_\odot$~Mpc$^{-3}$.
The massive GNS galaxies are collectively 
10-100 times more abundant than SMGs, which have space densities of
$10^{-5}-10^{-6}$ Mpc$^{-3}$ at $z\sim2-3$ (Blain et al. 2002). 
Rather, the number density is in agreement with published values
(Daddi et al. 2005; 2007) for other passively evolving and star-forming 
galaxies at $z\sim2$. 

How does our sample break down in terms of
the typical color-selection methods, which are usually used to identify massive
high redshift galaxies?
About 63\%  (104/166) of this final sample 
is listed in existing catalogs for DRG (Papovich et al. 2006), $BzK$ (Daddi et al. 2004),
or ERO (Yan et al. 2004) galaxy populations.
There are 8, 9, and 43 sources that are uniquely listed in one of the DRG, $Bzk$,
or ERO galaxy catalogs, respectively.  An additional 44 sources are listed in two or
more of these catalogs.
About 37\% (62/166) sources were not previously identified as DRG, ERO, or $BzK$
galaxies.

What are the selection biases impacting our sample?
General biases in the selection of massive galaxies in the GNS
survey have been discussed in Conselice et al. (2011), and we
only discuss below the points relevant for our sample.

The 60 GNS pointings were selected to include massive galaxies
identified via three color methods (DRG, $BzK$, and
IERO). Combining all three color criteria, rather than using any single
one, is already a step forward compared to many earlier studies
because  no single criterion  would isolate a 
complete sample of massive galaxies (e.g., van Dokkum et al. 2006;
Conselice et al. 2011).  These three criteria all pick massive
galaxies with red observed colors, but due to the range
of criteria involved, they  can pick both red dusty systems and red
evolved stellar populations.  

Another key step that makes our study less biased towards 
a specific type of massive galaxy is that our working sample 
at $z=1-3$ is neither limited to nor defined by the original 
color-selected massive galaxies. Rather,
it is derived from all galaxies within the survey area that are
bright enough to be mapped with NIC3/F160W and  for which a 
reliable stellar  mass and  photometric redshift could be determined
by Conselice et al. (2011),  as outlined in $\S$\ref{datamassz}.  
The first potential  bias in this final sample  is  introduced by 
excluding  galaxies that are  undetected by NIC3/F160W.
The second potential  bias is  introduced by 
excluding detected galaxies for which no reliable stellar 
mass and photometric  redshift could be determined.  For instance,
ultra-dusty galaxies, may not be detected in enough of the optical
bands to allow a photometric redshift to be reliably estimated. 

\begin{figure}[]
\centering
\scalebox{1.00}{\includegraphics{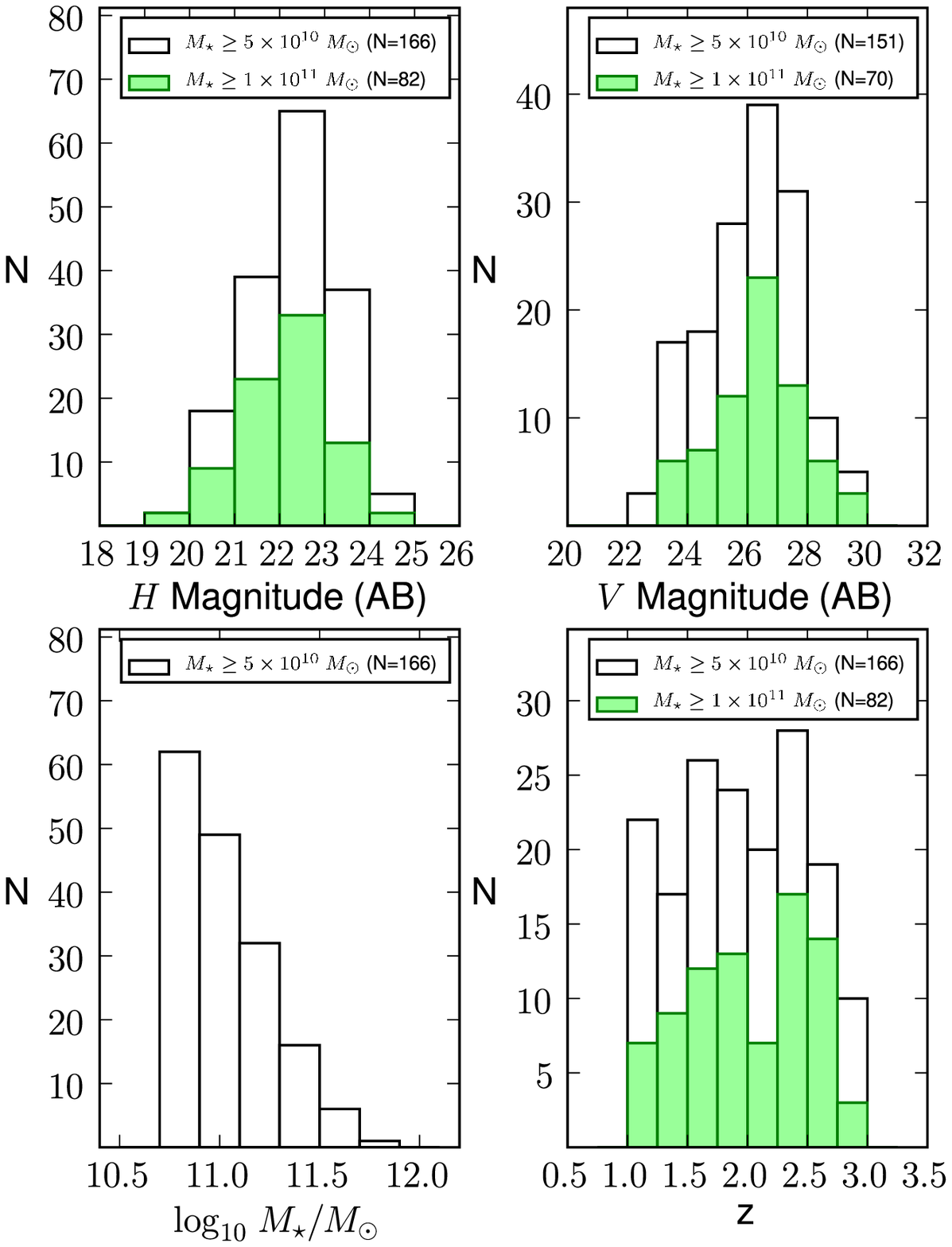}}
\caption{The distribution of apparent $H$ (F160W), $V$ apparent magnitude,
stellar mass, and redshift for the final,
complete sample of 166 galaxies with $M_\star \geq 5\times 10^{10}$ $M_\odot$ and 
redshift $z=1-3$. 
\label{maghist}}
\end{figure}

\begin{figure}[]
\centering
\scalebox{0.85}{\includegraphics{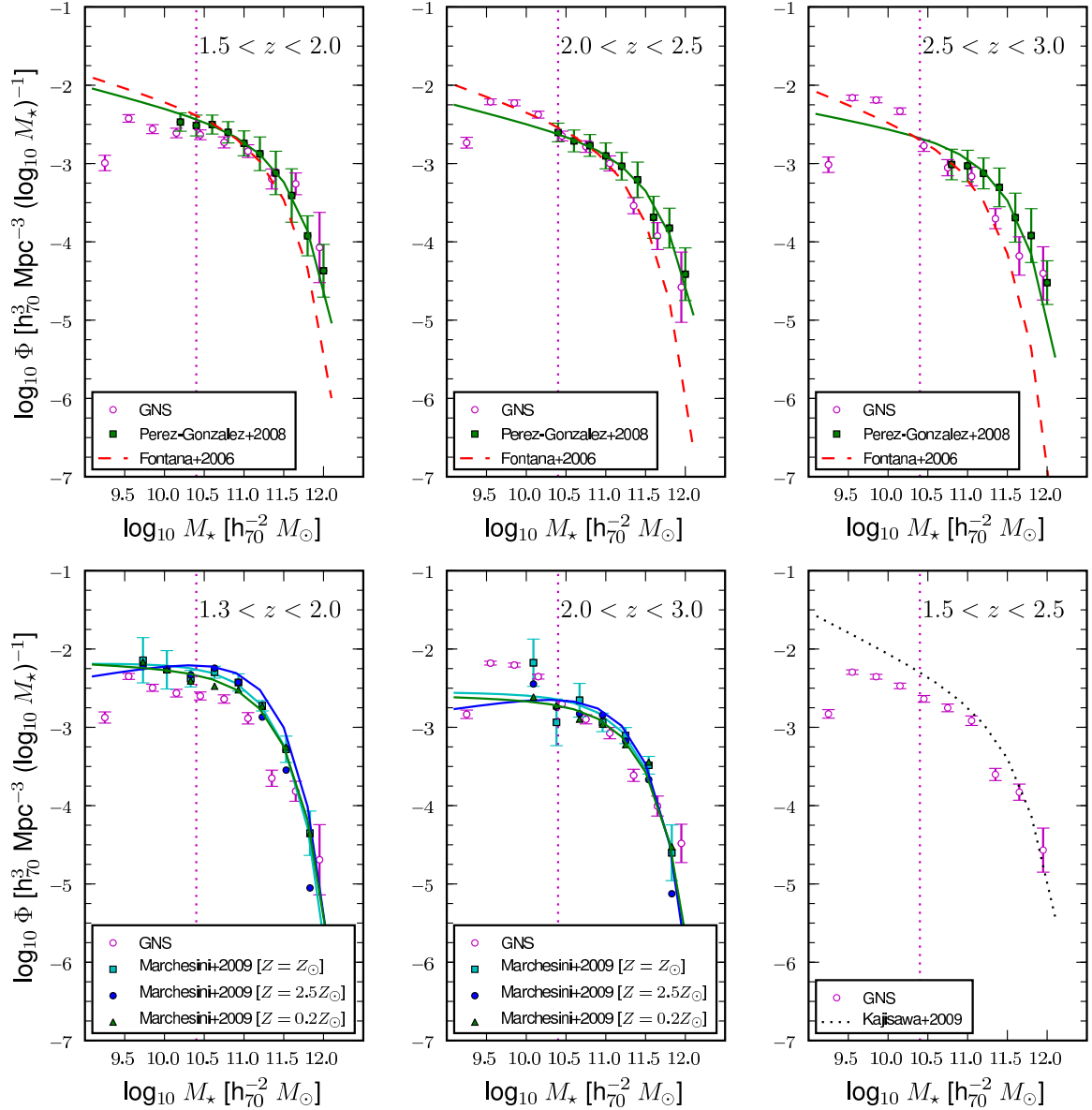}}
\caption{
We compare the galaxy stellar mass functions for GNS over $z=1-3$ with
those from other studies that are based on $K$ or IRAC-selected samples
(Kajisawa et al. 2009; Marchesini et al. 2009; P{\'e}rez-Gonz{\'a}lez et 
al. 2008; Fontana et al. 2006).
The vertical line in each plot marks the mass cut
($M_\star \geq 5\times 10^{10}$ $M_\odot$) for the GNS-based sample 
used in this paper. 
We include the data points with error bars from the other studies,
where available, along with each Schechter function fit.
Some studies (Kajisawa et al. 2009; Marchesini et al. 2009) present 
results for multiple sets of SED-modeling assumptions, and in these
cases we show the results for the assumptions that most closely match
those used for  GNS by Conselice et al. (2011).  For Kajisawa et 
al. (2009) , we show the mass function calculated with Bruzual \&
Charlot (2003) stellar templates.  
For Marchesini et al. (2009), we show  the stellar mass functions
calculated with Bruzual \& Charlot 2003 templates, metallicities of 
0.2, 1, and 2.5 $Z_\odot$, a Kroupa IMF,
and a Calzetti extinction law,  but in the above plot, we scale 
their mass functions by $+0.2$ dex along the
$x$-axis to convert their Kroupa IMF to a Salpeter IMF. 
For the GNS mass functions, in
comparison, the best metallicity is determined on a galaxy-by-galaxy  
basis from a set of
discrete values spanning 0.005 to 2.5 $Z_\odot$. 
The error bars for  Marchesini et al. (2009)
take into account the uncertainties due to cosmic variance, Poisson 
error, photometric redshifts,
and stellar SED templates. The error bars from P{\'e}rez-Gonz{\'a}lez et 
al. (2008) account for
Poisson error and uncertainty in photometric redshifts.  In comparison, 
the error bars on the
GNS mass functions show only Poisson error. 
\label{massfunc}}
\end{figure}

We assess the impact of the second bias by estimating how many 
massive galaxies we might miss due to the lack of available 
photometric redshift and stellar 
masses. Of the 8300 sources detected by GNS, 1076 have no 
photometric redshift and stellar mass measurements.  Most 
 (68\%)  of these 1076 sources are fainter ($H>25$) than our sample of 
massive galaxies (Figure~\ref{maghist}).  Among GNS objects as 
bright ($H<25$) as our sample of massive galaxies, only 8.5\%, or 349/4083 have
no redshift or stellar mass measurements. Furthermore, not all 8.5\% of these
bright ($H<25$) sources will be massive, so that this fraction represents
an upper limit on the sources  we might not include in our sample due to
the lack of a photometric redshift or stellar mass measurements.

We next discuss the impact of the first potential bias and the
type of objects the GNS survey might not detect.
It is relevant to ask whether we might miss galaxies with blue 
observed colors.   We believe this is not the case for the 
following reasons.  As discussed above, our working sample is
not strongly biased against galaxies with blue observed
colors because it is not limited to  those  massive galaxies
selected by the three color methods (DRG, $BzK$, and
IERO) that preferentially pick galaxies with red observed colors.  
Secondly, Conselice et al. (2011)  explicitly show that many galaxies with 
blue observed ($z-H$) colors, which would have been undetected 
by these color selections, do get included in this final sample of 
massive galaxies for  the GNS survey.  Nearly all known Lyman Break
Galaxies or BX/BM  objects  (Reddy et al. 2008) at $z=2-3$  in the GNS 
fields are detected by the GNS NIC3/F160W imaging (Conselice et al. 2011).

\begin{figure}[]
\centering
\scalebox{1.00}{\includegraphics{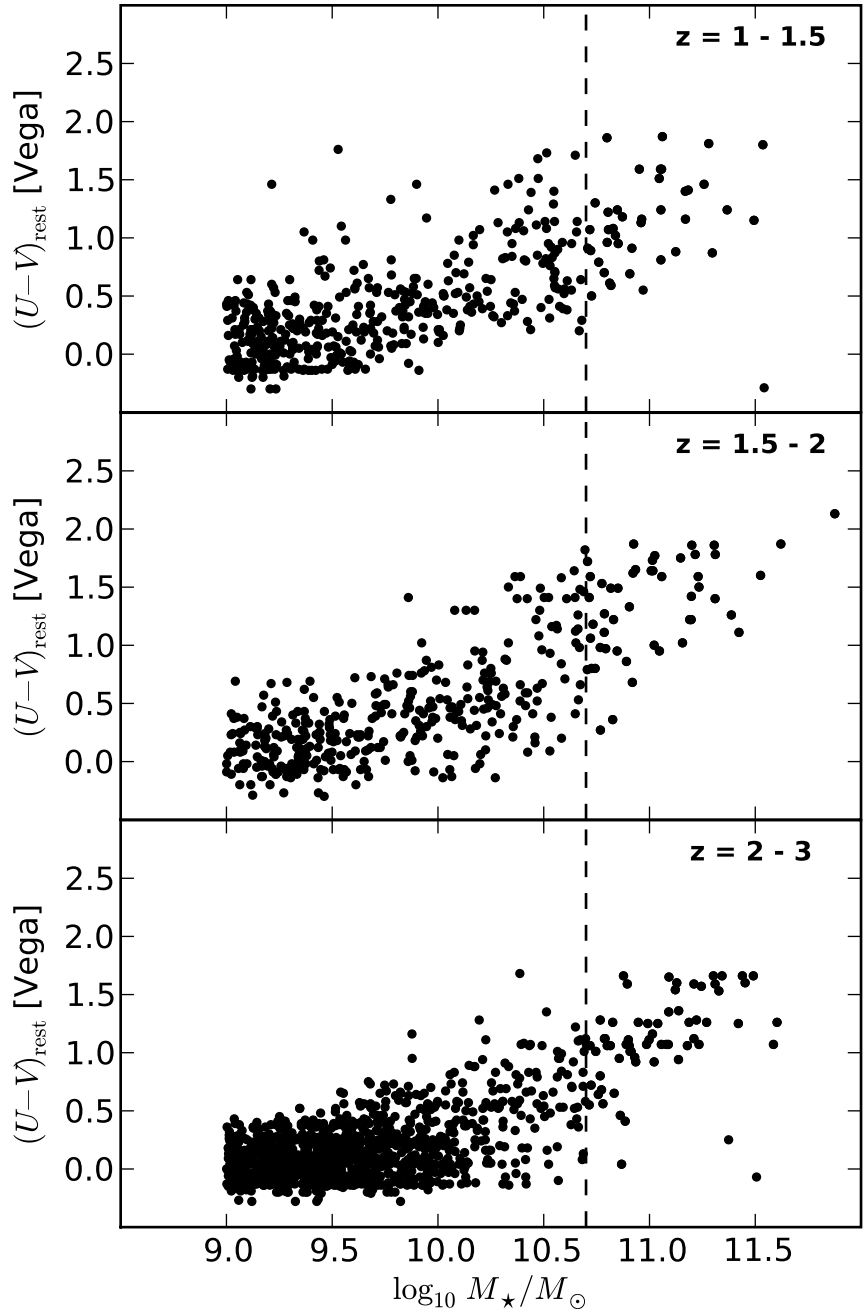}}
\caption{For all galaxies detected in the GOODS-NICMOS Survey (GNS) over 
$z=1-3$, the rest-frame $U-V$ color is plotted against $M_\star$ for different 
redshift bins.
Blue systems are preferentially at low masses, while the most massive
($M_\star \geq 1\times 10^{11}$ $M_\odot$)
galaxies are preferentially red.
The vertical line denotes $M_\star = 5\times 10^{10}$ $M_\odot$, the mass cut we
adopt for our final sample of 166 galaxies.
\label{massuv}}
\end{figure}

In terms of rest-frame colors, rather than observed colors, it is also
important to note that the galaxies detected by GNS at $z=1-2$ or
$z=2-3$ include systems with both blue and red  rest-frame $U-V$
colors. The rest-frame
$U-V$ color ranges from about $-0.4$ --  2.1   for galaxies  in the
stellar mass range $M_\star \sim 10^9$ -- $10^{12}$ $M_\odot$ (Figure
\ref{massuv}).  The systems with blue rest-frame $U-V$ colors are 
preferentially at low masses, while GNS  galaxies with 
$M_\star \geq 1\times 10^{11}$ $M_\odot$  at $z=2-3$ have
preferentially red rest-frame $U-V$ colors, in the range of 1.0 to 1.7.
These inherently red rest-frame $U-V$ colors  of 
the  massive  galaxies at  $z=2-3$ could be due to a 
combination of old stellar populations and dusty young star-forming
regions.  We checked that the colors are consistent with stellar population 
synthesis models (based on Bruzual \& Charlot (2003) and assuming a Chabrier 
IMF, an exponentially declining star formation history with a 100 Myr e-folding
time).  We find that even without dust extinction $U-V$ color rises rapidly. 
Assuming solar 
metallicity, $U-V$ is already $\sim1$ at an age of 0.5 Gyr and reaches $\sim1.6$ 
at 2 Gyr. For the case with dust extinction and an optical depth of 1, $U-V$ is 
$\sim1.1$ after 0.5 Gyr and $\sim1.8$ after 2 Gyr.

\section{Structural Properties of Massive Galaxies}\label{sprop}
\subsection{Structural Decomposition}\label{decomp}
We characterize the massive GNS galaxies with structural decomposition.
Ideally, one would like to fit multiple components (bulge, disk,
bar, nuclear point source, etc.) in the decomposition, but the $0\farcs3$ diameter
(or full width half-maximum)  of the PSF  (corresponding to $\sim2.4$ kpc
at $z=1-3$) prevents such detailed decompositions\footnote{For 
the more extended galaxies multiple components (e.g., bulge and
disk) decomposition was attempted with limited success and this is
discussed in \S\ref{sdisc0}.}.  Instead, we choose to fit the 2D light 
distributions with only single S\'ersic (1968) $r^{1/n}$  profiles, which have the form
\begin{equation}
\rm{ I(r)=I_e \ exp \left( -b_n \left[ \left( r\over{r_e} \right)  ^{1/n} -1 \right]  \right) },
\end{equation}
where $\rm{I_e}$ is the surface brightness at the effective radius $r_e$ and
$\rm b_n$ is a constant that depends on S\'ersic index $n$.
Knowledge of the PSF is important for deriving structural parameters.  
We model the PSF (Appendix~\ref{decomppsf}) while taking into account  both the 
variation in PSF with position on the NIC3 field and  the dependence on the 
\texttt{drizzle} algorithm.  We find a range in PSF FWHM of $\sim0\farcs26-0\farcs36$.

It is clear that a  single S\'ersic profile is not a complete
indicator of overall galaxy structure. For instance,
in detailed images of nearby
galaxies, the best-fit index $n$ for a single S\'ersic profile does not always
correlate with the bulge S\'ersic index obtained with 2D bulge-disk or
bulge-disk-bar decomposition (Weinzirl et al. 2009).
However,  the single S\'ersic index $n$ is on average a  good
way to separate  disk-dominated galaxies from   the class of luminous
spheroidal and bulged-dominated galaxies (see \S\ref{redshift}),  and  in studies of
high-redshift galaxies  the criterion $n \lesssim 2$  is often
used to separate spirals or disk galaxies from ellipticals
(e.g., Ravindranath et al. 2004; Bell et al. 2004; Jogee et al. 2004;
Barden et al. 2005;  Trujillo et al. 2007; Buitrago et al. 2008).

The NIC3/F160W images of the 166 sample galaxies were fit with a
single S\'ersic component using GALFIT (Peng et al. 2002).
In each image, objects that were near, but not blended with, the primary 
source were masked out.  For the fraction ($\sim15\%$) of the primary galaxies
that were blended or overlapping with another galaxy identified 
in the source extraction catalog, the blended sources
were each fitted simultaneously with a separate S\'ersic profile.
Some fraction of primary galaxies appeared morphologically disturbed 
($\sim8\%$, see 
Figure~\ref{gnsmontage} and \S\ref{sressp}), 
but these were fitted with only a single S\'ersic profile as they
only counted as a single galaxy in the source extraction catalog.

Bandpass shifting causes the $H$-band central wavelength to move from
4000-8000 $\AA$ over $z=1-3$. The $z=1-2$ and $z=2-3$ bins used in 
Figure~\ref{fnre}, for example, correspond to 5333-8000 $\AA$ ($I$-band) and 
4000-5333 $\AA$ ($B$-band), respectively. 
Even with the bandpass shifting, comparing the structural parameters ($n$, $r_e$)
measured in these two bands to each other and to parameters of $z\sim0$ galaxies 
measured in rest-frame $B$ is a vast improvement over previous studies forced to 
compare the rest-frame UV at $z>1$ to the rest-frame optical at $z<1$.
The systematic effects resulting from $H$-band changing
from $B$ to $I$-band over $z=1-3$ are small, as can be inferred from studies of
nearby galaxies.  Graham (2001) presents bulge-disk decompositions of local 
$z\sim0$ galaxies based on  images in the $B$ and $I$ bands.
The median ratio in $B$-band/$I$-band disk scalelength
is 1.13, so that the disks are measured to be slightly larger
in the B-band. If similar errors apply here, then
the bias $r_e$ due to bandpass shifting is on the order of 10\%.

Another important consideration is the effect of potential AGN on the
structural fits. When fitting high resolution images of nearby galaxies,
it is well known that fitting a galaxy that hosts a point source
with a single S\'ersic component will lead to an
artificially high  S\'ersic index $n $ (typically $n>4$; e.g.,
Weinzirl et al. 2009; Pierce et al. 2010). 
If a point source is added to the S\'ersic model,
the  index $n$ of the S\'ersic component falls to more reasonable
values. In the case of the massive GNS galaxies at $z=1-3$,
we expect that
the low resolution  ($0\farcs3$, corresponding to  2.5 kpc at $z\sim2$)
of the NIC3/F160W images will reduce the effect of potential point
sources on the structural decomposition. However, for completeness,
we have fitted all the galaxies  at $z=1-3$ in which a potential
AGN was identified  via a variety of techniques ($\S$\ref{agn})
with both a  S\'ersic component and a point source.  
The fractional luminosity of the point source components, 
or PSF/Total light ratio, ranges from 1-46\%, with a median of 10\%.
As expected, including the point source produces generally small
changes in  ($n$, $r_e$) and goes in the direction of lowering
$n$ and enlarging $r_e$. Overall, our results  
are not biased by the presence of AGN. 
In the rest of the paper, we therefore choose to use the structural
parameters for a  single S\'ersic component fit.

\subsection{Derived Structural Properties at $z=2-3$}\label{sressp}
\begin{figure}[]
\centering
\scalebox{0.80}{\includegraphics{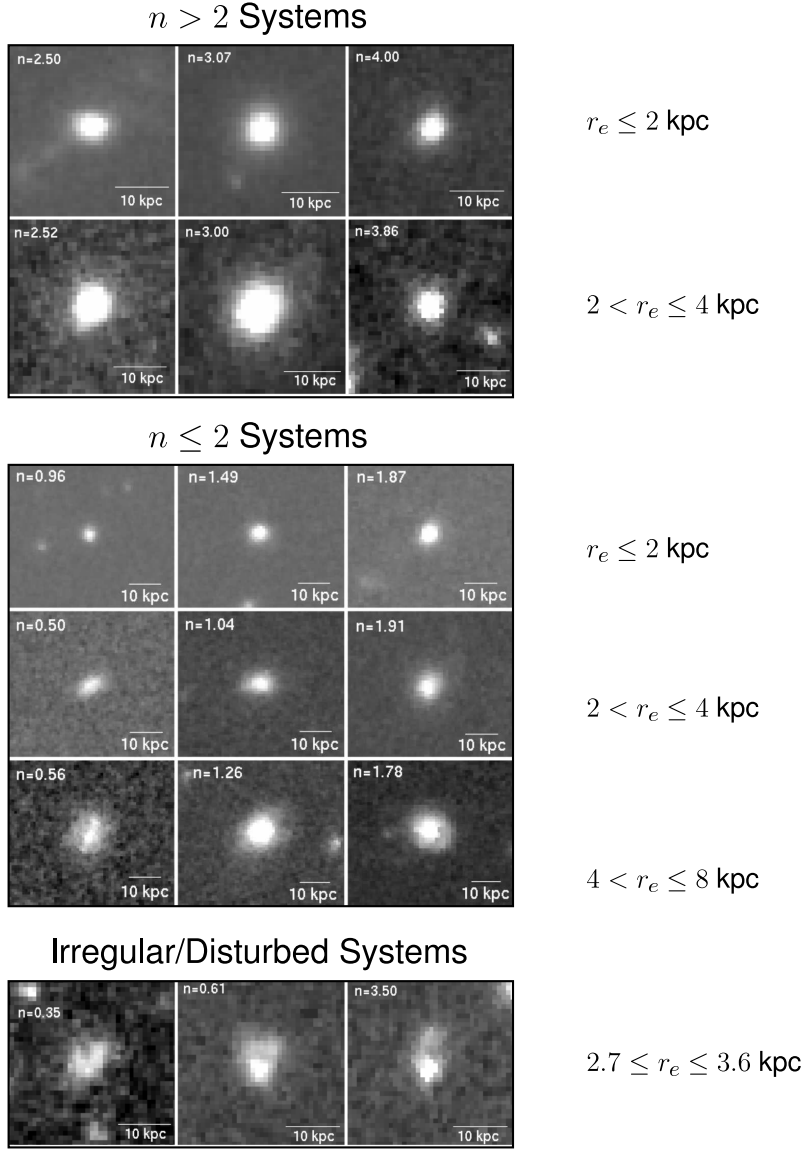}}
\caption{
NICMOS F160W images for representative GNS galaxies
with $M_\star \geq 5\times 10^{10}$ $M_\odot$ at $z=2-3$. 
The S\'ersic index $n$ and effective radius $r_e$ referenced 
here are based on fitting single  S\'ersic  components to the
NICMOS images, as described in  \S\ref{sprop}.
The top panel shows example systems with S\'ersic index $n>2$
and half-light radii $r_e \le 4$ kpc.
The middle panel shows examples with $n\leq 2$ and $r_e \leq 8$ kpc.
The majority ($\sim82\%$; Table~\ref{tabmgc}) of the massive 
GNS galaxies have $r_e \le 4$~kpc. In such systems, structural 
features are generally hard to discern due to resolution effects, 
so that systems appear fairly featureless (top 4 rows).
In the small fraction of  massive galaxies at $z=2-3$ with 
large $r_e > 4$~kpc, one can discern some structural features
such as an elongated bar-like feature or a combination of a 
central condensation surrounded  by a more extended lower surface 
brightness component, reminiscent of a bulge and disk (row 5).
The bottom panel (row 6)
contains systems that appear morphologically disturbed. 
\label{gnsmontage}}
\end{figure}

The results of the  structural fits to the NIC3/F160W images of the 166
sample galaxies are shown in Table~\ref{tabmgc},  Figure~\ref{gnsmontage}, 
Figure~\ref{fnre}, and Figure~\ref{mgchist}.

Figure~\ref{gnsmontage}  shows  examples of massive 
($M_\star \geq 5\times 10^{10}$ $M_\odot$)  galaxies at $z=2-3$
with different  ranges of S\'ersic index $n$  and effective radius $r_e$.
The majority ($\sim82\%$; Table~\ref{tabmgc}) of the massive 
GNS galaxies at $z=2-3$ have $r_e \le 4$~kpc.  In such systems, structural 
features are generally hard to discern due to resolution effects, 
so that systems appear fairly featureless (top 4 rows of Figure~\ref{gnsmontage}).
In the small fraction of  massive galaxies at $z=2-3$ with 
large $r_e > 4$~kpc, one can discern some structural features
such as an elongated bar-like feature or a combination of a 
central condensation surrounded  by a more extended lower surface 
brightness component, reminiscent of a bulge and disk (5th row).
Row 6 contains morphologically disturbed systems.  The fraction
of such systems is small, only $\sim8\%$, but this is a lower limit
given redshift-dependent effects such as degraded physical resolution and
surface brightness dimming.

The lower two rows of 
Figure~\ref{fnre}  shows  the rest-frame optical
S\'ersic index $n$ and effective radius $r_e$ for the samples
of massive galaxies at $z=1-2$ and  $z=2-3$. 
For comparison,  the top row of Figure~\ref{fnre} also shows
the  rest-frame optical
structural parameters for $z\sim$~0  galaxies of similar stellar
mass taken from   Allen et al. (2006), who performed a 
single component S\'ersic fit to $B$-band images of  galaxies
in the Millennium Galaxy Catalogue (MGC), a large
ground-based imaging and spectroscopic survey over 37.5~deg$^2$
(Liske et al. 2003; Driver et al. 2005).
It is clear from  Figure~\ref{fnre}, Figure~\ref{mgchist}, and 
Table~\ref{tabmgc}  that the massive galaxies at  $z=2-3$
are strikingly offset toward lower ($n$, $r_e$) compared to the
massive $\sim$~0 galaxies.

Firstly, we find that  {\it the majority 
($64.9 \pm 5.4\%$  for  $M_\star \geq 5\times 10^{10}$ $M_\odot$, and
$58.5 \pm 7.7\%$ for  $M_\star \geq 10^{11}$ $M_\odot$)  of  massive
galaxies  at $z=2-3$  have  low $n\le$~2, while the fraction at $z\sim$~0 is five times lower. }
We will  present evidence in $\S$\ref{sdisc0}  that  most of the massive 
systems with a  low $n\le$~2  harbor a massive disk component, 
so that  our results point
to  the {\it predominance of disk-dominated systems among massive galaxies
at $z=2-3$.}

Secondly, we also find that massive galaxies at $z=2-3$  typically
have smaller $r_e$ than  massive galaxies at  $z\sim 0$.
{\it In particular,  $\sim$~40\%   
($39.0  \pm 5.6$\%  for  $M_\star \geq 5\times 10^{10}$ $M_\odot$ and 
$39.0  \pm 7.6$\%  for  $M_\star \geq 1\times 10^{11}$ $M_\odot$ )
 of massive galaxies at $z=2-3$ are ultra-compact  ($r_e \le 2$~kpc),  
compared to less than one percent at $z\sim$ 0.}
The massive ultra-compact  ($r_e \le 2$~kpc),   galaxies at $z=2-3$ 
have few  counterparts among $z\sim 0$ massive galaxies.

The population of galaxies with low $n\le2$  and the population of 
ultra-compact ($r_e \le 2$~kpc) galaxies  show limited overlap.
Only $28.0\pm6.4\%$ of the systems with  low $n\le2$   are  ultra-compact
and the remaining majority 
($72.0 \pm 6.3 \%$  for $M_\star \geq 5\times 10^{10}$ $M_\odot$, and 
$75.0 \pm 8.8 \%$ for  $M_\star \geq 10^{11}$ $M_\odot$)
are extended  ($r_e > 2$~kpc).
Conversely, among the ultra-compact  ($r_e \le 2$~kpc) systems,   nearly 
half   ($46.7 \pm 9.1 \%$  for $M_\star \geq 5\times 10^{10}$ $M_\odot$, and 
$37.5 \pm 12.1\%$   for  $M_\star \geq 10^{11}$ $M_\odot$)   have
low $n\le2$.

Figure~\ref{figsbr}  further  illustrates the striking difference
between massive galaxies  at $z=2-3$ and $z\sim~0$ 
by comparing their  effective radius $r_e$ and 
their mean rest-frame optical surface brightness  
$<\mu_e>$  within  $r_e$.
The value of $<\mu_e >$   was measured from the 
extinction-corrected rest-frame $B$-band light
within $r_e$  and is defined as:
\begin{equation}
\rm{ \mu_e = B_{\rm{corr}} + 2.5 \rm{log}_{10}(2\pi r_e^2) - 10 \rm{log}_{10}(1+z)}
\end{equation}
where $B_{\rm{corr}}$ is the extinction-corrected, rest-frame apparent $B$ 
magnitude and $-10 \rm{log}_{10}(1+z)$ and is the correction for surface 
brightness dimming.
The MGC galaxies at $z\sim0$ are corrected only for Galactic extinction, while
for the GNS galaxies the correction includes Galactic and internal extinction.
{\it 
The  mean rest-frame optical surface brightness   can be  2.0  to 6.0
magnitudes  brighter for the massive galaxies at $z=2-3$
than for $z\sim 0$ massive galaxies.  } This  is  due to 
their smaller sizes and likely differences in the age of the 
stellar populations. 
The  high mean rest-frame optical surface brightness  
of the massive  galaxies  at  $z=2-3$  
translates into high mean stellar mass densities, and 
suggests that  highly dissipative events played an 
important role in their formation (see $\S$\ref{discuss}).

It is worth noting that the use of deeper images for the
$z\sim$~0 galaxies could make the large offset in 
($n$, $r_e$) at $z=2-3$ versus $z\sim0$ even stronger.
The MGC $B$-band images have a median sky background of 22 mag/arcsec$^2$.
Low surface brightness halos
may be detected around some of the $z\sim0$
galaxies in deeper exposures.  This is true for some massive elliptical
and cD galaxies, and in these cases the ($n$, $r_e$) are significantly
boosted if the halo is region is also fitted (Kormendy et al. 2009).

How do these results compare with earlier studies?
While many of the earlier studies focused on small samples,  
this work is a step forward because of the improved number statistics  
that come with an unbiased and complete sample of massive galaxies.
The observed {\it apparent}  size evolution in our data
generally agrees with 
results reported in other studies of massive galaxies (e.g.,
Daddi et al. 2005; Trujillo et al. 2007;
Zirm et al. 2007; Toft et al. 2007;
Buitrago et al. 2008; van Dokkum et al. 2008; 2010; Williams et al. 2010).

The ratio in $r_e$ of high-redshift galaxies with respect to $z\sim0$
galaxies, or $r_e/r_{e,z\sim0}$, can be modeled as a power law in 
redshift of the form
$\alpha (1+z)^\beta$, where $\alpha$ and $\beta$ are constants.
Using the $z\sim0$ massive ($M_\star \geq 5 \times 10^{10}$ $M_\odot$) MGC
galaxies as the normalization, we measure $\alpha$ and $\beta$ for different
subsamples of the massive galaxies and summarize the results in Table~\ref{beta}.
For all galaxies the slope $\beta$ is -1.30 for a fit over $z=0-3$.  
For disk-like $n\le2$ 
galaxies $\beta$ is also -1.30, and for $n>2$ galaxies $\beta$ is -1.52.
For non-AGN host galaxies with $SFR_{\rm IR}$ detected above the $5\sigma$ 
detection limit
(see \S\ref{sfrate}), $\beta$ is -1.21, while for non-AGN host galaxies not 
detected by  $Spitzer$ the slope is substantially steeper (-1.67).

These results are comparable to the findings of earlier studies.
Buitrago et al. (2008) show for massive ($M_\star \geq 10^{11}$ $M_\odot$) 
galaxies over $z=0-3$ that $\beta$ varies from -0.8 for $n<2$ disk-like galaxies 
to -1.5 for $n>2$ spheroidal galaxies. Williams et al. (2010) find $\beta$ is -0.88 
for all massive ($M_\star \geq 6.3\times 10^{10} M_\odot$) galaxies over $z=0.5-2$.
van Dokkum et al. (2010) find a slope
of -1.27 for massive ($M_\star \geq 10^{11}$ $M_\odot$) galaxies over z=0-2,
which is a good match to our slope (-1.30) for massive 
($M_\star \geq 5 \times 10^{10}$ $M_\odot$) galaxies of all $n$ 
over $z=0-3$.
Compared to massive $z\sim0$ galaxies, the implied mean size evolution is
a factor of $\sim4$ from $z=2-3$ and a factor of $\sim3$ from $z=1-2$.
In order to determine whether this apparent size evolution 
is real,  one needs to address a number of systematic effects, as
outlined in the next section.

\subsection{Impact of Systematic Effects on Structural Properties }\label{sstr}
In the previous section we found that the massive galaxies at  $z=2-3$
 are strikingly offset toward lower ($n$, $r_e$) compared to the
massive $\sim$~0 galaxies.
It is relevant to ask whether the large fraction of low  ($n$, $r_e$)
systems we observe among massive galaxies at $z=2-3$, compared to
massive galaxies at $z\sim$~0 is real or due to a number of systematic
effects. We address the most important effects in the main text and include
the others in Appendix \ref{appendix1}. We consider the issues listed below:

\begin{enumerate}
\item 
Is it possible that  the distribution of ($n$, $r_e$) for massive
galaxies at  $z \sim 0$  and at $z=2-3$ is intrinsically similar,
but  that  some selection effects at $z=2.5$ is  making us preferentially
detect  the compact low $n$ systems, thereby causing an artificial excess
of the latter?
We argue that this is very unlikely  because even if we take all the
massive compact low $n$  systems at $z \sim 0$, and appropriately scale them
for the difference in number density between $z\sim 0$ and $z=2.5$, we still
would fall way short of reproducing the observed number densities of compact
low $n$ systems.  The number density of massive  ($M_\star \geq 1\times 10^{11}$
$M_\odot$) galaxies at $z=2.5$ is approximately 30\% that at $z\sim 0$. If we take
the most compact ($r_e \le 2$~kpc) and low $n\leq2$ systems at $z\sim~0$,  and scale
this number by 30\%, we find a much lower number density ($2.8\times10^{-6}$ 
gal Mpc$^{-3}$), than the observed no density ($5.0\times10^{-5}$ gal 
Mpc$^{-3}$) at $z=2.5$ for such compact systems.

\item
 Can redshift-dependent systematic effects cause structural parameters,
   such as the high S\'ersic index $n$ of  massive galaxies
   at $z\sim$~0,  to `degrade' into the regime of low $n\le 2$
    values, measured in  the $z=2-3$ systems.  We address this
    issue in $\S$\ref{redshift}.

\item
How robust are our fits to the NIC3/F610W images of the
   $z=2-3$ galaxies?  Could some
   of the galaxies with a best-fit  S\'ersic index $n\leq2$ have
   similarly good fits with much higher $n$? We show in
   Appendix \ref{decompcoupling} and Appendix \ref{smonte} that this is unlikely.  
   We are confident that the fraction of $n\le2$ systems is not being
   overestimated.

\item
  Can the offset in ($n$, $r_e$) between the $z=2-3$ galaxies
   and the $z\sim$~0  galaxies be caused by systematic differences
   between the fitting techniques applied by us to the NIC3/F610W images 
   of $z=2-3$ galaxies and the fitting techniques used by  Allen
   et al. (2006) on the $B$-band images of the massive  $z\sim$~0
   galaxies in MGC?  We conduct additional tests (see  Appendix \ref{smgcts}) 
   and conclude that this is also not  the case. 

\end{enumerate}

\subsubsection{Artificial Redshifting}\label{redshift}

We next investigate whether redshift-dependent systematic effects could
potentially cause  the offset in  ($n$, $r_e$) shown in  Figure~\ref{fnre}
between massive galaxies at  $z\sim$~0 and  $z=2-3$, by causing
the  ($n$, $r_e$)  of  massive $z\sim$~0 galaxies to  `degrade' into the
regime of low $n\le 2$ and low $r_e$ exhibited by  the $z=2-3$ systems.

Ideally one would investigate this question by artificially redshifting
the entire MGC subsample of 385 massive $z\sim$~0 galaxies shown in
Figure~\ref{fnre} out to $z\sim2.5$, and re-decomposing the redshifted
galaxies. However, this is  extremely time consuming, and, furthermore,
many of the galaxies do not have high quality SDSS  $ugriz$ images
which are needed for redshifting software (FERENGI; Barden et al.
2008) to work.  
We therefore decide to  artificially redshift a smaller, but 
{\it representative} sample S1 of 255 galaxies.  S1 consists of 42 massive 
($M_\star \geq 5 \times 10^{10}$ $M_\odot$) MGC galaxies combined with 213 
nearby ($z<0.05$) massive galaxies having high quality and well-resolved 
SDSS imaging. We ensure the  ($n$, $r_e$) of the 255 galaxies in S1 match 
those of the entire subsample of MGC galaxies shown in Table~\ref{tabmgc}, 
Figure~\ref{fnre2}, and Figure~\ref{redshifthisto}.
We also ensure that the distribution of Hubble types
of sample S1 matches those of the MGC subsample; the MGC subsample contains
$\sim66\%$ E/S0 galaxies versus $\sim34\%$ Spirals, while sample S1 is 
$\sim64\%$ E/S0 galaxies and $\sim36\%$ Spirals.
Many of the galaxies in S1 are well studied and include
E, S0, and Sabc galaxies from Barden et al. (2008),
E galaxies in Kormendy et al. (2009), as well as S0s and bulge-dominated
spirals from Eskridge et al. (2002).

We used FERENGI (Barden et al. 2008) to artificially redshift the
SDSS  $ugriz$  images (tracing rest-frame UV-to-optical light)
of $z\sim 0$ galaxies,  out to $z=2.5$, and re-observe
them  with the NIC3 F160W filter to the same depth as the GNS survey.
During this process, FERENGI  mimics the effects of surface brightness 
dimming, instrumental resolution, transmission efficiency, and PSF effects.  
It also  corrects for other  geometrical effects of cosmological redshift by
appropriately re-binning input images for the desired redshift  and platescale. 

Specifically,  during artificial redshifting, as is  standard convention,  
FERENGI  assumes surface brightness dimming at the rate of
$(1+z)^{-4}$  for the  bolometric luminosity of the full redshifted 
rest-frame optical SED.  For galaxies where only part of this
redshifted  rest-frame optical SED falls within the NIC3/F160W 
filter bandwidth, 
the  observed  flux per unit wavelength 
$f_{\lambda}$ relates to the rest-frame  luminosity per unit
wavelength  at redshift   $z$ via  a $(1+z)^{-3}$  dependence 
 (e.g., Weedman 1986).  
The exact surface brightness dimming  in such a case will be set
by the integral of  $f_{\lambda}$ over  the filter-detector  response
function and  depends on the  detailed  shape of the 
SED   (e.g.,  Hogg 1999;  Hogg et al.  2002). In practice, when 
using  the  FERENGI software,  the relevant degree of surface 
brightness dimming  is  automatically applied when FERENGI  
convolves the redshifted images with the NIC3 F160W PSF 
and then re-observes the redshifted SED   with the NIC3 
F160W ($H$)  filter-detector, while taking into account the  
filter-detector characteristics, such as  bandwidth and transmission  
efficiency.  
An exposure time of three-orbits  (8063 seconds) and a 
resolution of  $0\farcs2$/pixel is assumed to mimic the GNS survey.
A sky background equal to the mean sky background of the GNS NIC3
images  (0.1 counts/second) was added to the redshifted images.
Poisson noise, sky noise, and read noise (29 $e^-$ for NIC3) were
then added to the  redshifted images.

During artificial redshifting of local galaxies, it is standard
procedure to  incorporate surface brightness evolution  
(Barden et al. 2008)  because galaxies at higher redshifts have
been observed to have higher mean surface brightness 
after applying the standard correction for the geometrical effect 
of  cosmological surface  brightness  dimming.
For instance, Lilly et al. (1998) find that surface brightness 
for disk-dominated galaxies of similar properties
increases on average by 0.8 magnitudes by $z=0.7$. 
Barden et al. (2005) find from the GEMS ACS survey that 
galaxies with $M_V \lesssim -20$ show a 
brightening of $\sim1$ magnitude in rest-frame $V$-band by $z\sim1$. 
Labb{\'e} et al. (2003) find a disk-like galaxy with spectroscopic redshift 
$z=2.03$ to have a rest-frame $B$-band surface brightness $\sim2$ 
magnitudes brighter than nearby galaxies.  
Finally in our own study, the mean surface brightness within $r_e$ of massive
galaxies at $z=2-3$ is 
2 to 6 magnitudes higher than that of massive galaxies at $z\sim$~0,
with a mean offset of $\sim4.5$ magnitudes (Figure~\ref{figsbr}).

In our experiment of artificially redshifting  massive galaxies from 
$z\sim$~0 to $z=2.5$,  we applied a conservative value of 2.5 magnitudes 
of surface brightness evolution. This value is motivated by several
considerations:
a)  2.5 magnitudes of surface brightness evolution  is on the conservative
     side as many of the massive galaxies at $z=2.5$ show even more
     evolution  (Figure~\ref{figsbr}).  Thus, using this value will not 
     lead to overoptimistic recovery of faint features during the 
     experiment;
b)  The adopted 2.5 magnitudes of evolution out to $z=2.5$  corresponds
     to one magnitude of brightening per unit redshift. This rate of
     brightening  is comparable to those seen in studies out to $z\sim2$ 
     (Lilly et al. 1998; Barden et al. 2005; Labb{\'e} et al. 2003);
c) Using the Bruzual \& Charlot (2003) models, one can show
    that the passive evolution of a single stellar population
    from $z=2.5$ to $z=0$, assuming an exponentially 
    declining star formation history  associated with
     an e-folding time of 100 Myr, will lead  the rest-frame
    $B$ luminosity to decline  by 2.5 to 3 magnitudes, depending
    on the chosen metallicity.  

While we believe that 2.5 magnitudes  of surface brightness evolution is 
a conservative and reasonable value to use during the artificial 
redshfiting experiment,  for the sake of completeness, we have also tested
the effect of applying a surface brightness evolution (brightening)
of 0, 1.25, 2.5, and 3.75 magnitudes  between $z \sim0$ and $z=2.5$.
There is a discernible difference in the recovered morphology and structural
parameters between 0 and 1.25 magnitudes of brightening, but less difference
between 1.25, 2.5, or 3.75 magnitudes of brightening.  More details on 
the use of zero  surface brightness evolution  are given in point 4 at
the end of this section.

After artificially redshifting S1 from $z\sim 0$ to
$z=2.5$, we fit both the original galaxy images and their redshifted
counterparts with single S\'ersic profiles.
We compare the rest-frame optical structural parameters
in the original and redshifted images
in order to assess the  influence of redshift-dependent systematic effects
(e.g., surface brightness dimming, loss of spatial resolution) and see
how well the structural parameters are recovered. 
We also compare the
redshifted distribution of ($n$, $r_e$) to the one actually observed
in the GNS massive galaxies to assess whether they are similar.
Note that  the structural parameters are measured at $z\sim$~0
from $g$-band images, while at $z=2.5$ they are measured from the
artificially redshifted images in the NIC3/F160W band so that
all  parameters are measured  in the rest-frame blue optical
light, thereby avoiding bandpass shifting problems.
Our main results are outlined below.

\begin{enumerate}
\item
Figure~\ref{fnre2} shows the  ($n$, $r_e$) distribution obtained by redshifting 
the sample S1 (magenta points in row 1) of 255 $z\sim 0$ massive galaxies 
to $z\sim$~2.5 (blue points in row 2). This redshifted distribution of  
($n$, $r_e$)  is still significantly offset from those observed in the massive 
GNS galaxies at $z=2-3$ (red points in row 2).

This difference is shown more quantitatively in Figure~\ref{redshifthisto}
where results in discrete bins of $n$ and $r_e$ are compared. The massive 
galaxies at $z=2-3$ (red line) includes $64.9 \pm 5.4$\% of systems with low 
$n\le 2$, while the corresponding fraction for the redshifted sample 
(blue line) is $10.6 \pm 1.9$\%. Similarly, for the $r_e$ distribution
of the massive galaxies at $z=2-3$, $39.0 \pm 5.6$\% have
$r_e \le 2$~kpc, while the redshifted sample has $1.2\pm 0.7$\%.
We therefore conclude that cosmological and instrumental effects
are not able to account for the large offset  shown in
Figure~\ref{fnre2} and  Figure~\ref{redshifthisto}  between
the ($n$, $r_e$) distributions  of the massive galaxies at $z=2-3$
and those at $z\sim$~0.

\item
It is very interesting to look at how the structural parameters
of galaxies of different morphological types change during the
redshifting.
Figure \ref{redshift1x1earl}
compares the rest-frame optical structural parameters  in massive
E, S0, and spirals at $z\sim$~0  to
the structural parameters  recovered after these galaxies were
artificially redshifted.

From Figure \ref{redshift1x1earl}, one can see that $r_e$ is
recovered to better than a factor of 1.5 for the vast majority of
redshifted E/S0 and spirals of early-to-late Hubble types.
In the case of a small fraction of $z\sim$~0 galaxies with highly 
extended halos or disks and associated large $r_e$, the recovered 
$r_e$ at $z=2.5$ can be nearly a factor of
two lower than the original $r_e$   at $z\sim$~0.  Inspection of the
surface brightness profiles shows that this effect primarily happens
because surface brightness dimming prevents the outer lower surface
brightness components of the galaxies from being adequately recovered
after redshifting.   

It is striking that even after redshifting out to $z=2.5$,
practically none of the massive $z\sim 0$
galaxies fall into the regime of $r_e \le 2$~kpc  (shown as shaded areas)
inhabited by  the ultra-compact systems, which make up $\sim$ 40\% 
($39.0  \pm 5.6$\%  for  $M_\star \geq 5\times 10^{10}$ $M_\odot$ and
$39.0  \pm 7.6$\%  for  $M_\star \geq 1\times 10^{11}$ $M_\odot$)
of the massive  galaxies   at $z = 2-3$ (see $\S$\ref{sressp}).
{\it Thus, these massive ultra-compact ($r_e \le 2$~kpc) systems at
  $z=2-3$ appear to truly have no  analogs among $z \sim 0$ massive
galaxies, in terms of their size, structure, and optical surface brightness.}

The top row of Figure \ref{redshift1x1earl}  shows the distribution of
S\'ersic index $n$  before and  after redshifting out to $z=2.5$.
The recovered  S\'ersic index $n$ can be lower or higher than the
original $n$  at $z\sim 0$,  but is recovered to better than a factor
of two in all cases.
The shaded area in the plots  represents the regime of $n\leq2$ where
the majority
($64.9 \pm 5.4 \%$   for  $M_\star \geq 5\times 10^{10}$ $M_\odot$ and
$58.5 \pm 7.7$\%  for  $M_\star \geq 1\times 10^{11}$ $M_\odot$)
of massive  GNS galaxies at $z=2-3$  lie (Table~\ref{tabmgc}).
It is interesting to note that massive E and S0s, which are
spheroid-dominated and bulge-dominated systems, do not typically
lie in the $n\leq2$ regime,  before or after redshifting.
In contrast, a large fraction  of $z\sim 0$   spirals with intermediate-to-late
Hubble types\footnote{The Hubble types are based on the bulge-to-total light ratio ($B/T$), which
we  measured with bulge-disk and bulge-disk-bar decomposition
of $z\sim$~0  $g$-band images.}  populate the  $n\leq2$ regime,
both before and after redshifting.
Disk features on  large and small scales (e.g., outer disk or disky pseudobulge)
lead to an overall  single S\'ersic index $n\leq 2$  for the entire galaxy.
{\it It is possible that similar disk features are responsible at least
in part, for the low $n \le 2$  values shown by the majority
($\sim65\%$)  of the massive GNS galaxies at $z=2-3$.}
We discuss this  point further in \S\ref{discuss}.

\item 
One important question is whether the use of deeper images of the 
$z\sim$~0 galaxies would change the conclusion of the redshfiting 
experiment.    
In the present experiment, we used SDSS $g$-band images, which 
have an exposure time of 54 seconds and a typical  sky background of 
22 mag/arcsec$^2$.  
Deeper exposures of nearby galaxies may potentially detect 
an outer low surface brightness  halo (if such a halo exists), 
which is missed in the  SDSS images, and  in that case 
lead us to measure {\it  larger} ($n$, $r_e$)  at {\it $z\sim$~0}    
with a S\'ersic fit.   Such halos can be found in very  local massive
elliptical and cD galaxies, where the measured ($n$, $r_e$) can increase
significantly  if the halo is included in the fit (Kormendy et al. 2009).
However, such low  surface brightness  halos will be dimmed
out and not recovered during  the artificial redshifting of these 
deep images,  so that  the ($n$, $r_e$) parameters  recovered 
at $z=2.5$  will be similar to those we presently obtain
from the SDSS images.
The net effect will be that using deeper images of local massive
galaxies during the artificial redshifting  will at most raise the 
($n$, $r_e$)  at {\it $z\sim$~0}, but not at $z=2.5$. Thus 
the   difference  in the ($n$, $r_e$) at $z\sim0$ compared $z=2.5$
will be unchanged  (for systems without halos) or amplified (for
systems with  such halos).  Our overall conclusion from 
the redshifting experiment regarding degradation of the profiles to
$n\le 2$ and $r_e\le 2$ kpc would remain unchanged or be even 
stronger.

\item
Finally, as one additional  test, we repeated the redshifting experiment 
assuming zero surface brightness evolution, rather than 2.5 magnitudes of 
brightening, out to $z=2.5$. Even in this case there is still a large offset 
in the ($n$, $r_e$) distributions of the redshifted sample S1 compared to the 
massive GNS galaxies. Specifically, the fraction of systems with low $n\le 2$ 
($22.0 \pm 2.6$\%) is still significantly less than that for massive GNS 
galaxies at $z=2-3$ ($64.9 \pm 5.4$\%).  Likewise, there are still few systems 
with $r_e \le 2$~kpc ($1.6 \pm 0.8$\%) compared to the high fraction 
($39.0 \pm 5.6$\%) found at $z=2-3$. Thus, even without surface brightness 
evolution it is still true that cosmological and instrumental effects
are not able to account for the large offset between massive galaxies at 
$z=2-3$ versus $z\sim$~0.
\end{enumerate}

\section{Star Formation Activity}\label{sfrate}
\subsection{Matching GNS Galaxies to MIPS 24~$\mu$m Counterparts}\label{sfmatch}

The $Spitzer$ GOODS Legacy Program (Dickinson et al. 2003a; Dickinson et al. in 
preparation)
provides deep $Spitzer$ MIPS 24~$\mu$m observations of the GOODS fields.
In the discussion below, we only consider  MIPS
24~$\mu$m  counterparts with  $f_{24\mu m} \geq 30$ $\mu$Jy, 
the $5\sigma$ flux limit. The MIPS images have a PSF diameter of
$6\arcsec$  ($\sim$~42 kpc at $z=1-3$), versus the NIC3/F160W PSF of  $0\farcs3$.
MIPS 24~$\mu$m  counterparts of the massive GNS galaxies were identified by selecting
the closest MIPS 24~$\mu$m source within a maximum matching radius of $1\farcs5$.
We initially find  84/166 massive GNS galaxies with MIPS 24~$\mu$m  
counterparts with $f_{24\mu m} \geq 30$ $\mu$Jy and  further refine 
these matches below.

There are several potential problems with the above procedure.
Firstly,  it allows for the situation 
where a given MIPS 24~$\mu$m source could
be matched  to several massive GNS galaxies. This would happen if some 
massive GNS galaxies were crowded within a radius of a few arcseconds so that
the MIPS source would be within $1\farcs5$ of all of them.
This situation occurs for 2/84 ($\sim2\%$) of
massive galaxies with a MIPS counterpart.  We reject these two cases,
reducing the number of unique and secure matches from 84 to 82.

A second possible caveat is that within the large MIPS 24~$\mu$m PSF of $6\arcsec$ 
diameter, there may be several other NIC3/F160W sources,  in addition to the main
massive GNS galaxy to which the MIPS source is matched.  These extra NIC3/F160W
sources may even be  lower mass galaxies 
not in our sample  of massive  ($M_\star  \ge  5\times 10^{10}$
$M_\odot$)  galaxies.
In such a scenario, all the extra NIC3 sources could potentially contribute to the 
MIPS 24~$\mu$m flux, and  assigning all the 24~$\mu$m flux of the MIPS counterpart 
to the nearest massive GNS galaxy would overestimate the 24~$\mu$m flux of this galaxy.
In order to assess the extent of this potential problem, we proceed as follows.
For the MIPS  24~$\mu$m counterpart assigned previously to each  massive GNS galaxy,
we determine how many extra NIC3/F160W sources with
$M_\star \geq 10^9$ $M_\odot$, in addition to the massive GNS galaxy, lie within a 
circle of  diameter  $6\arcsec$  (i.e., the PSF diameter)
centered on the MIPS source.
Of the 82 massive GNS galaxies with a secure MIPS 24~$\mu$m counterpart,
30 involve cases where there are extra NIC3 sources, along with
the massive GNS galaxy, inside the MIPS PSF diameter.

Next, we estimate the relative expected contributions of the massive GNS galaxy
and the extra NIC3/F160W sources to the overall 24~$\mu$m flux by using
the stellar mass ratio 
of the  main massive GNS galaxy (e.g., $M_{\star 1}$) and of the 
contaminating source (e.g., $M_{\star 2}$),
scaled by a function that takes into account
the different redshifts of the two sources.
Specifically, for the two sources with stellar mass $M_{\star 1}$ and
$M_{\star 2}$, having redshifts $z_1$ and $z_2$ and luminosity distances
$D_{L 1}$ and $D_{L 2}$, the stellar mass ratio $M_{\star 1}/M_{\star 2}$
is scaled by $((1+z_2)\times D_{L 2}^2)/((1+z_1)\times D_{L 1}^2)$.
In 8 of 30 cases, the contribution of the extra NIC3  
contaminating  sources to the overall 24~$\mu$m flux 
is $>20\%$  that of the main GNS galaxy,  and 
spans $\sim40\%$ to $\sim126\%$.
We reject these latter 8 cases rather than try to correct for the 
contamination, which in all cases is distributed across two or more
nearby galaxies.
For the remaining 22 cases,
the contamination by extra NIC3/F160W sources is  $<20\%$
and  we deem that our afore-described procedure of assigning all the
24~$\mu$m flux of the MIPS counterpart to the massive GNS galaxy
is reasonable.

Therefore, in summary,  74/166 ($44.6 \pm 3.9$\%) massive
($M_\star \geq 5\times 10^{10}$ $M_\odot$) GNS
galaxies have a  reliable MIPS 24~$\mu$m  counterpart (with $f_{24\mu m} \geq
30$ $\mu$Jy)  whose entire flux is assigned to the massive GNS galaxy.
In contrast, 82/166 ($49.4 \pm 3.9$\%), massive GNS galaxies do not have a 
reliable MIPS counterpart with $f_{24\mu m} \geq 30$ $\mu$Jy and in these cases
we can only measure upper limits on their SFR.
Table \ref{tabsfrprop} lists the fraction of massive GNS
galaxies with a MIPS 24~$\mu$m  counterpart as  a function of redshift.

\subsection{Star Formation Rates} \label{sfrmeasure}
In order to estimate the SFR, the total IR luminosity  ($L_{\rm IR}$)
over  8--1000~$\mu$m is first estimated from the observed 24~$\mu$m flux 
(corresponding to rest-frame wavelengths of 6--12~$\mu$m over $z=1-3$) by 
using SED templates from Chary \& Elbaz (2001). Using solely 24 $\mu$m flux 
density to measure $L_{\rm IR}$ works well for inferred 
$L_{\rm IR} \lesssim 10^{12}$ $L_\odot$ galaxies at $z\sim2$, but $L_{\rm IR}$ is 
overestimated by a factor of $\sim3$ in more luminous galaxies 
(e.g., Papovich et al. 2007). Early results from Herschel  (e.g.,
Elbaz et al.  2010;  Nordon et al. 2010; D. Lutz, private communication)
suggest that at $z>1.5$, the SFRs extrapolated from 24~$\mu$m fluxes may 
overestimate the true SFR, typically by a factor of 2 to 4 and possibly as 
much as a factor of 10. This overestimate could be due to a rise in the 
strength of PAH features, changes in the SEDs, or AGN contamination  at 
$z>1.5$. Murphy et al. (2009) find that estimates of $L_{\rm IR}$ from 
24~$\mu$m flux density alone are incorrect because the templates used are 
based on local galaxies with smaller PAH equivalent widths than galaxies of 
similar luminosity at high-redshift. We account for this discrepancy by 
making a correction for galaxies with inferred 
$L_{\rm IR} > 6\times 10^{11}$ $L_\odot$ using 
\begin{equation}
\rm{log_{10}(L_{\rm IR}) = 0.59\times \rm{log}_{10}(L_{\rm IR}^{24}) + 4.8}, 
\end{equation}
where
$L_{\rm IR}^{24}$ is the infrared luminosity inferred solely from 24 $\mu$m flux
density (R. Chary, private communication).
The upper-left and upper-right panels of Figure \ref{sfrf24} show the distribution
of 24 $\mu$m flux and the inferred $L_{\rm IR}$.

The obscured star formation rate  can be calculated using the expression 
\begin{equation}
\rm{SFR_{\rm IR}~=~9.8\times 10^{-11} L_{\rm IR}}
\end{equation}
from Bell et al. (2007). This 
calculation is based on a Chabrier IMF (Chabrier 2003) and assumes that the 
infrared emission is radiated by dust that is heated primarily by massive
young stars. Uncertainties in the SFR estimates are a factor of $\sim2$ or
higher for individual galaxies. 

If an AGN is present, then SFR$_{\rm IR}$ only 
gives an upper limit to the true SFR.  In $\S$\ref{agn}, we adopt several 
techniques to identify  AGN candidates in the sample and  estimate the mean
SFR for galaxies with and without a candidate AGN (see Table~\ref{tabsfrprop}).
The upper-right panel of Figure~\ref{sfrf24} shows $L_{\rm IR}$ for AGN and 
non-AGN, and the bottom panels show SFR$_{\rm IR}$.
The AGN candidates dominate the tail of highest $L_{\rm IR}$ and SFR$_{\rm IR}$.
Among the HyLIRGs\footnote{HyLIRGs are defined to have $L_{IR} \geq 10^{13}$ $L_\odot$},
9/11 ($\sim82\%$) turn out to be AGN. 
After excluding the AGN candidates, the mean $L_{\rm IR}$ is a factor of $\sim8$ 
times lower, while the mean  SFR$_{\rm IR}$ is reduced a factor of 
$\sim1.5$ to $\sim2.7$, and the difference rises with redshift 
(Table~\ref{tabsfrprop}).

How do our measurements of SFR$_{\rm IR}$ compare with UV-based SFR derived in other
studies of high-redshift galaxies?  The left panel of Figure~\ref{sfrmassz} plots 
SFR$_{\rm IR}$ versus 
$M_\star$  for the massive ($M_\star \geq 5\times 10^{10}$ $M_\odot$) GNS 
galaxies at $z=2-3$ with 24~$\mu$m flux above the $5\sigma$ limit (30~$\mu$Jy). 
We demonstrate  that the SFR derived at $z=2-3$ for non-AGN are in approximate 
agreement with the UV-based SFR from Daddi et al. (2007). 
Drory \& Alvarez (2008) parameterize SFR as a function of mass and redshift 
for a wide range in stellar mass ($M_\star \sim 10^9 - 10^{12}$ $M_\odot$).
In the right panel of Figure~\ref{sfrmassz},
the black line shows average SFR versus redshift for a 
$5\times 10^{10}$ $M_\odot$ galaxy as calculated by Drory \& Alvarez (2008).  
The mean SFR$_{\rm IR}$ for massive non-AGN GNS galaxies, with SFR$_{\rm IR}$
above the $5\sigma$ limit, are higher by a 
factor of $\sim1.5-4$ over $z=1-3$, with the offset worsening with redshift.
This disagreement with mean
SFR$_{\rm IR}$ is not just a bias caused by the requirement that 
SFR$_{\rm IR}$ exceed the $5\sigma$ limit, which selects the most intense
star-forming systems at each redshift.  Even if the upper limits on SFR$_{\rm IR}$
are included, our SFR$_{\rm IR}$ do not show the same break and flattening
seen at $z\sim2$ by Drory \& Alvarez (2008). Finally,
Bauer et al. (2011) measure dust-corrected UV-based SFR (SFR$_{\rm UV,corr}$)
for galaxies in GNS over $1.5<z<3$.  Among massive 
($M_\star \geq 5\times 10^{10}$ $M_\odot$) galaxies, SFR$_{\rm UV,corr}$ can differ
by as much as a factor of 10, but for higher SFR$_{\rm IR}$ the difference is typically
a factor of $\sim2-3$.   

\subsection{Relation Between Star Formation and Structure}\label{sstmsf} 
Figure \ref{sfrmassn}  shows the distribution of
SFR$_{\rm IR}$  among systems of different $n$.
On the LHS panel,  galaxies with SFR$_{\rm IR}$  below
the 5$\sigma$  detection limit are shown as downward pointing arrows.
The potential AGN candidates identified in
$\S$\ref{agn} are coded separately as $\Sigma_{\rm SFR_{IR}}$ is
likely overestimating the true SFR in the galaxy.
For the histograms on the RHS panel, the y-axis shows
the fraction of massive GNS galaxies in each redshift bin, while
on the x-axis, we plot the actual value of SFR$_{\rm IR}$ for
systems with SFR$_{\rm IR}$ above the 5$\sigma$ detection limit
(indicated by the vertical line), and the upper limit for 
the other systems.

The massive galaxies at $z=1-3$ display several interesting 
relations between their star formation activity and  
structure, as characterized by the   S\'ersic index $n$.
Firstly, among the non-AGN massive ($M_\star \geq 5\times 10^{10}$ 
$M_\odot$) galaxies   at $z=2-3$,  
the fraction of  galaxies with low $n\leq 2$  having 
SFR$_{\rm IR}$ high enough to produce a 24 $\mu$m flux
above the  5$\sigma$ detection limit  
is  (53.4$\pm 10.9$\%), which is significantly higher than the corresponding fraction 
($15.4\pm 10.0\%$)  for systems with $n>2$.
Secondly, among the non-AGN massive ($M_\star \geq 5\times 10^{10}$ $M_\odot$)
galaxies at $z=2-3$ with  SFR$_{\rm IR}$  above the  5$\sigma$
detection limit, 
{\it the majority  (84.6 $\pm 10.0$\%)  have low $n\leq 2$,  
while  none have  $n>4$. }
The corresponding numbers  for  the redshift bin $z=1-2$
are $67.7\pm 8.0$\% and $11.8\pm 5.5$\%, respectively.
Thirdly, the RHS panel of  Figure \ref{sfrmassn}  shows that
the high SFR tail in each redshift bin  is populated primarily by
$n\leq 2$  systems. While the  $n\leq 2$ disky systems have a
wide range of SFR$_{\rm IR}$ (21 to 626 $M_\odot$ yr$^{-1}$ at $z=1-2$, 
53 to 1466 $M_\odot$ yr$^{-1}$ at $z=2-3$), 
{\it they include the systems of the highest SFR at both $z=1-2$ and
  $z=2-3$.}
Thus, the systems with low $n\le2$  seem to be more actively
star-forming than the systems of high $n>3$. 

Most  ($72.0\pm6.3\%$ of  systems with  low  $n\le2$ are extended  ($r_e>2$ kpc) 
so that  a  relation is also expected between SF activity and size.
We thus investigate next the relationship between SFR and effective
radius $r_e$.
The distribution of  SFR$_{\rm IR}$  for different $r_e$ ranges
is shown in Figure \ref{sfrmassr}.
The same convention as for Figure \ref{sfrmassn}  is adopted, with
upper limits being plotted for galaxies with  SFR$_{\rm IR}$  below
the  5$\sigma$ detection limit, and  only non-AGN systems being plotted
on the RHS panel.
We find that among the non-AGN massive ($M_\star \geq 5\times 10^{10}$
$M_\odot$) galaxies at $z=2-3$,  the fraction of  ultra-compact 
($r_e \le 2$ kpc) objects with SFR$_{\rm IR}$ above the 5$\sigma$ detection limit  
is only   $15.0\pm 8.0$\% compared  to the fraction  (32.4 $\pm 8.0$\%) 
for the whole sample. Thus, among non-AGN massive galaxies over  $z=2-3$, 
{\it the ultra-compact ($r_e\le2$ kpc) galaxies show a deficiency 
by a factor of $\sim 2.2$ of systems with  SFR$_{\rm IR}$ 
above the  detection limit,   compared to the whole sample.}
At $z=1-2$, the deficiency is a factor of $\sim 3.5$.
Furthermore,  as illustrated by the RHS panel of Figure \ref{sfrmassr}, 
although there are some  ultra-compact ($r_e\le2$ kpc) galaxies with
high SFR$_{\rm IR}$, on average,  the mean SFR$_{\rm IR}$  of  the
$z=2-3$  and   $z=1-2$ is  significantly lower than that of more
extended galaxies.   

\section{Constraints On Cold Gas Content}\label{scold}
The high estimated SFR$_{\rm IR}$ found in \S\ref{sfrate} suggest that
copious cold gas reservoirs are present to fuel the star formation.
For the massive GNS galaxies with SFR$_{\rm IR}$ measurements above the
5$\sigma$ detection limit, 
we assume half of SFR$_{\rm IR}$ lies within the circularized rest-frame 
optical half-light radius $\left ( r_c = r_e \times \sqrt{b/a} \right )$
from  single  component S\'ersic fits, and thereby estimate that the
deprojected SFR per unit area as
\begin{equation}
\Sigma_{\rm SFR_{I\rm R}} = \frac{0.5 \times \rm SFR_{IR}}{\pi \times r_c^2}.
\end{equation}
In galaxies that AGN host candidates, $\Sigma_{\rm SFR_{IR}}$ is 
likely overestimating the true SFR in the galaxy (see $\S$\ref{sfrate}).
If   potential AGN candidates are included, 
$\Sigma_{\rm SFR_{IR}}$ ranges from 
$\sim 0.10 - 360.8$ $M_{\odot}$ yr$^{-1}$ kpc$^{-2}$, with a
mean value of  $\sim19.4$ $M_{\odot}$ yr$^{-1}$ kpc$^{-2}$ over $z=1-3$.
After  excluding the potential AGN candidates  
$\Sigma_{\rm SFR_{IR}}$ ranges from 
$\sim 0.24 -360.8 $ $M_{\odot}$ yr$^{-1}$ kpc$^{-2}$, with a
mean value of  $\sim 14.8$ $M_{\odot}$ yr$^{-1}$ kpc$^{-2}$.
This range is comparable to that seen in BzK/normal galaxies,
ULIRGS, and submillimeter galaxies (e.g., see Daddi et al. 
2010b).


We use a standard Schmidt-Kennicutt law (Kennicutt 1998), with a power-law 
index of 1.4  and a normalization factor of $2.5\times10^{-4}$,
to estimate the cold gas surface density from $\Sigma_{\rm SFR_{IR}}$.
The results are uncertain by at least a factor of $\sim2.5$ because
different relations between molecular gas surface density and SFR surface
density have been  suggested for  various types of star-forming systems
over a broad range of redshifts (Kennicutt 2008;  
Gnedin \& Kravtsov  2010; Daddi et al. 2010b; Genzel et al. 2010; 
Tacconi et al. 2010).  If  potential AGN candidates are included, the resulting 
implied cold gas surface density  
\begin{equation}
\Sigma_{\rm gas} = \left [ \frac{10^4\times \Sigma_{\rm SFR_{IR}}}{2.5} \right ]^{1/1.4}
\end{equation}
ranges from  $\sim 73 - 25,091$ $M_{\odot}$~pc$^{-2}$,
with a median value of $\sim907$ $M_{\odot}$~pc$^{-2}$ 
over $z=1-3$ (Figure~\ref{figfgas}).
The corresponding values after excluding AGN candidates are
$\sim 136- 25091 $ $M_{\odot}$~pc$^{-2}$, 
with a median value of $\sim 607 $ $M_{\odot}$~pc$^{-2}$ 
(Figure~\ref{figfgas}).
These values are again comparable to those observed in BzK/normal galaxies,
ULIRGS, and submillimeter galaxies (e.g., see Daddi et al. 
2010b).

In the subsequent discussion,  we only cite values obtained after  
excluding AGN candidates, but Figure~\ref{figfgas} also shows  
the values for the full sample of galaxies.
Next we estimate the cold gas fraction relative to the baryonic mass
within $r_c$.
For each galaxy, we use the above cold gas surface density
to estimate the total cold gas mass
within the circularized rest-frame optical half-light radius,
\begin{equation}
M_{\rm gas}(r_c) = \Sigma_{\rm gas} \times \pi \times r_c^2. 
\end{equation}
$M_{\rm gas}$ ranges from
$3.4\times10^9 - 1.0\times 10^{11}$~$M_{\odot}$,
with a mean value of $1.9\times10^{10}$ $M_{\odot}$ (Figure~\ref{figfgas}).
The  baryonic mass ($M_{\rm Baryon}$) within $r_c$  is taken to be the sum of
cold gas mass and stellar mass within $r_c$, and we assume that the latter term is
half of the total stellar mass of the galaxy.

The cold gas fraction    ($f_{\rm gas}(r_c)$) 
within the circularized rest-frame optical half-light radius
$r_c$  is   defined as  
\begin{equation}
f_{\rm gas}(r_c) \equiv M_{\rm gas}/\left [M_{\rm gas}+M_\star \right ].
\end{equation}
The cold gas fraction ($f_{\rm gas}(r_c)$) ranges from  $6.5 - 65.4$\%,  
with a mean of $\sim23\%$ over $z=1-3$ (Figure~\ref{figfgas}).
Figure~\ref{figfgas-z} shows how  $f_{\rm gas}(r_c)$ varies as a
function of  stellar mass and redshift, both with and without the AGN candidates.
For  galaxies with $5\times 10^{10}$ $M_\odot \leq M_\star < 10^{11}$ $M_\odot$ 
above the 5$\sigma$ detection limit,  the mean    $f_{\rm gas}(r_c)$ 
(without AGN candidates) rises from  $\sim 19\%$ to $\sim 25\%$ to $\sim 41\%$ 
across the three redshift bins.  In comparison, for $M_\star \geq 10^{11}$ $M_\odot$ 
galaxies, the mean cold gas fraction is $\sim 14\%$ to $\sim 23\%$.  The $1\sigma$ 
error bars are large and there is considerable overlap between the two mass ranges.
{\it Still, the highest cold gas fractions within the circularized rest-frame optical 
half-light radius at a given redshift are found among the less massive galaxies,} 
consistent with downsizing.

Our inferred cold gas fractions  ($f_{\rm gas}(r_c)$) within the
circularized rest-frame  optical half-light radius  $r_c$  
can be higher or  lower  than the  total cold gas fraction of the
galaxy, depending on whether the molecular gas is centrally
concentrated or extended, respectively.   While bearing this
caveat in mind, we note that  our inferred values for   
$f_{\rm gas}(r_c)$ are consistent with previous direct measurements 
of the total cold gas fraction at  high-redshift.
Daddi et al. (2008, 2010a) report gas fractions of 50-65\% in massive
($M_\star \sim 4\times 10^{10} - 1\times10^{11}$ $M_\odot$)
IR-selected $BzK$ galaxies at $z\sim1.5$.
Tacconi et al. (2010) also measure cold gas fraction from CO observations
of high-redshift galaxies at $z=1.1-2.4$. 
For stellar masses spanning $M_\star \sim 3\times 10^{10} - 3.4\times10^{11}$
$M_\odot$, they find cold gas fractions in the range of $\sim14-78$\%.

\section{AGN in massive galaxies at $z=1-3$}\label{agn}
\subsection{Frequency of AGN}\label{agnsummary}
We use  a variety of techniques (X-ray properties, IR power-law, IR-to-optical excess, 
and mid-IR colors) to identify Active Galactic Nuclei (AGN) among the massive GNS 
galaxies because selection based on X-ray emission alone may fail at high redshift 
in the case of Compton-thick AGN where much of the soft X-ray emission is Compton 
scattered or absorbed by thick columns of gas ($N_H \gg 10^{24}$~cm$^{-2}$; 
Brandt et al. 2006). 
We briefly summarize here and in Table~\ref{tabagnfrak} the number of AGN 
identified by each of the selection methods\footnote{The mid-IR selection criteria of 
Lacy et al. (2004) and Stern et al. (2005) were investigated but considered unreliable. 
Contamination from high-redshift star-forming galaxies drastically reduces their 
accuracy (e.g., Donley et al. 2008).  Applying these methods at $z=1-3$ would 
add more false-positives than true AGN.}.

\begin{enumerate}
\item
X-ray counterparts to the massive GNS sources were searched for in the CDF-N 
and CDF-S catalogs of Alexander et al. (2003) and Luo et al. (2008), as well 
as the ECDF-S catalogs of Lehmer et al. (2005). A total of 33/166 massive GNS 
galaxies had counterparts within $1\farcs5$ across all catalogs.  

\item
Following Alonso-Herrero et al. (2006) and Donley et al. (2008), we look for 
AGN power-law emission over  $z=1-3$ using SEDs from the IRAC bands at 3.6, 
4.5, 5.8, and 8.0 $\mu$m.  The IRAC SEDs were fit with a power-law SED
($f_\nu \propto \nu^{\alpha}$). There are only 3/166 sources 
with power-law index $\alpha \le -0.5$ that are considered power-law galaxies 
(PLGs) and obscured AGN candidates. 

\item
Heavily obscured AGN may be present in highly  reddened, IR-excess galaxies.
Fiore et al. (2008) identify obscured AGN candidates in IR-bright, 
optically faint, red galaxies over $z=1.2-2.6$ using the criteria 
$f_{24\mu m}/f_R \geq 1000$ and $R-K>4.5$. We search for such IR-bright, 
optically faint systems with  $f_{24\mu m}/f_R > 1000$ in our sample of 
massive galaxies. $R$-band flux is determined by linear interpolation between 
the $ACS$ $V$ and  $i$-band fluxes.
We find 25 sources meeting this criteria, of which 16 are new AGN
candidates not identified via the above two methods.
\end{enumerate}

Among the 166 massive GNS galaxies at $z=1-3$, the AGN fraction is 49/166 or 
$29.5 \pm 3.5\%$. When the results are broken down in terms of redshift, the 
AGN fraction rises with redshift, increasing from $17.9 \pm 6.1\%$ at 
$z=1-1.5$ to $40.3 \pm 8.8 \% $ at $z=2-3$. The percentage of 
AGN among all massive GNS galaxies is higher than at $z\sim1$, where it is 
reported that less than 15\% of the total 24~$\mu$m emission at $z <1$ is in 
X-ray luminous AGN (e.g.,  Silva \etal 2004; Bell \etal 2005; 
Franceschini \etal 2005; Brand \etal 2006). 

\subsection{Relation Between AGN Activity and Structure}\label{agnstru}
We summarize the properties of the AGN host candidates and 
discuss their implications in terms host galaxy structure.

Figure \ref{agnnr} shows the single S\'ersic  $n$ versus $r_e$.  
Most ($80.6 \pm 7.9\%$) of the AGN  hosts at $z=2-3$  have $r_e > 2$ kpc and are 
not ultra-compact.  
AGN appear to be found preferentially in the more extended galaxies.
Indeed, at $z=2-3$, the AGN fraction in ultra-compact galaxies is $\sim2.7$ times 
lower than in extended galaxies ($20.0 \pm 16.3\%$ versus $53.2 \pm 10.0\%$).
At $z=1-2$ the deficiency is a factor of $5.6$.
Thus, the ultra-compact galaxies are more quiescent in terms of both AGN activity
and SFR activity (see \S\ref{sfrate}).

Furthermore, a significant fraction of these AGN
($64.6 \pm 10.7\%$) have disky ($n\leq2$) morphologies.
Over half  ($58.2 \pm 11.6\%$) of the AGN candidates are both disky
and not ultra-compact.   Similar statistics apply over $z=1-2$.
The disky nature of AGN hosts at $1.5<z<3$ has been measured previously
by Schawinski et al. (2011). From decomposition of the rest-frame
optical light for 20 AGN imaged with $HST$ WFC3, they measure a mean S\'ersic
index of 2.54 and a mean effective radius of $3.16$ kpc.  Their results 
for ($n$, $r_e$) are consistent with our results for $z=2-3$ in 
Table~\ref{tabagnfrak} and Figure \ref{agnnr}. 
Furthermore, Kocevski et al. (2011, in prep.) find from
visual classification of rest-frame optical morphologies that 
$51.4^{+5.8}_{-5.9}$ of X-ray selected 
AGN ($L_X \sim 10^{42-44}$ erg s$^{-1}$) at $1.5<z<2.5$ reside in galaxies
with visible disks; only $27.4^{+5.8}_{-4.6}$ reside in pure spheroids.

If the disky AGN host candidates  host massive black holes,  
then massive black holes are present in galaxies that
are not dominated by a massive spheroid.
In the local Universe,  nearly all massive galaxies are believed to host a 
central supermassive  black hole  
(Kormendy 1993; Magorrian et al. 1998;  Ferrarese
\& Merritt 2000; Gebhardt et al. 2000; Marconi \& Hunt 2003), and the black
hole mass is tightly  related to the bulge stellar velocity dispersion
(Ferrarese \& Merritt 2000; Gebhardt  et al. 2000). This has led to
the suggestion that the black hole and bulge or spheroid probably grew in tandem
(e.g., Cattaneo \& Bernardi 2003; Hopkins et al. 2006). The presence at $z=2-3$
of luminous and potentially massive black holes in high mass galaxies that do
not seem to have a prominent bulge or spheroid may be at odds with this picture.

\section{Discussion}\label{discuss}
\subsection{Do Massive Galaxies With $n\le2$ at $z=2-3$ Host Disks?}\label{sdisc0}
We have shown in \S\ref{sressp} that  the majority
($64.9\% \pm 5.4\%$ for $M_\star \geq 5\times 10^{10}$ $M_\odot$, and
$58.5\% \pm 7.7\%$ for $M_\star \geq 10^{11}$ $M_\odot$)  of  massive galaxies
at $z=2-3$ have low $n\le$~2, while the fraction at $z\sim0$ is five
times lower.  
We also demonstrated  via artificial redshifting experiments and extensive 
tests ($\S\ref{sstr}$ and the Appendix) 
that this difference  between $z=2-3$ and $z\sim 0$ is real  and not 
driven primarily by systematic effects.
Furthermore, most ($\sim72\%$) of these with low $n\leq2$  
massive galaxies at $z=2-3$ are extended with  $r_e > 2$ kpc, rather than being
ultra-compact.

What is the nature  of the large population of galaxies with low
$n\leq2$   at $z=2-3$?  We  present below different lines of evidence 
which suggest that many of
these massive galaxies at $z=2-3$ with $n\leq2$, particularly 
the extended ($r_e> 2$~ kpc) systems,  likely   host  a significant 
disk component.

\begin{enumerate}
\item
Some insight into the interpretation of $n\leq2$ values can be gleaned by 
considering  massive galaxies at $z\sim 0$. As discussed in
$\S$\ref{redshift} and illustrated in Figure \ref{redshift1x1earl},
massive E and S0s, which are spheroid-dominated and bulge-dominated
systems, are predominantly associated with $n>2$, both at $z\sim 0$
and after artificially redshifting to $z=2.5$. In contrast, spiral galaxies 
of intermediate to low bulge-to-total ratios, often have an overall low 
S\'ersic index $n\leq2$  (Figure \ref{redshift1x1earl}) because they have a 
disk component, such as an  outer disk or a central disky
pseudobulge (e.g., Kormendy \& Kennicutt 2004; Jogee  1999; Jogee et
al 2005),  which contributes significantly to the total blue light of the galaxy.
An extension of these arguments to $z=2-3$ suggests the large
fraction $\sim65\%$
of massive galaxies 
at $z=2-3$ with low $n\le$~2  is driven, at least partially, by the presence 
of an outer disk or central disky pseudobulge.

\item
We next consider the relationship between disk structure and 
projected ellipticity  $e$. 
The top panels of Figure~\ref{figba}  show the deconvolved ellipticity 
$e = 1-b/a$
determined by GALFIT for the massive
($M_\star \geq 5\times 10^{10}$ $M_\odot$)
galaxies at $z=2-3$ with  $n\le2$ and  $n>2$.
The lower left and right panels of Figure~\ref{figba}
show the distributions of deconvolved  ellipticity  
determined by GIM2D of similarly massive 
spiral   (Sabc and Sd/Irr)   and 
E/S0\footnote{The MGC catalog assigns the 'E/S0'  Hubble type
and  unfortunately does not allow us to identify Es separately.}
galaxies in the  MGC catalog.

The projected ellipticity distribution of 
massive galaxies at $z=2-3$ with $n\le2$  is quite 
different from  that of $z\sim 0$ 
 massive  E/S0 galaxies.
For local  E/S0s, the distribution of $e$ drops sharply at $e>0.35$ 
and there are few systems at $e>0.5$. In contrast,  for the
massive galaxies at $z=2-3$ with $n\le2$, the  $e$  distribution 
continues to rise out to
$e\sim 0.5$. There is also a significant fraction  ($\sim58\%$) of
systems with $n\le2$ having  $e$ above 0.5, specifically 
in the range of 0.5 to 0.75.   In effect, 
a  Kolmogorov-Smirnov (KS) test ( Table~\ref{kstest})  shows
that the  galaxies at $z=2-3$ with $n\le2$  
have a 0\% KS-test probability of coming from the same distribution 
as local massive E/S0s in MGC. 
These comparisons suggest that  {\it  the  massive galaxies 
at $z=2-3$ with $n\le2$  are very  different from local 
bulge-dominated and spheroid-dominated E/S0s.}

Among the massive systems with $n\le2$   at  $z=2-3$, 
$28.0\pm6.4\%$ are  ultra-compact  ($r_e \le 2$~kpc).  Thus, 
our conclusion complements the results of van der Wel et al. (2011) 
who analyze  WFC3 images of 
a small sample of 14 massive ($M_\star \geq 6\times10^{10}$
$M_\odot$), quiescent, and compact  ($r_e \le 2$ kpc) galaxies at
$1.5<z<2.5$ and report  that  most  ($65\pm15\%$) are 
disk-dominated systems
They find that 5 of 14 galaxies are flat in projection and have an
ellipticity  $\geq0.45$. 

What is the nature of the massive galaxies at $z=2-3$ with $n\le2$? 
Figure~\ref{figba}  and the KS tests in Table~\ref{kstest}  show that the 
massive galaxies at $z=2-3$ with $n\le2$  are more 
similar  to $z\sim 0$  massive  Sd/Irr  (KS 
probability of 23.5\% and  $D = 0.317$)  and  to $z\sim 0$  massive 
Sabc  spirals (KS
probability of 4.8\% and  $D = 0.221$)  than to $z\sim 0$  massive E/S0s.
However, the similarity to massive late-type spirals at $z\sim 0$  is
clearly limited, since most massive galaxies at $z=2-3$  with $n\le2$   
have smaller half-light radii  ($r_e$ primarily below 7 kpc;  
Figure~\ref{fnre})  than any of the  $z\sim 0$  massive systems. 
It is possible that they host   less extended  and thicker disks 
than present-day  massive spirals.

Another possibility is that the massive galaxies at $z=2-3$ 
with $n\le2$   might be related to clump-cluster and 
chain galaxies (Cowie et al. 1995;  van den Bergh et al. 1996;
Elmegreen et al. 2005, 2009a, 2009b). Such galaxies very 
often  host disk structures (Elmegreen et al. 2009a), and 
many of them appear  to represent  a population of highly 
clumped disk galaxies viewed at different  orientations (Elmegreen et al. 2008;
Elmegreen et al. 2005).   While clumpy disks may be among the massive GNS galaxies
with low $n\leq2$, we cannot identify them due to resolution effects.
Finally, we note that in principle a low S\'ersic index could be the
result of a merger that has not fully coalesced.   However,  as noted
in \S{\ref{sressp}} most 
massive GNS galaxies do not visually appear  to be made of multiple 
distorted systems in early phases of mergers.  Artificial  redshifting
of present-day interacting  systems  show that our GNS images should 
be able to resolve systems in early phases of merging, such as  NGC4568 
and  NGC 3396,  but would be unlikely to resolve late merger
phases, such as Arp 220 into two separate systems.

\item
Another line of evidence for  massive galaxies at $z\sim 2 $ 
with  potentially  thick disks comes from the SINS survey
 (Genzel et al. 2008; Shapiro et al. 2008;
F{\"o}rster Schreiber et al. 2009), which provides  ionized gas kinematics of
$z\sim2$ star-forming galaxies and finds examples of clumpy, turbulent, and
geometrically thick systems having high velocity dispersions
($\sigma \sim30-120$ km/s). About $\sim 1/3$ of such systems show rotating
disks kinematics. Furthermore,  F{\"o}rster Schreiber et al. (2011)  find from 
$HST$ NIC2 imaging that five star-forming galaxies with
rotating disk kinematics are well-characterized with shallow $n\le1$
S\'ersic profiles.
Compared to these SINS galaxies, the massive GNS galaxies at $z=2-3$ are
more massive on average. 

\item
In this work ($\S$\ref{decomp}), we fitted the NIC3/F160W images of the 
massive galaxies at $z=2-3$  with  single S\'ersic components, rather than 
separate bulge and disk components because the low resolution (PSF FWHM of 
$0\farcs3$ corresponding to $\sim2.4$ kpc at $z=1-3$) of the images prevent 
reliable  multiple component decomposition for all the galaxies, particularly 
the fairly compact ones. However, for the galaxies with large  $r_e \ge 4$~kpc 
we attempted a bulge-plus-disk decomposition following the techniques 
outlined in Weinzirl et al. (2009). The decomposition was reliable only for 
the more extended systems within this group and yielded bulge-to-total light
ratios below 0.4, indeed suggesting the presence of a significant disk
component among  massive galaxies  at $z=2-3$ with $n\le2$.  

\item
It is also interesting to note that most ($\sim$ 72\% for 
$M_\star \geq 5\times 10^{10}$ $M_\odot$) of these 
massive galaxies at $z=2-3$ with low $n\leq2$ are extended ($r_e > 2$~kpc) 
rather than ultra-compact systems. This is in itself does not prove that disk 
components exist in low $n\leq2$ systems, but it is suggestive of
such a picture. 
Furthermore, we found in  $\S$\ref{sstmsf} that at $z=2-3$, 
the $n\leq 2$ disky systems have a wide range of SFR$_{\rm IR}$ and include
systems of the highest SFR$_{\rm IR}$. This result is generally consistent 
with the idea 
that the systems with $n\leq2$ are  actively star-forming and host copious 
amounts of gas (\S\ref{scold}), which tends to settle in disk-like 
configurations.

\item
For completeness, we note that in principle the presence of a 
massive disk component is not the only way to produce a low 
S\'ersic index  $n\leq2$ in massive galaxies
at $z=2-3$.  For the ultra-compact ($r_e \leq 2$ kpc)
massive galaxies with $n\leq2$, it has been argued that such systems
could be somewhat like a massive elliptical, which has a
bright high surface brightness central component surrounded by
a very extended low surface brightness envelope. If the low surface
brightness envelope is somehow not detected by the NIC3/F160W
images, then the latter could yield a lower $n\leq2$, as the
wings of the surface  brightness profile  would be effectively
clipped.     However, this scenario does not seem likely since 
our artificial redshifting experiments ($\S$\ref{redshift})
show  that  $z\sim 0$ massive  Es are not degraded into 
ultra-compact systems.  Furthermore, Szomoru et al. (2010) confirm
the  absence of  a low surface brightness halo in an
ultra-compact, massive galaxy at $z=1.9$
from extremely deep ($H \sim 28$ mag arcsec$^{-2}$)  WFC3 
imaging.

\end{enumerate}

In summary, based on all the above tests and arguments, we 
conclude that the massive galaxies at $z=2-3$ with $n\leq2$, 
particularly the more extended systems with  $r_e> 2$~ kpc, 
likely   host  a massive disk component, which contributes
significantly to the  rest-frame blue light of the galaxies.

\subsection{Formation of Massive Galaxies By $z=2-3$}\label{sdisc1}
How do the massive galaxies with ultra-compact ($r_e \le2$ kpc) and  
low $n\le2$ disky structures form by $z=2-3$?
The surface brightness in the rest-frame
$B$-band of the massive galaxies at $z=2-3$ is on average 4.5 magnitudes
brighter than massive $z\sim0$ galaxies (Figure~\ref{figsbr}).
This implies a large mass surface density of young-to-intermediate-age
stars had to built up in less than a few Gyr.  Implied stellar
mass surface densities exceed  several   $10^{10} M_\odot$~pc$^{-2}$
even for conservative  mass-to-light ratios.
This implies that {\it rapid and  highly dissipative gas-rich events
must have led to the formation of these massive galaxies by
$z=2-3$.}
Both gas accretion and wet major mergers  at $z>2$ are
likely to have played an important role because at such
high redshifts,
the short dynamical timescales associated with mergers,
and the   short cooling time associated with gas accretion
imply that  both mechanisms would lead to
a rapid buildup of  cold gas. The latter can in turn lead to
rapid star formation and
dense stellar remnants (e.g., Wuyts et al. 2009; Wuyts et al. 2010;
Khochfar \& Silk 2011; Bournaud et al. 2011).

A further constraint on the formation  pathway is provided by the 
structure of the massive galaxies at $z=2-3$. We have shown in 
\S\ref{sressp} that  as much as $\sim 65$\% of the massive galaxies 
at $z=2-3$ have a low $n \leq 2$, and we further argued in \S\ref{sdisc0}
that most of  these systems  with $n\leq2$ at $z=2-3$ likely host a 
massive disk component. 
Major mergers of low-to-moderate gas 
fraction (e.g., $\le 30\%$)  will typically produce merger remnants with a 
de Vaucouleurs type profile and a S\'ersic index $n > 3$ 
(Naab, Khochfar, \& Burkert 2006; Naab \& Trujillo 2006).
Mergers with moderate-to-high gas fractions are expected to produce 
lower S\'ersic $n$ that are still in general  $>2$.  
For instance,  Figure~14 of Hopkins et al. (2009) show the S\'ersic 
index of major merger remnants for a range of orbits and a range
of progenitors with gas fractions spanning from 10-100\%.
Although some massive ($M_\star \geq 10^{11} M_\odot$) 
remnants with $n\sim1$ arise in mergers with $f_{\rm{gas}}\geq80\%$,
most  remnants 
of gas-rich ($f_{\rm{gas}}\geq40\%$) mergers  have a  S\'ersic index
$n>2$. 
Furthermore, Rothberg \& Joseph (2004)  find from $K$-band imaging of 
52 merger remnants  that  $\sim 51\%$ (26/51) have $n>3$,  $\sim37\%$ 
(19/51) have $n\sim 2-3$, and only  a small fraction ($\sim12\%$,
6/51) have $n\sim 1-2$. 
Thus, when considering isolated gas-rich major mergers, namely
those not fed by cold streams,  it is  challenging
to produce a population of merger remnants where $\sim65\%$
of the systems have $n\leq2$.

The challenge of producing a large population of disky
($n\leq 2$) systems with high SFRs from isolated gas-rich major mergers
may be an  indication that the accretion of cold gas along cosmological
filaments ((Birnboim \& Dekel 2003; Kere{\v s} et al. 2005; Dekel \& Birnboim 2006;
Dekel et al. 2009a; Dekel et al. 2009b; Kere{\v s} et al. 2005; Kere{\v s} et al. 2009;
Brooks et al. 2009; Ceverino et al. 2010)
may be particularly important in the build-up of massive galaxies by $z=2-3$.
As merger remnants at $z>2$ acquire gas via cold-mode accretion,  a gas disk is
expected to form  (Khochfar \& Silk 2009a; Burkert et al. 2010).
Depending on the angular momentum of the accreted gas, it can settle
into a compact disk component or into an outer extended disk.
Burkert et al. (2010)  discuss a scenario where turbulent
rotating disks  can form,  segregating into compact
($r_e \sim 1-3$ kpc)  dispersion-dominated  ($1 \leq v/\sigma \leq 3$)
systems  and more extended ($r_e \sim 4-8 $ kpc),  rotation-dominated
($v/\sigma >3$) disks.
The formation of a gas disk via cold-mode accretion and its
subsequent conversion into  a stellar disk, would 
lower the overall S\'ersic index of the massive galaxies at $z=2-3$,
making them more in line with the observed values.

However, many key questions remain unanswered.  Can theoretical
models account for  the observed  fractions  of   massive
galaxies with low $n\le2$, as well as the
fraction of galaxies with ultra-compact  ($r_e > 2$~kpc) sizes?
Can the relation between structure, SFR, and AGN activity  discussed
in  \S\ref{sstmsf}  and \S\ref{agnstru}, as well as the range in SFR
at a given stellar mass,  be accounted for?
We will address these questions in a future paper  (Jogee et
al., in preparation) where we  perform detailed comparisons
to different theoretical scenarios.

\subsection{
Transformation of Massive Galaxies at  $z=2-3$ Into Present-Day E and S0s}\label{sdisc2}

Next we discuss the transformation of  massive galaxies at 
$z=2-3$  into their more massive present-day descendants,
which are  primarily E and S0s.
During this transformation, the massive galaxies will need 
to significantly increase  $n$  since 
the majority  ($\sim 65$\%) of  massive
galaxies  at $z=2-3$  have  low $n\le$~2, while the corresponding fraction 
among massive systems at $z\sim$~0 is five times lower 
(Table~\ref{tabmgc} and Figure~\ref{fnre}).    
Similarly, the galaxies will also need to significantly raise
$r_e$,  since approximately 40\% of massive  galaxies  at $z = 2-3$ are in 
the form of ultra-compact ($r_e \le 2$~kpc)  galaxies compared
to less than 1\% at $z\sim$ 0 (Table~\ref{tabmgc} and Figure~\ref{fnre}).    
In general, the massive $z=2-3$ galaxies must experience a substantial
growth in $r_e$ by up to a factor of $\sim6$, a dimming in rest-frame
optical  surface brightness within $r_e$  by up to 6 magnitudes 
 (Figure~\ref{figsbr}),  and their $n$ must increase to $n>2$. 
An  increase in ($n$, $r_e$) and a dimming in  $\mu_e$    can 
be achieved via several pathways.

A natural pathway to produce large changes in ($n$, $r_e$,
$\mu_e$)  is a dry  major merger of two disk systems. This produces 
a  remnant with $n\sim 4$,  a higher  $r_e$,  and a lower
surface brightness within  $r_e$ than the progenitors (Naab, Khochfar, 
\& Burkert 2006; Naab \& Trujillo 2006; Naab et al. 2009).
In this case, the change in $n$ is  produced by the transformation 
of galaxies with disks  into systems dominated by spheroids or
bulges.  This type of transformation must  take place from $z=2-3$ to 
$z\sim 0$  in many of the  massive galaxies  because 
$\sim 65$\% of them  at $z=2-3$ have $n \leq 2$,  which we argued 
is indicative  of a massive disk in many cases (\S\ref{sdisc0}). In 
contrast the E/S0s at $z\sim$0  are dominated by spheroids or bulges.

Other lines of evidence  support the idea that  dry major mergers play
a role in  making the most massive $z\sim$~0 ellipticals.
The most massive local ellipticals are found to
harbor cores (missing light), which are believed to be scoured by binary black
holes that form in dry major mergers  (Kormendy et al. 2009).  
From a study of  the tidal features associated with bulge-dominated early-type
galaxies, van Dokkum (2005)  concludes that today's most luminous ellipticals
form through mergers of gas-poor, bulge-dominated systems.  
Kriek et al. (2008)  focus on massive red-sequence galaxies at $z\sim 2.3$
with little or no ongoing star formation, finding that the changes
in color and number density of galaxies  
on the high-mass end ($M_\star \geq 1 \times 10^{11}$ $M_\odot$)
of the red sequence from $z\sim2.3$ to the present are better
explained by a combination of passive evolution and  red mergers 
that induce little star formation, rather than by passive evolution alone.  

While dry major mergers play a role in the evolution of massive
galaxies,  it remains debated whether they can account for the full
size and mass evolution of massive galaxies. 
From a  theoretical standpoint, the predicted dry major  
merger  rate appears to be too low.  
From simulations, Khochfar \& Silk (2009b) find that 
only between $10\%-20\%$ of massive ($M_\star > 6.3 \times 
10^{10}$ $M_\odot$) galaxies have had a dry major merger in 
the last Gyr at any redshift $z<1$. 
Hopkins et al. (2010) find from semi-empirical models that the 
importance of major mergers in bulge formation scales with galaxy 
stellar mass.  Namely, an $L_*$ galaxy with $M_\star \sim 10^{11}
 M_\odot$ at $z=0$ will experience only one dry major merger at 
$z<2$.
Shankar et al. (2010)  calculate that the frequency of dry 
mergers increases with final stellar mass, and they find that by $z=0$
massive ($M_\star > 10^{11} M_\odot$)  early-type galaxies undergo 
on average $<1$ dry major merger since their formation.   

From an observational  standpoint,  direct measurements of the dry 
major merger rate at $z<1$ are highly uncertain.
Bell et al. (2006) suggest that present-day spheroidal galaxies with 
$M_{\rm V}<-20.5$ on average have 
undergone anywhere between 0.5 and 2 dry major  mergers since 
$z \sim 0.7$. The analysis carries large uncertainties as it is
based on a small number ($\sim 6$) of  observed dry major mergers.  
Several observational studies report that between 16\%  to  35\%
of massive ($M_\star > 2.5 \times 10^{10}$ $M_\odot$) galaxies have
undergone a major merger  since  $z\sim 0.8$  (e.g.,  Jogee 
et al. 2009;   Lotz et al. 2008; Conselice et al. 2009),  but  it
should be noted  that most of the major mergers in the above studies  
are star-forming systems, and  there are very few  dry major mergers.  
Robaina et al. (2010)   find  that  galaxies with  
$M_\star > 1 \times 11^{10}$ $M_\odot$ have undergone, on average, 
only 0.5  mergers since $z\sim$ 0.7 involving progenitor galaxies
that are both more massive than $M_\star > 5 \times 10^{10}$ $M_\odot$.  
Hammer et al. (2009)  focus on starbursts with disturbed  
ionized gas morphologies and kinematics at $z\sim$~0.65, and they
argue based on modeling that $\sim 6$ Gyr ago  46\% of the galaxy 
population was involved in major mergers,  most of  which were gas-rich. 
Kaviraj et al. (2011) find that theoretically and empirically
determined major merger rates at $z<1$ are too low by factors 
of a few to account for the fraction of disturbed  systems they 
find among  morphologically classified  early-type  massive 
($M_\star > 1 \times 10^{10}$ $M_\odot$) galaxies at $0.5<z<0.7$. 
They suggest 
that the overall evolution of massive early type galaxies,
particularly the low-level star formation activity, may be 
heavily influenced by minor merging at late epochs.
At higher redshifts $1<z<2$,  higher major merger   rates are 
reported than at $z<1$  (e.g., Conselice et al. 2003), but the
frequency of dry major mergers is claimed to be low (Williams
et al. 2011).


An alternate pathway that could be at least as important as major
mergers consists of  consecutive dry minor mergers or accretion of externally
formed stars such that stellar mass is cumulatively added to the
outskirts of a compact galaxy (e.g., Naab et al. 2009; Feldman et al. 2010).
Naab \& Trujillo (2006) show that successive minor mergers can, on average,
raise the S\'ersic index of the merger remnant about as effectively as
major mergers. 
Furthermore, it is claimed from simulations and analytical arguments
that dry minor mergers produce a much larger  increase in size 
($r_e$)  and  a larger fall in average stellar mass densities within 
$r_e$   than do dry major mergers  (Naab et al.  2009; Bezanson et
al. 2009).
Shankar et al. (2011) find in simulations that massive
($M_\star \geq 10^{11}$ $M_\odot$) $z\sim0$ galaxies grow primarily 
by dry minor mergers, especially at $z<1$.
Oser et al. (2010; 2011) use cosmological simulations 
to study  40 individual massive galaxies 
with present-day stellar masses of  $M_\star > 6.3 \times 10^{10}$
$M_\odot$.  
They find that massive galaxies at $z>2$ are dominated by
``in situ'' star formation fueled by in-falling cold gas within the galaxy.
As cold-mode accretion becomes inefficient at $z\approx2$,
accretion of externally created stars (i.e., stellar satellites)
dominates at $z<2$.
For galaxies of present-day stellar mass  $M_\star > 6.3 \times
10^{10}$  $M_\odot$,   the average number-weighted merger 
mass-ratio is  $\sim$  1:16,  while the average mass-weighted merger 
mass-ratio is  $\sim$  1:5. In other words,  the  mass growth since $z
\sim2$  is dominated by minor mergers with a mass ratio of 1:5.
The importance of stellar accretion increases with galaxy mass and
toward lower redshift,  and it substantially  raises  the  galaxy
stellar mass  and size.
For systems with present-day stellar mass  $M_\star > 6.3 \times
10^{10}$  $M_\odot$,  a size evolution of up to a factor
of $\sim 5-6$  occurs from $z=2$ to $z\sim 0$.
However,  one strong caveat of these simulations is that all their 
massive ($M_\star > 1 \times 10^{11}$ $M_\odot$) galaxies 
at $z=2$  are ultra-compact ($r_e \leq2$ kpc),  while observations 
(see Fig.~\ref{fnre}) show a large fraction of such massive 
galaxies at $z=2$ are extended  ($r_e = 3-10$ kpc), 
with a wide range in star formation rate.
The increase of size and mass induced by minor mergers in these
simulations is  qualitatively in agreement with  our 
results on size evolution for the ultra-compact systems   
and also with  the  inside-out growth
reported by van Dokkum et al. (2010) from stacking deep rest-frame
$R$-band images of massive galaxies over the redshift range of 0.6
to 2.0.

However, many questions remain unresolved.  While dry minor mergers 
appear to be effective at inducing significant evolution in mass and
size from $z\sim 2$  to $z \sim 0$ in the simulations  of  Oser et
al. (2010; 2011), it is unclear if they can really drive the large
change  in Sersic index $n$ required by the observations. 
Furthermore, these simulations focus only on
ultra-compact  ($r_e \le 2$~kpc) galaxies, and are not representative of 
the large dominant population of more extended galaxies at $z=2-3$.
Finally, it is not clear whether minor mergers can account for the changes
in  effective surface brightness between $z=2-3$ and $z\sim0$.  
We will evaluate these issues more thoroughly with a detailed
comparison to models in a subsequent paper (Jogee et al. in 
preparation).

\section{Summary}\label{summary}
We present a study of the structure, activity, and evolution of 
massive galaxies at $z=1-3$ 
using  deep  
($5\sigma$ limiting magnitude of $H$=26.8 AB for 
an extended source of diameter $0\farcs7$), 
high resolution   (PSF$\sim 0\farcs3$)  NIC3/F160W images  
from the GOODS-NICMOS Survey (GNS),  along with complementary ACS,  
$Spitzer$ IRAC and MIPS, and Chandra X-ray data. 
One of the strengths of our study is that the NIC3/F160W data
provide rest-frame optical imaging over $z=1-3$ for 
one of the largest 
(166 galaxies with $M_\star \geq 5 \times 10^{10}$ $M_\odot$ 
and 82 with $M_\star \geq 10^{11}$ $M_\odot$), 
most diverse, and relatively  unbiased samples of massive
galaxies  at $z=1-3$  studied to date.
Our main results are summarized below.

\begin{figure}[]
\centering
\scalebox{1.00}{\includegraphics{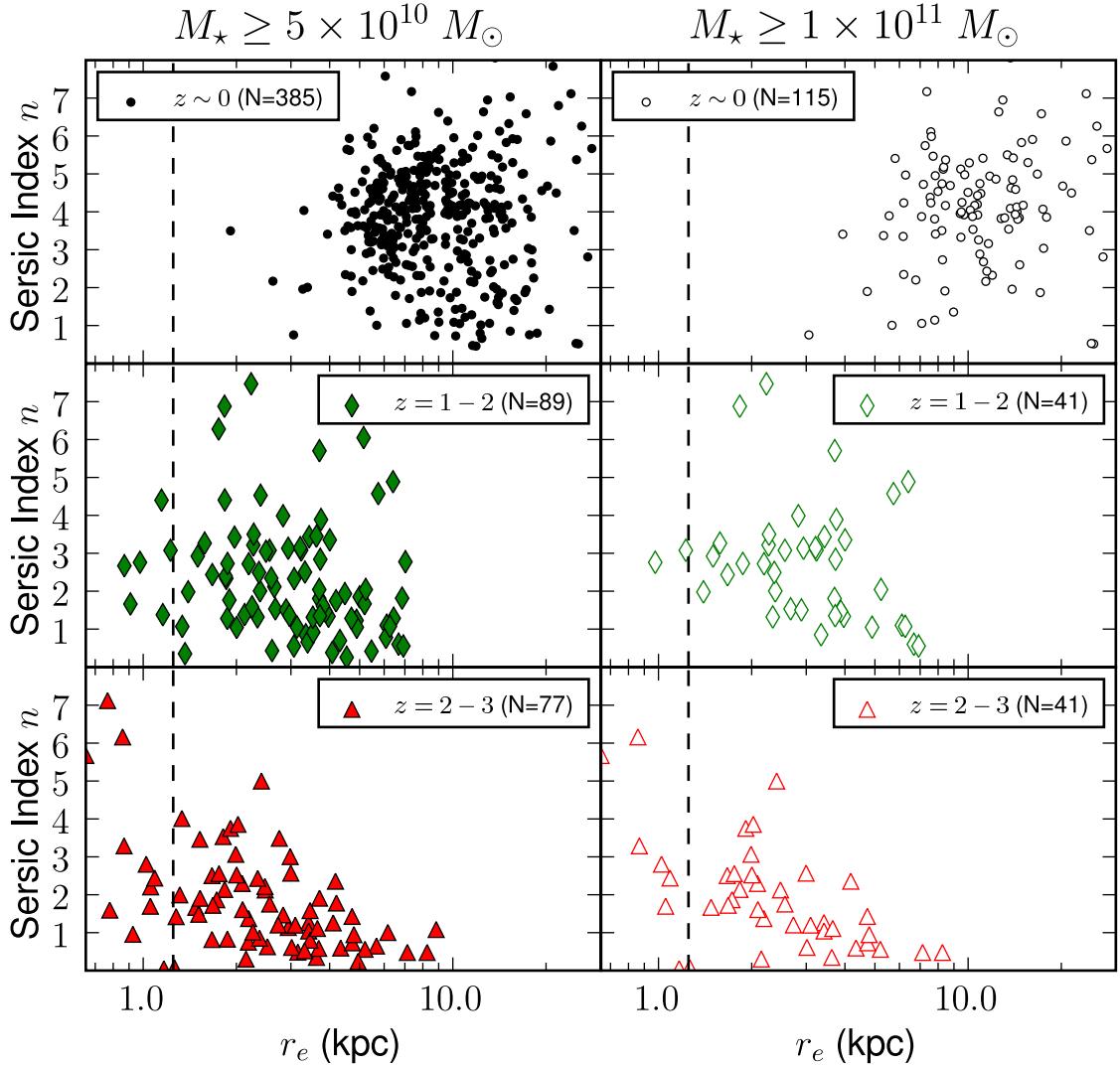}}
\caption{
The $B$-band S\'ersic index $n$ and effective radius $r_e$
derived from  single S\'ersic profile fits to massive
($M_\star \geq 5\times 10^{10}$ $M_\odot$) galaxies
are plotted for the three redshift bins  listed in  Table~\ref{tabmgc}.
In the top row, the black points represent fits to
$z\sim 0$ galaxies by Allen et al. (2006) on $B$-band images of
galaxies from the  Millennium Galaxy Catalog  (Liske et al. 2003).
The lower two rows are based on our fits to the NIC3
F160W images of massive GNS galaxies at $z=1-2$ and  $z=2-3$.
Note that the massive galaxies at  $z=2-3$  are strikingly 
offset toward lower ($n$, $r_e$) compared to the massive 
$z\sim$0 galaxies, and have five times more low $n\le$~2 disky 
systems (see also Figure~\ref{mgchist}).
The black dashed line represents the typical half-width half 
maximum  of the NICMOS3 PSF at $z=1-3$ of $\sim1.2$ kpc. 
\label{fnre}}
\end{figure}

\begin{figure}[]
\centering
\scalebox{0.80}{\includegraphics{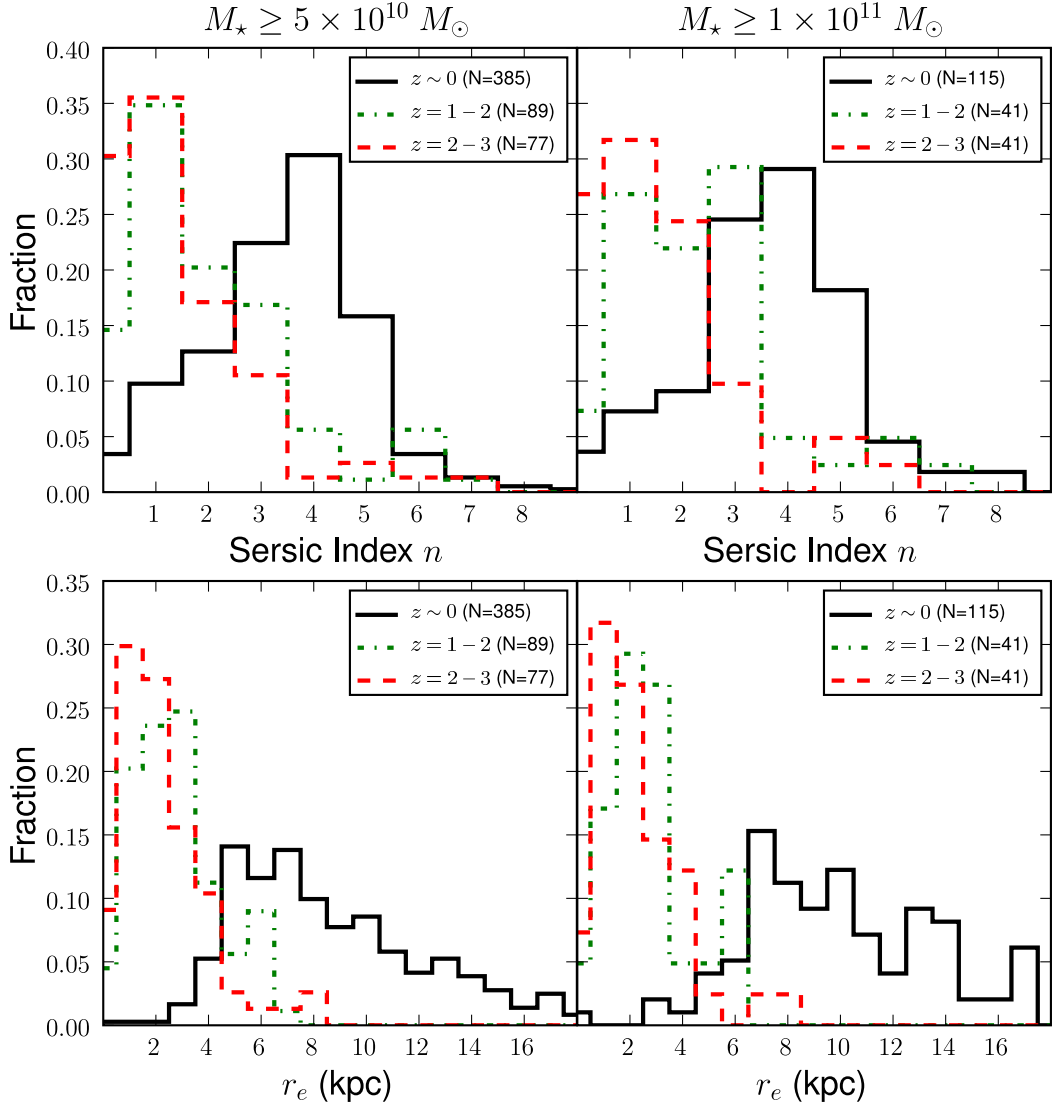}}
\caption{
Left column: The distributions of rest-frame optical  S\'ersic index and effective 
radius $r_e$ based on single S\'ersic profile fits  to massive 
($M_\star \geq 5\times 10^{10}$ $M_\odot$) 
galaxies are plotted for the three  redshift bins  listed in  Table \ref{tabmgc}:  
at $z\sim 0$ (solid line), based on the fits of Allen et al. (2006) on 
$B$-band images of 
galaxies from the  Millennium Galaxy Catalog  (Liske et al. 2003), and 
at $z=1-2$ (dash-dotted line) and  $z=2-3$ (dashed line), based on  our  fits to the NIC3
F160W images of massive GNS galaxies.
Note that a significant fraction ($39.0 \pm 5.56\%$) of 
massive ($M_\star \geq 5\times 10^{10}$ $M_\odot$) 
galaxies at  $z=2-3$ have $r_e \le 2$~kpc, compared to only $0.52 \pm 0.37\%$ 
at $z\sim 0$.  Note also that most ($64.9\pm5.4\%$) of massive galaxies at 
$z=2-3$ have low $n\le2$ (disky) structures compared to only $13.0\pm1.7\%$ at
$z\sim0$. Right column: Same as left column but for the mass range
$M_\star \geq 1\times 10^{11}$ $M_\odot$.
\label{mgchist}}
\end{figure}

\vspace{3mm}
{\it  1.  Structure of massive galaxies at rest-frame optical
  wavelengths: }
We  analyze the rest-frame optical 
structure of the massive galaxies by fitting single S\'ersic 
profiles to the 2D light distribution in the NIC3/F160W images.
We find that 
the rest-frame optical structures of the massive galaxies
are very different at $z=2-3$ compared to $z\sim$~0,
with  their  S\'ersic index $n$ and half-light radius  $r_e$  
being strikingly offset toward
lower values compared to $z\sim$~0.
(Table~\ref{tabmgc} and Figure~\ref{fnre}). 
Through extensive tests and artificial redshifting experiments 
we conclude that the  offset in ($n$, $r_e$) between
massive  galaxies at $z=2-3$ and   $z\sim0$  is real and not 
primarily driven by 
systematic  effects related  to the fitting techniques 
instrumental effects, or  redshift-dependent effects 
(e.g., cosmological surface brightness dimming and the loss of
spatial resolution).  In effect, we find 
{\it a  large population of ultra-compact ($r_e \le 2$~kpc) systems}, 
as well as 
{\it a  dominant population of systems with low $n\le2$ 
disky morphologies} at $z=2-3$. We further describe these 
populations below.

\begin{figure}[]
\centering
\scalebox{1.00}{\includegraphics{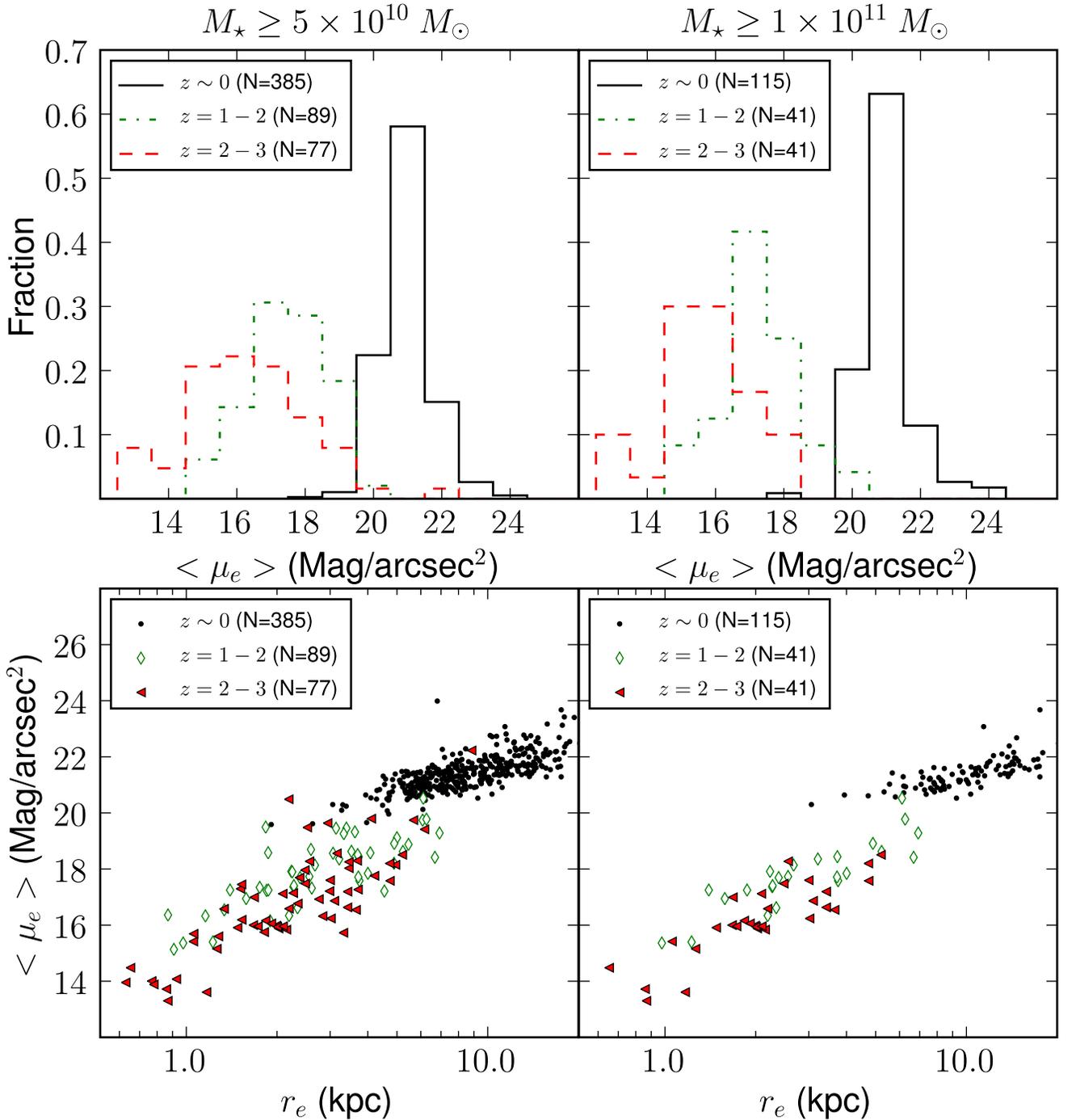}}
\caption{
Left column: The top panel shows mean extinction-corrected rest-frame $B$-band 
surface brightness within the effective radius ($<\mu_e>$) for massive
($M_\star \geq 5\times 10^{10}$ $M_\odot$) galaxies for the three redshift bins listed in Table \ref{tabmgc}.  The solid line
is for $z\sim0$ MGC galaxies.  The dash-dotted line ($z=1-2$) and the
dashed line ($z=2-3$) are based on our fits to the NIC3 F160W images of massive GNS galaxies. The GNS galaxies at $z=2-3$
have a mean surface brightness of 16.8 mag/arcsec$^2$ and are systematically brighter than the $z\sim0$ MGC galaxies, which have a mean
surface brightness of 21.3 mag/arcsec$^2$. In the bottom panel, surface brightness within the effective radius is plotted
against effective radius $r_e$ for the same redshift bins. Right column:  The same plots are repeated for galaxies with
$M_\star \geq 1\times 10^{11}$ $M_\odot$. 
Surface brightness is calculated with the extinction-corrected rest-frame $B$-band 
light and is defined as
$<\mu_e> = B_{\rm{corr}} + 2.5 \rm{log}_{10}(2\pi r_e^2) - 10 \rm{log}_{10}(1+z)$,
where $B_{\rm{corr}}$ is the extinction-corrected, rest-frame $B$ apparent magnitude.
\label{figsbr}}
\end{figure}

\begin{figure}[]
\centering
\scalebox{1.00}{\includegraphics{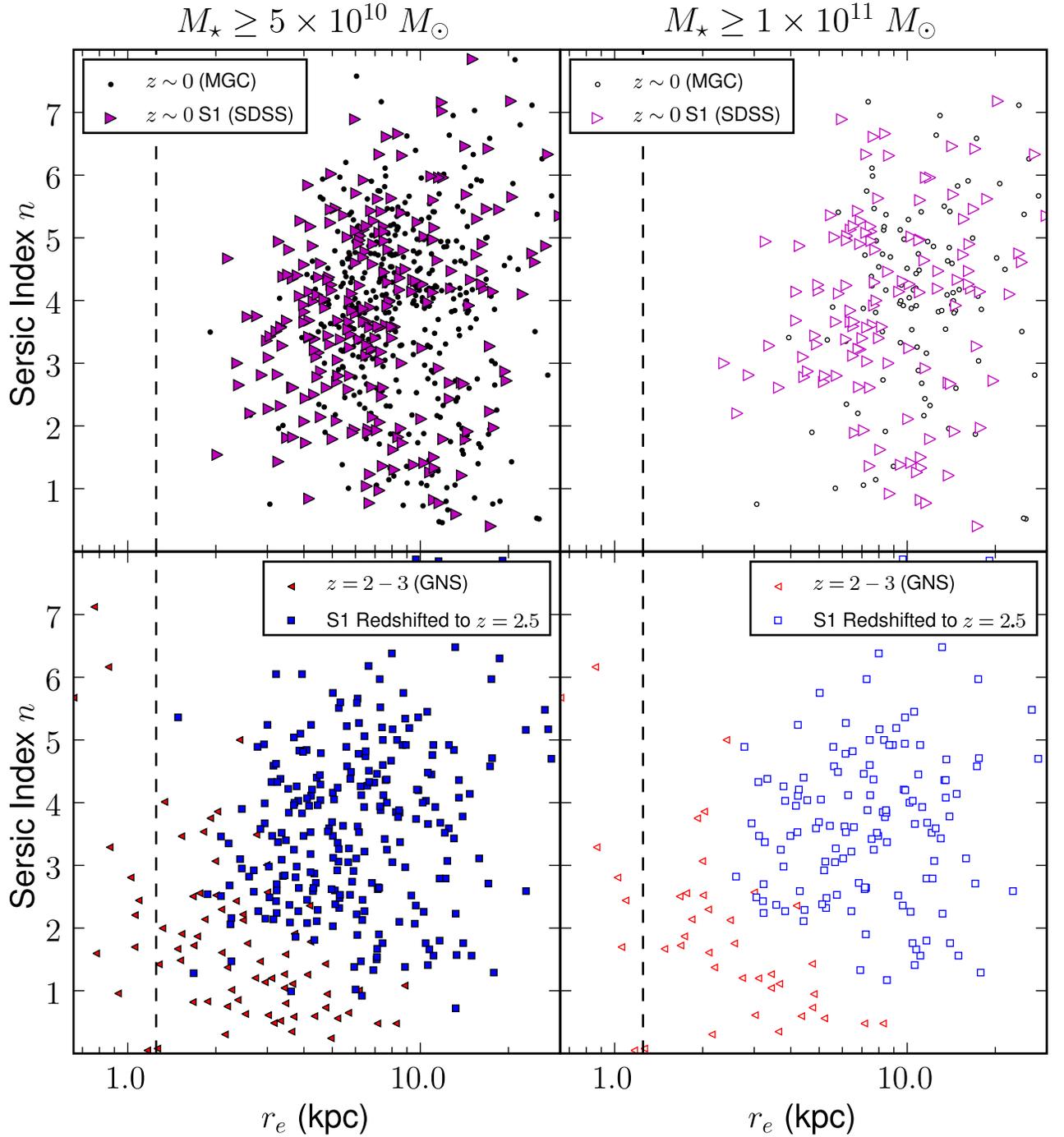}}
\caption{
Top row: The black points show the massive 
($M_\star \geq 5\times 10^{10}$ $M_\odot$) $z\sim$~0
galaxies from MGC described earlier in Figure~\ref{fnre}.
The magenta points denote the SDSS-based
sample S1 of 255 representative massive 
($M_\star \geq 5\times 10^{10}$ $M_\odot$) $z\sim$~0 galaxies used
in the redshifting  experiment. Note the ($n$, $r_e$) distribution
of S1 covers the same parameter space as that of the MGC sample.
This is also shown quantitatively in Figure~\ref{redshifthisto}.
Row 2: We show as blue squares the ($n$, $r_e$) distribution
obtained after redshifting S1 to $z=2.5$ and 're-observing' it with 
NIC3/F16W as in the GNS survey. We assume a surface brightness evolution of 
2.5 magnitudes and brighten each redshifted galaxy by this amount.
The actual observed ($n$, $r_e$) distributions of the massive galaxies at 
$z=2-3$ in the GNS survey are significantly offset toward lower values 
compared to the redshifted galaxies. The black dashed line represents the 
typical half-width half max of the NICMOS3 PSF at $z=1-3$ of $\sim1.2$ kpc.
\label{fnre2}}
\end{figure}

\begin{figure}[]
\centering
\scalebox{1.00}{\includegraphics{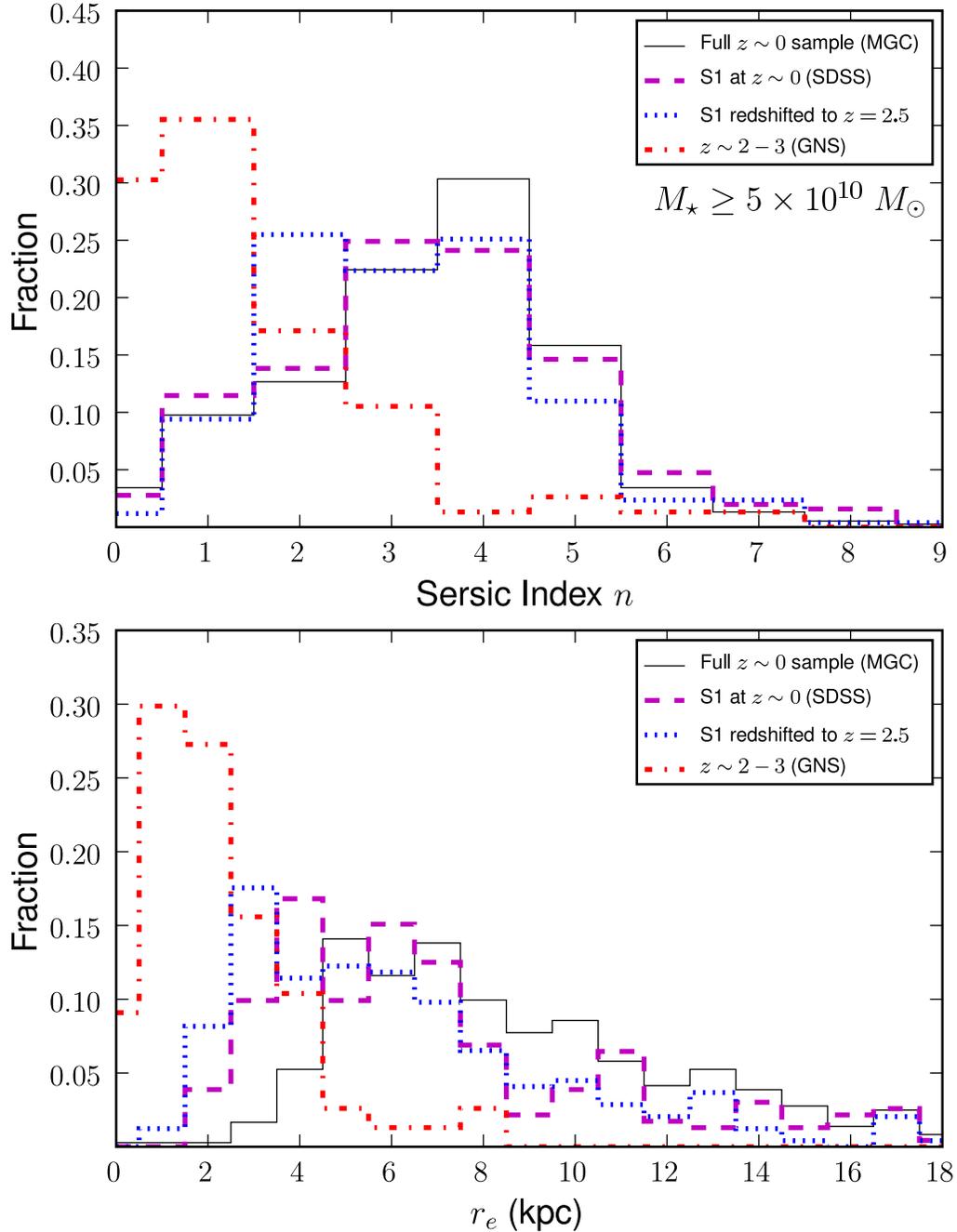}}
\caption{
This figure illustrates the same information as in Figure~\ref{fnre2} 
but in more quantitative terms.  It
shows the $n$ and $r_e$ distributions for
the full MGC sample of massive $z\sim$~0 galaxies (black line) and
the representative sample S1 of 255 galaxies used 
in the  redshifting experiment  (magenta line).
Sample S1 does a good job of matching the full MGC sample and is
typically within $\pm10\%$ for a given bin.
We also contrast the  ($n$, $r_e$) values after redshfiting 
S1 to $z=2.5$ (blue line) with the actual distribution  
observed in the massive the GNS galaxies at $z=2-3$ 
(red line). While $64.9\pm5.4\%$ and $39.0\pm5.6\%$
of the massive $z=2-3$ galaxies have $n\le2$ and $r_e\le2$~kpc,
respectively, the corresponding fractions for the redshifted sample are
$10.6\pm1.9\%$ and $1.2\pm0.7\%$. The results shown here are for galaxies with 
$M_\star \geq 5\times 10^{10}$ $M_\odot$, but a similar
result is obtained for $M_\star \geq 1\times 10^{11}$ $M_\odot$.
\label{redshifthisto}}
\end{figure}

\begin{figure}[]
\centering
\scalebox{1.00}{\includegraphics{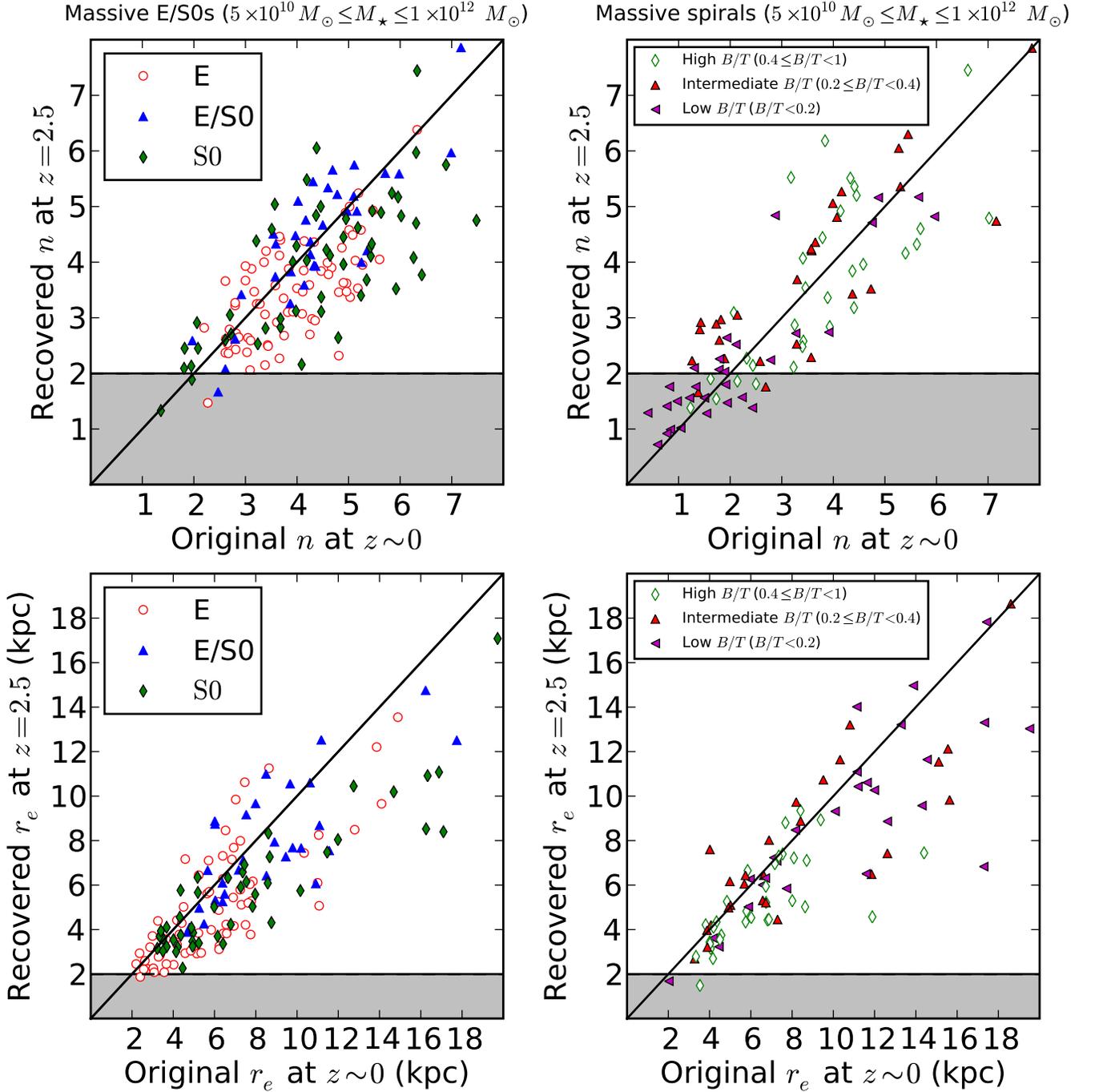}}
\caption{
Left column:
The panels compare the rest-frame optical structural parameters 
(S\'ersic index $n$ and effective radius $r_e$) of
massive ($M_\star \geq 5\times 10^{10}$ $M_\odot$) elliptical and S0 
galaxies  at $z\sim$~0  to the structural parameters  recovered after 
these galaxies were artificially redshifted to $z=2.5$, brightened
by 2.5 magnitudes in surface brightness, and re-observed  
with  NIC3/F160W.
At $z\sim$~0,  the structural parameters were measured  from $g$-band 
images, while at $z=2.5$ they are measured from the artificially 
redshifted images in the NIC3/F160W band, so that all parameters are 
measured  in the rest-frame blue optical light. The black lines represent 
equality, while the shaded area represents the regime of  $n \le 2$ 
and $r_e \le 2$~kpc, where $64.9 \pm5.4\%$ and $39.0 \pm 5.6\%$, 
respectively, of massive GNS galaxies at $z=2-3$ lie 
(Table \ref{tabmgc} and Figure~\ref{fnre2}). 
{\it The plots show that the S\'ersic index $n$ and effective radius $r_e$
of the massive  $z\sim$~0  E and S0s may be lower or higher after redshifting 
out to $z=2.5$, but they do not, in general, drop to values as low 
as $n \le 2$ and $r_e \le 2$~kpc, and avoid the shaded area.}
Right column: Same as left column, but this time for
massive ($M_\star \geq 5\times 10^{10}$ $M_\odot$)   $z\sim$~0 \
spiral galaxies. The galaxies are coded by bulge-to-total  light ratio
($B/T$). $B/T$ was measured with bulge-disk and bulge-disk-bar
decomposition of the $z\sim$~0  $g$-band images.
 The top plot shows that  it is mainly  massive $z\sim$~0
late-type  spirals of low $B/T$ that yield  S\'ersic index
$n$  as low as $n \le 2$ after redshifting, and populate
the shaded area where $64.9 \pm 5.4$\% of massive GNS galaxies
at $z=2-3$ lie.
However, as shown by this lower plot, the local massive spirals have much
larger $r_e$ ($r_e\gg2$~kpc) and after artificial redshifting avoid
the shaded area where $39.0 \pm 5.6\%$ of the massive GNS
galaxies at $z=2-3$ lie.
\label{redshift1x1earl}}
\end{figure}

\begin{figure}[]
\centering
\scalebox{1.00}{\includegraphics{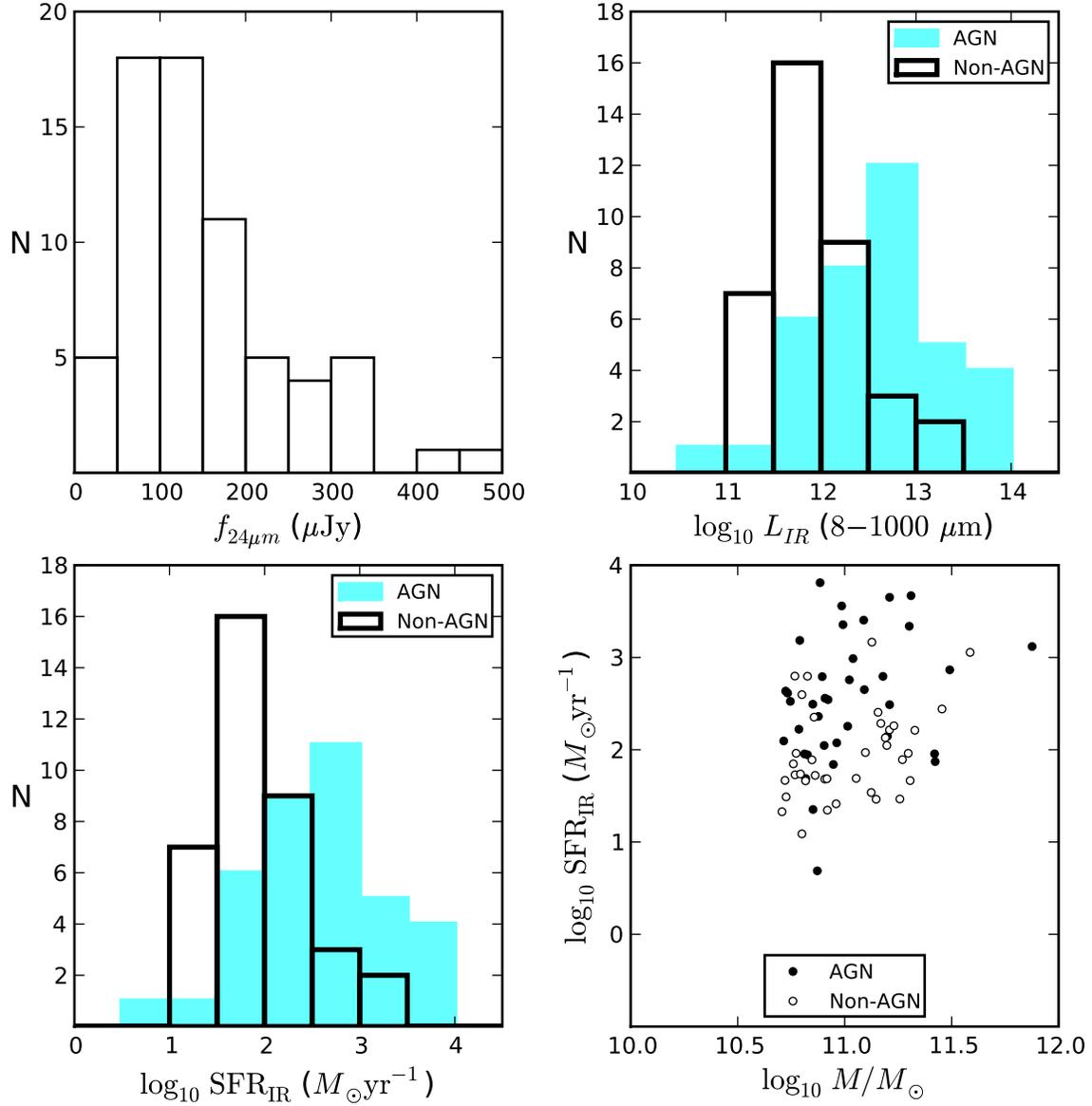}}
\caption{
Top left: The $f_{24\mu m}$ distribution for the massive
($M_\star \geq 5\times 10^{10}$  $M_\odot$) GNS
galaxies with reliable MIPS 24 $\mu$m counterpart.
Upper right: The inferred $L_{IR}$ distribution over 8--1000 $\mu$m.
Lower left: The inferred SFR$_{\rm IR}$ distribution based on $L_{IR}$,
which is estimated using the Chary \& Elbaz (2001) templates, with a correction
at L$_{IR} > 6\times 10^{11}$ $L_\odot$.
Lower right: SFR$_{\rm IR}$ versus $M_\star$. For sources containing
an AGN, the measured $L_{IR}$ and SFR$_{\rm IR}$ are upper limits. The upper right
and bottom panels use different coding for sources identified in \S\ref{agn}
as hosting an AGN.
\label{sfrf24}}
\end{figure}

\begin{figure}[]
\centering
\scalebox{1.00}{\includegraphics{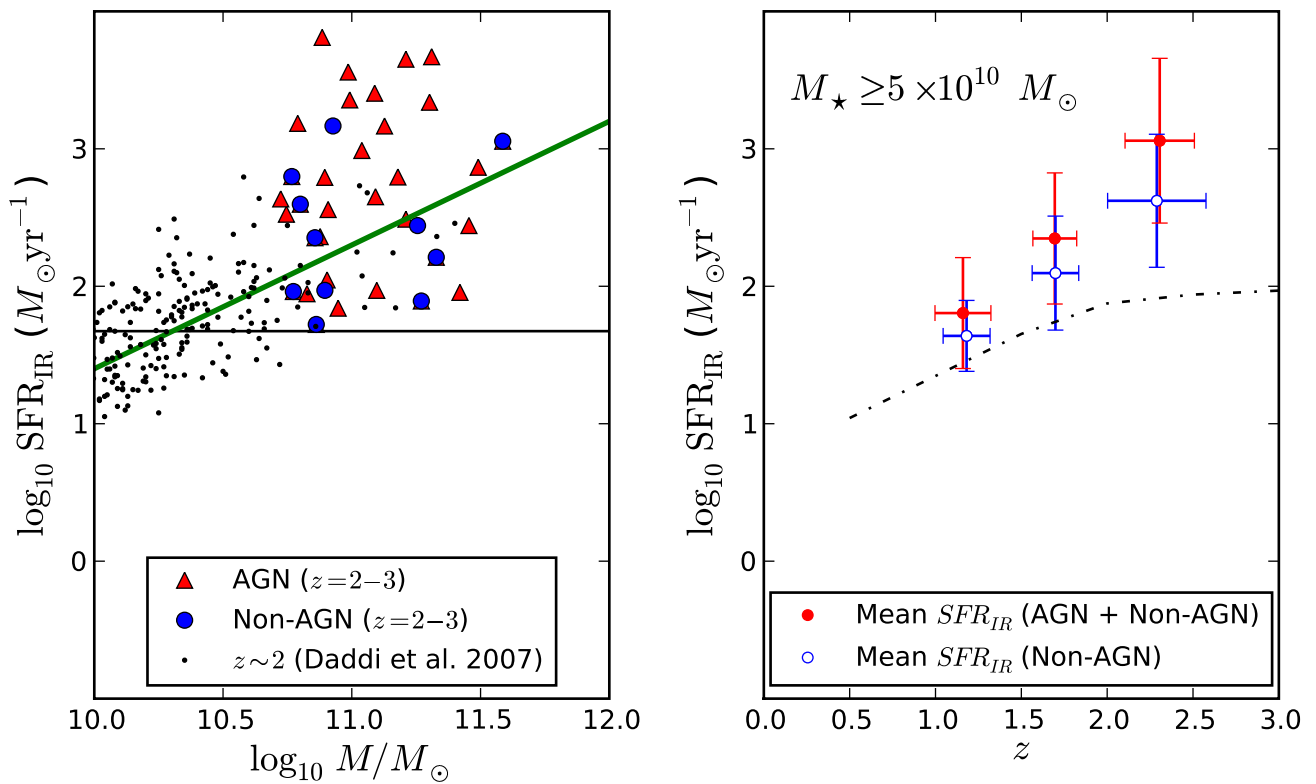}}
\caption{
The left-hand panel shows SFR$_{\rm IR}$ versus $M_\star$ at $z=2-3$.  The AGN 
candidates are coded as triangle symbols, and their SFR$_{\rm IR}$ likely overestimate their true SFR. The completeness limit in SFR$_{\rm IR}$
(corresponding to the limiting 24~$\mu$m  flux of $\sim$30~$\mu$Jy) is shown as a black solid line. The black dots represent SFR from UV measurements by
Daddi et al. (2007) for $z\sim 2$; the diagonal green line is their
corresponding SFR-mass correlation at $z\sim 2$. The right-hand panel 
shows mean SFR$_{\rm IR}$ in the different redshift bins for sources
with SFR$_{\rm IR}$ above the detection limit. The error bars are the
$1\sigma$ standard deviation around the mean.
The black line shows average UV-based SFR versus redshift for a galaxy with
$5\times 10^{10}$ $M_\odot$, as calculated by Drory \& Alvarez (2008).
\label{sfrmassz}}
\end{figure}

\begin{figure}[]
\centering
\scalebox{0.9}{\includegraphics{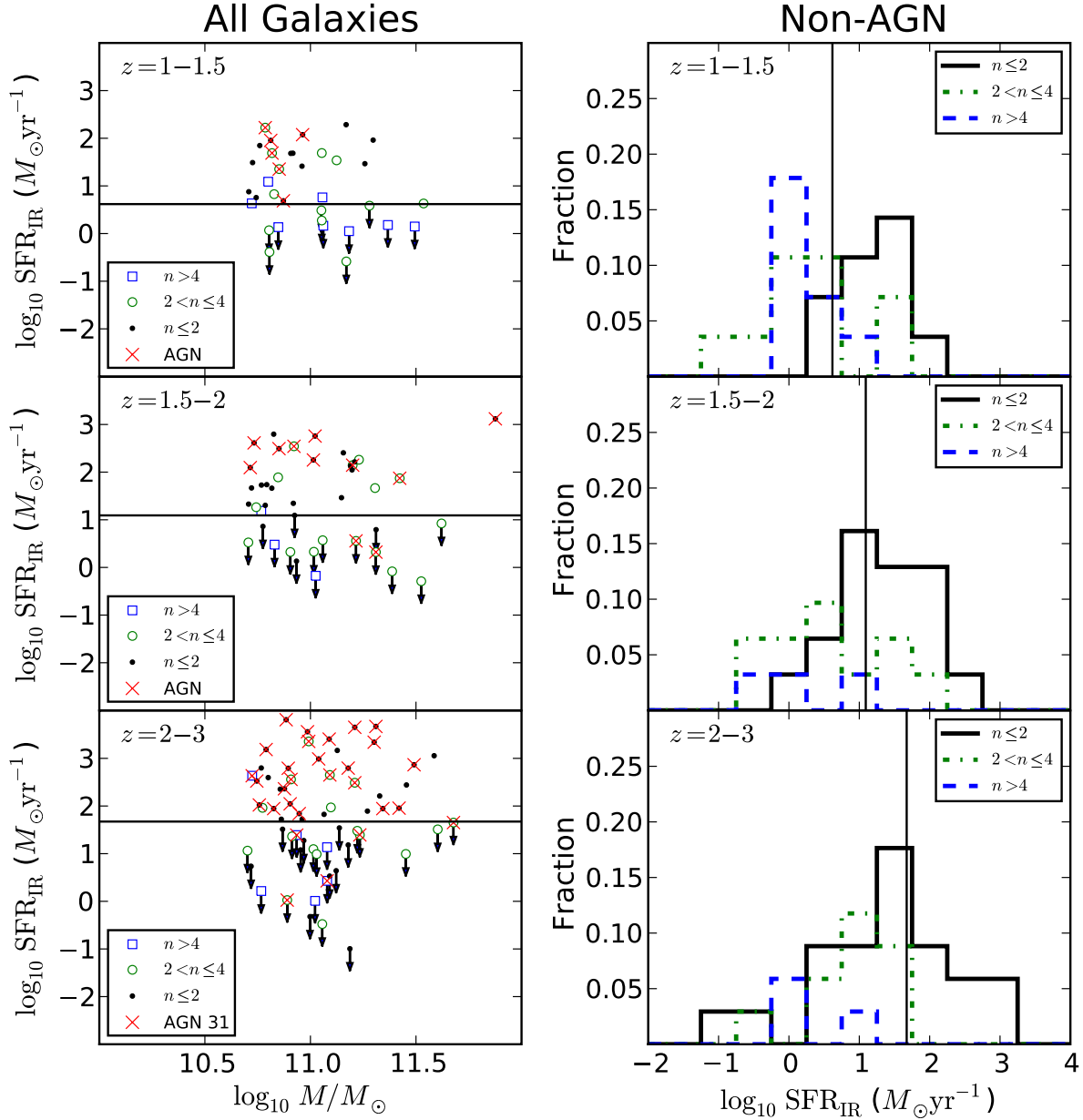}}
\caption{
Left column:  SFR$_{\rm IR}$  is plotted versus $M_\star$, for all galaxies
 with $M_\star \geq 5\times 10^{10}$  $M_\odot$, in different
redshift bins spanning 1-1.4 Gyr in cosmic time.
Data are sorted according to the S\'ersic index $n$ calculated in \S\ref{decomp}.
AGN candidates (see \S\ref{agn}) are labeled with red x's.
Galaxies with  SFR$_{\rm IR}$  below the detection limit (shown as a horizontal line)
are shown with downward  pointing arrows because they are upper limits.
{\it At $z=2-3$ the majority ($84.6\pm10.0\%$) of massive 
non-AGN galaxies with SFR$_{\rm IR}$ above the detection limit have $n\le2$ 
(disky) structures.} Right column: For non-AGN sources, histograms show 
{\it the fraction of massive galaxies} in each redshift bin  with a given 
SFR$_{\rm IR}$ for separate ranges of $n$.  
The vertical black lines mark the  SFR$_{\rm IR}$  detection limit. 
For sources to the left of the line, we plot  upper limits for SFR$_{\rm IR}$.
{\it The high SFR tail   in each redshift bin  is 
populated primarily by systems with low $n\le2$ (disky) structures. }
\label{sfrmassn}}
\end{figure}

\begin{figure}[]
\centering
\scalebox{0.90}{\includegraphics{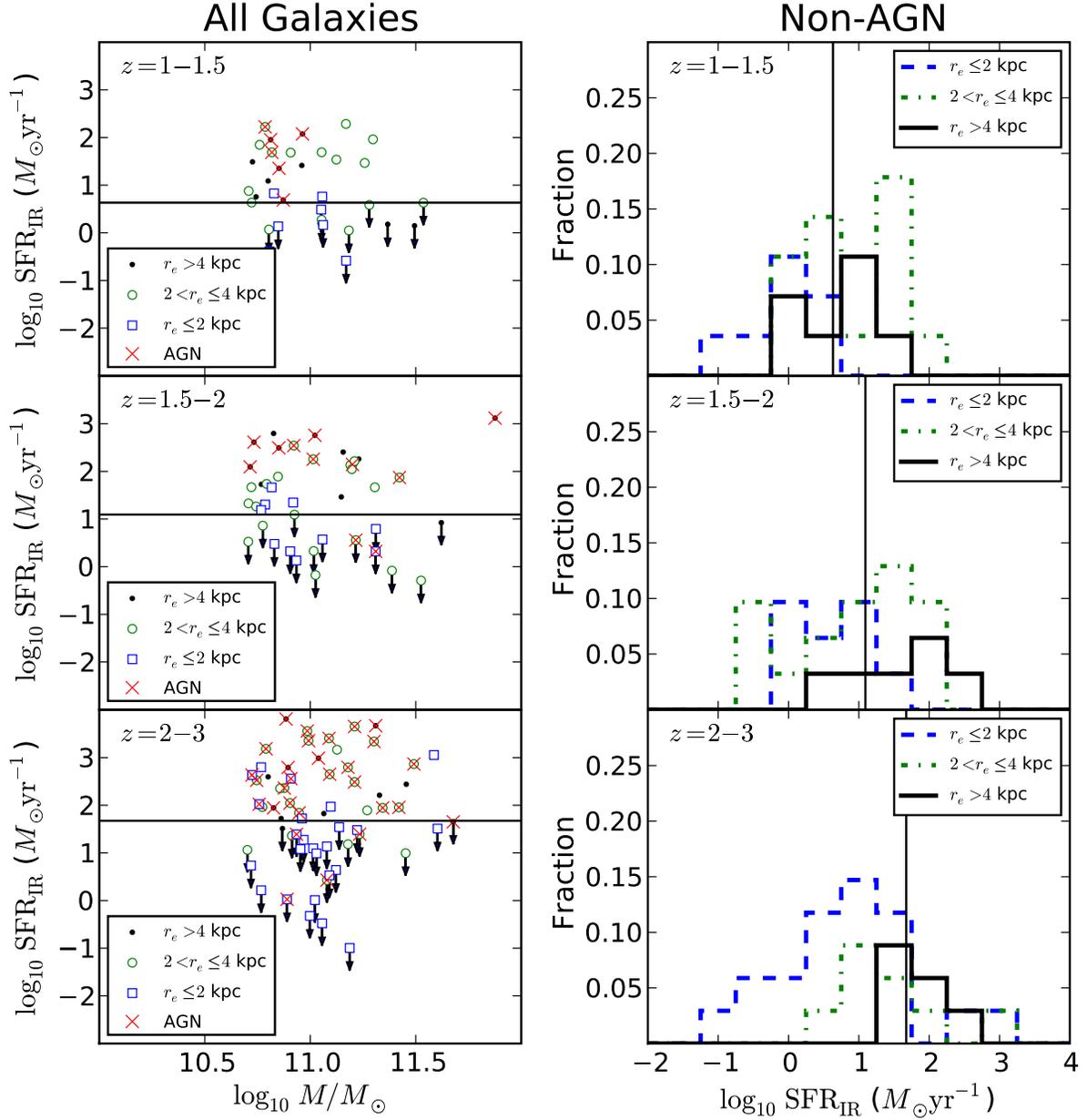}}
\caption{
Same as Figure \ref{sfrmassn}, but now the data are sorted by half-light
radius $r_e$.
Note that only a small fraction of the ultra-compact ($r_e\leq2$ kpc) galaxies
have SFR$_{\rm IR}$ above the 5$\sigma$ detection limit.  Some ultra-compact
galaxies have high SFR$_{\rm IR}$, but, on average, their mean SFR$_{\rm IR}$ 
are lower than in more extended systems.
\label{sfrmassr}}
\end{figure}

We find  that  approximately 40\%
($39.0  \pm 5.6$\%  for  $M_\star \geq 5\times 10^{10}$ $M_\odot$ and
$39.0  \pm 7.6$\%  for  $M_\star \geq 1\times 10^{11}$ $M_\odot$) of the 
massive  galaxies  at $z = 2-3$ are in the form of ultra-compact 
($r_e \le 2$~kpc)  galaxies compared to less than 1\% at $z\sim$ 0 
(Table~\ref{tabmgc} and Figure~\ref{fnre}). 
These ultra-compact galaxies are practically unmatched 
among $z\sim 0$ massive galaxies,  and their surface brightness in the 
rest-frame optical can be 4-6  magnitudes brighter (Figure~\ref{figsbr}).

Secondly, we find that the majority 
($64.9\% \pm 5.4\%$  for  $M_\star \geq 5\times 10^{10}$ $M_\odot$, and
$58.5\% \pm 7.7\%$ for  $M_\star \geq 10^{11}$ $M_\odot$)  of  massive
galaxies  at $z=2-3$  have  low $n\le$~2, while the corresponding fraction 
among massive systems at $z\sim$~0 is five times lower.
Most ($\sim72\%$) of these massive galaxies at $z=2-3$ with low $n\leq2$   
have $r_e > 2$ kpc, and therefore complement the ultra-compact galaxies. 
We further explore the meaning of a   S\'ersic index $n\le2$ at $z=2-3$, 
and present evidence that 
{\it most of 
the  massive galaxies with  $n\le2$ at $z=2-3$, particularly the 
extended ($r_e>2$ kpc) ones, likely host a 
prominent disk},   unlike the majority of massive galaxies at $z\sim 0$.
Our evidence is based on rest-frame optical morphologies,
ellipticities, artificial  redshifting experiments,   as well as 
bulge-to-total ratios  from bulge-plus-disk  decompositions of extended systems. 

\vspace{3mm}
{\it  2.  Star formation rates:}
We estimate star formation rates using IR luminosities (8-1000 $\mu$m) 
derived from  the $Spitzer$ 24 $\mu$m flux for massive GNS galaxies
having  a secure MIPS 24~$\mu$m counterpart and a 24~$\mu$m  flux 
exceeding the 5$\sigma$ detection limit of 30 $\mu$Jy.
AGN host candidates are excluded because the inferred IR luminosities 
overestimate the true star formation rate.

We find a strong link between galaxy structure and SFR. 
Among the non-AGN massive ($M_\star \geq 5\times 10^{10}$ 
$M_\odot$) galaxies   at $z=2-3$ with SFR$_{\rm IR}$  high 
enough to yield 
a 5$\sigma$  (30~$\mu$Jy) $Spitzer$  24 $\mu$m detection, 
the majority  (84.6 $\pm 10.0$\%)  have low $n\leq 2$. 
While the  $n\leq 2$ disky systems have a
wide range of SFR$_{\rm IR}$ (53 to 1466 $M_\odot$ yr$^{-1}$ at $z=2-3$), 
they include the systems of the highest SFR$_{\rm IR}$ at both $z=1-2$ and
$z=2-3$. In contrast, the massive 
ultra-compact objects at $z=2-3$ are less likely by a factor of $\sim 2.2$ to 
have SFR$_{\rm IR}$ above the detection limit, compared to the whole sample of 
non-AGN massive galaxies. 

\vspace{3mm}
{\it 3. AGN activity: } 
Using a variety of techniques (X-ray properties, IR power-law, and 
IR-to-optical excess) to identify AGN, we find that  49/166 ($29.5 \pm 3.5\% $)
of the massive galaxies at $z=1-3$ are AGN candidates. The AGN fraction rises 
with redshift, increasing from $17.9 \pm 6.1\%$ at $z=1-1.5$ to 
$40.3 \pm 8.8 \% $ at $z=2-3$ (Table~\ref{tabagnfrak}). 

We find a relationship between host galaxy structure and AGN activity that
complements the relationship between SFR and structure. 
Among massive galaxies at $z=2-3$,  AGN appear to be found preferentially 
in galaxies that are not ultra-compact, as evidenced by the fact that 
most ($80.6 \pm 7.9\%$) AGN  hosts have $r_e > 2$ kpc.
In fact, at $z=2-3$, the AGN fraction in ultra-compact galaxies is $\sim2.7$ 
times lower than in extended galaxies ($20.0 \pm 16.3\%$ versus 
$53.2 \pm 10.0\%$).  {\it Thus, ultra-compact galaxies  appear quiescent 
in terms of both  SFR and AGN activity.} In terms  of their S\'ersic index $n$, 
 a large fraction ($64.6 \pm 10.7\%$) 
of AGN hosts at $z=2-3$ have disky ($n\leq2$) morphologies.

\begin{figure}[]
\centering
\scalebox{0.95}{\includegraphics{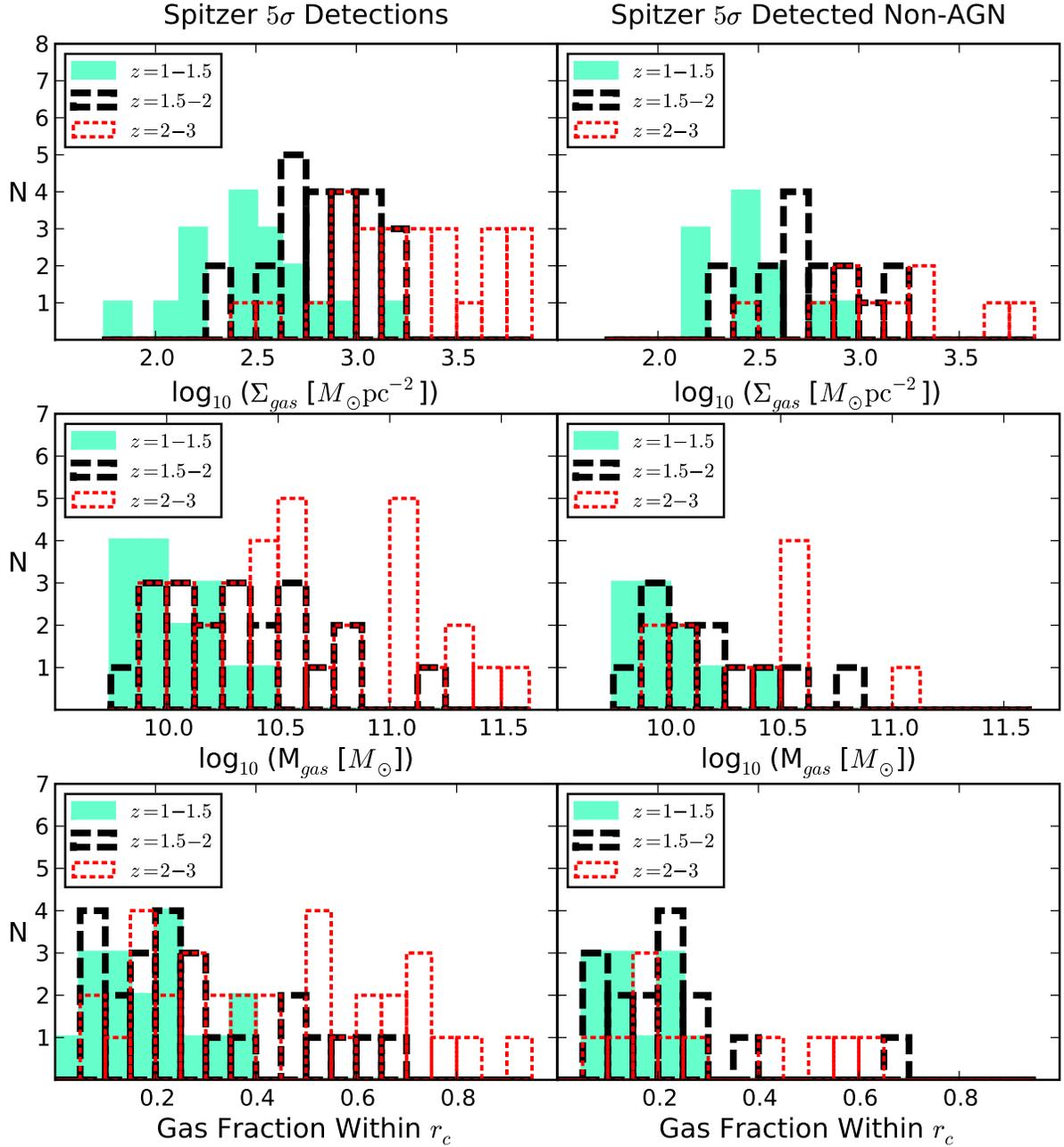}}
\caption{
Left column:
For galaxies with SFR$_{\rm IR}$ above the $5\sigma$ detection limit, the 
distributions of cold gas surface density ($\Sigma_{\rm gas}$), cold gas
mass $M_{\rm gas}$, and cold gas fraction ($f_{\rm gas}(r_c)$) within the 
circularized optical half-light radius $r_c$ are shown for different redshift 
ranges.  $\Sigma_{\rm gas}$ is calculated using a Schmidt-Kennicutt law 
(Kennicutt 1998) with power-law index 1.4 a normalization factor of 
$2.5\times10^{-4}$.  The cold gas fraction 
($f_{\rm gas}(r_c)\equiv M_{\rm gas}/(M_{\rm gas}+M_\star)$)
is calculated relative to the total baryonic mass within $r_c$.
Right column: Same as left column except that only non-AGN sources are shown.
\label{figfgas}}
\end{figure}

\begin{figure}[]
\centering
\scalebox{0.90}{\includegraphics{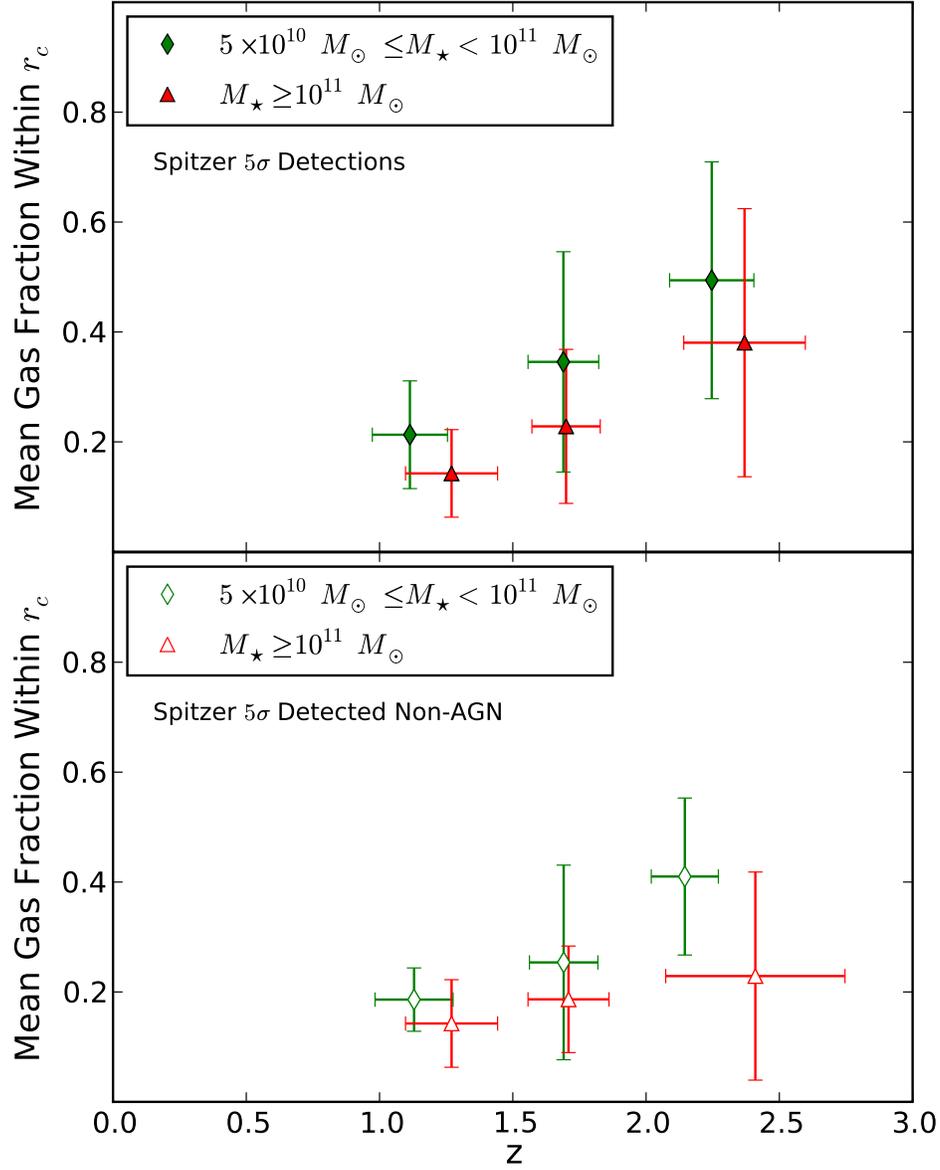}}
\caption{
Top:
For galaxies with SFR$_{\rm IR}$ above the $5\sigma$ detection limit,
the mean cold gas fraction 
($f_{\rm gas}(r_c)\equiv M_{\rm gas}/(M_{\rm gas}+M_\star)$)
within the circularized optical half-light radius $r_c$ is shown in three 
redshift bins for all galaxies with 
$5\times 10^{10}$ $M_\odot \leq M_\star < 10^{11}$ $M_\odot$ 
and $M_\star \geq 10^{11}$ $M_\odot$.
The error bars indicate the $1\sigma$ scatter in gas fraction and redshift.
Bottom: Same as the top except that only non-AGN sources are shown.
\label{figfgas-z}}
\end{figure}

\begin{figure}[]
\centering
\scalebox{1.00}{\includegraphics{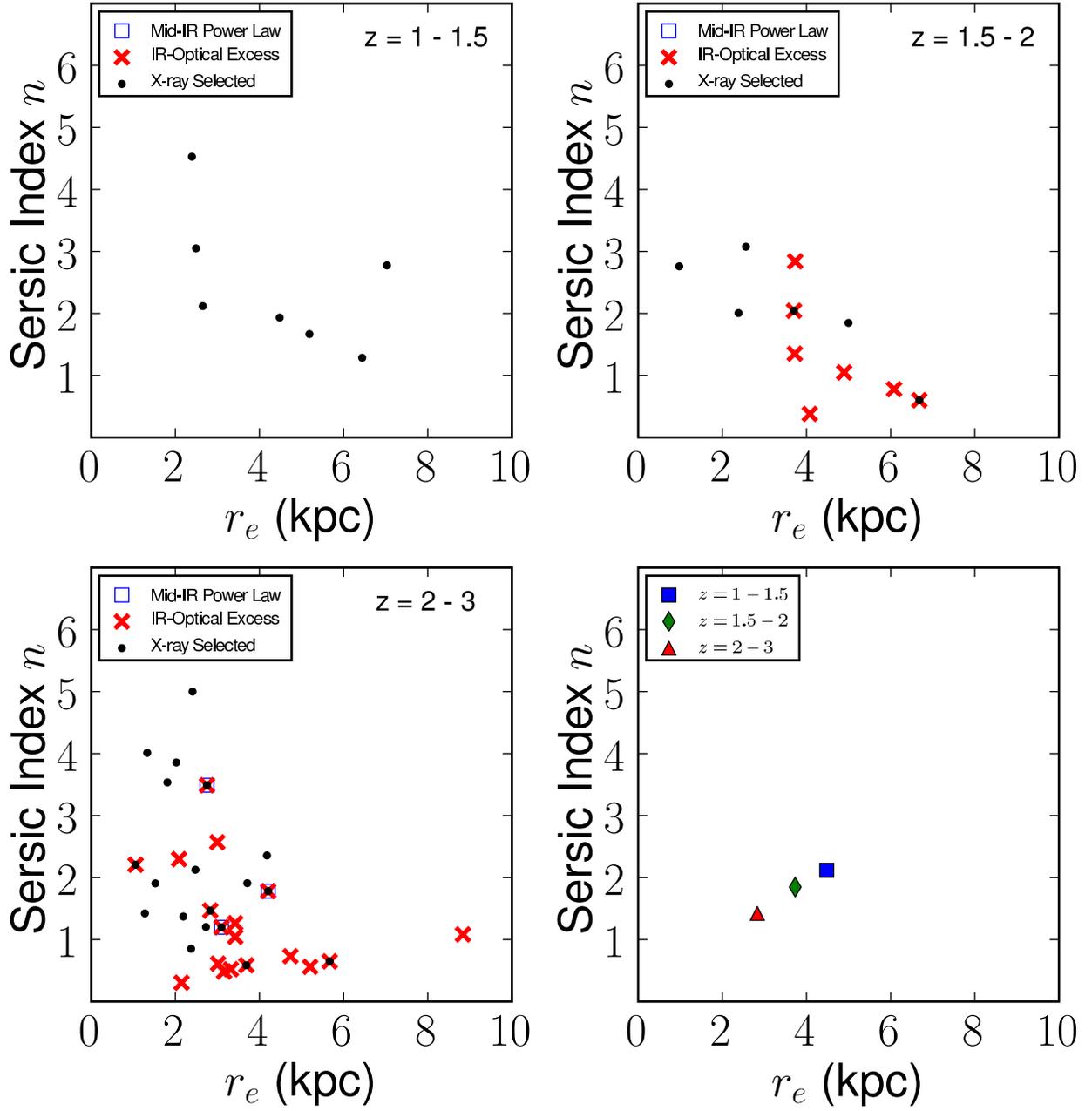}}
\caption{
The upper and lower-left panels show
single S\'ersic index $n$ versus effective radius $r_e$ for the 
49 AGN candidates selected
either based on X-ray properties, mid-IR power-law, or IR-to-optical
excess.  The lower-right panel shows the median S\'ersic index and $r_e$
in each redshift bin.
\label{agnnr}}
\end{figure}

\begin{figure}[]
\centering
\scalebox{1.00}{\includegraphics{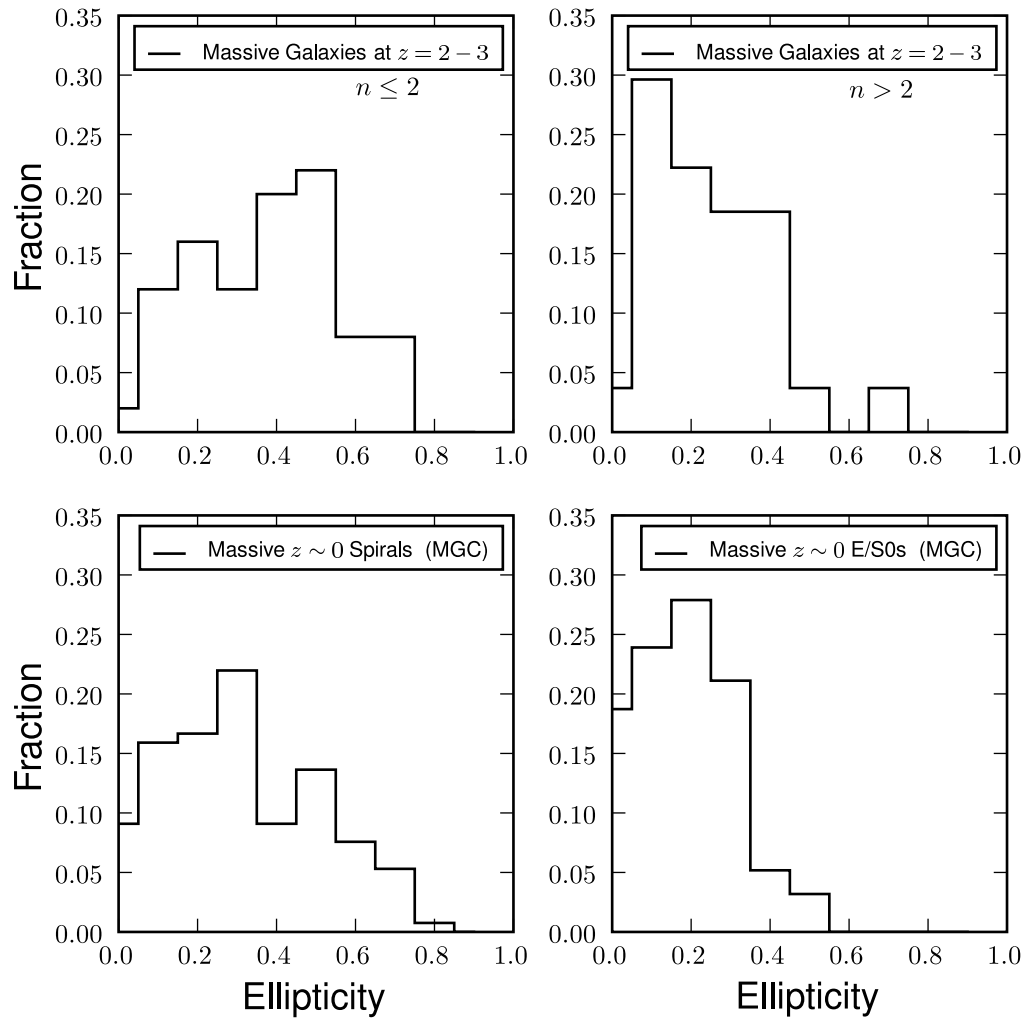}}
\caption{
 In the top panels, the deconvolved ellipticity ($1-b/a$) measured by
GALFIT is shown for massive ($M_\star \geq 5\times 10^{10}$ $M_\odot$)
GNS galaxies
at $z=2-3$ with $n\le2$ and $n>2$.  The bottom panels show the deconvolved
 ellipticity for similarly massive E/S0 and Spiral galaxies as
 measured with  GIM2D by Allen et al. (2006).
\label{figba}}
\end{figure}

\vspace{3mm}
{\it 4.  Cold gas content:}
We apply a  standard Schmidt-Kennicutt law (Kennicutt 1998) to the
SFR$_{\rm IR}$ of the non-AGN host candidates.
The high estimated SFR$_{\rm IR}$  suggest that
copious cold gas reservoirs are present. We estimate that the average cold gas
surface density in non-AGN hosts
ranges from $\sim 136$ to  $\sim 25,091$ $M_{\odot}$~pc$^{-2}$ at $z=1-3$,
with  a median value of $\sim607$ $M_{\odot}$~pc$^{-2}$ (Figure~\ref{figfgas}).
The implied cold gas fraction within the rest-frame optical half-light radius
ranges from $6.5\%$ to  $65.4\%$,
with a mean of $\sim41\%$ at $z=2-3$  (Figure~\ref{figfgas}).
The highest gas fractions at a given redshift are found among the less massive 
galaxies, consistent with downsizing.

\vspace{3mm}
{\it 5.  Formation of  massive galaxies by $z=2-3$:}
The massive galaxies at $z=2-3$ already have an average rest-frame
optical  surface brightness
within $r_e$  that can be up to 3-6 magnitudes brighter than $z\sim0$ 
massive galaxies.  The associated high  stellar mass densities 
imply that massive galaxies at  $z=2-3$ must have formed via 
rapid, highly  dissipative events at $z>2$.  
Both gas-rich major mergers and  gas  accretion at $z>2$ are viable as
their associated short  dynamical timescales and short gas cooling times 
at $z>2$ would  lead to a rapid buildup of mass. 
However, the large fraction ($\sim65$\%) of massive galaxies at $z=2-3$ with 
$n\leq2$ and disky morphologies suggest that  cold-mode accretion at
$z>2$  must have played an important role  in  the build-up of massive
galaxies  by  $z=2-3$,
since it may be challenging to  have  such a  large  fraction of 
of merger remnants with low $n \le 2$   from isolated  gas-rich 
major mergers.

\vspace{3mm}
{\it 6. Transformation of massive galaxies at  $z=2-3$ into present-day E and S0s:}
In order for  massive galaxies at $z=2-3$ 
to evolve into   $z\sim0$ massive systems (which are primarily
E and S0s), they need to radically change their rest-frame optical
structure and distributions of   ($n$, $r_e$). 
In particular they need to raise $n$  well above 2, increase $r_e$ by an average
factor of 3-4, 
and dim the average rest-frame optical surface brightness.
Dry major mergers can  induce  changes in galaxy size, 
S\'ersic index, and stellar surface density,  but they may be 
too rare  to account for all the needed evolution.  Successive dry minor mergers 
have been shown to influence galaxy size, S\'ersic index, and stellar surface 
density in  a similar direction.  
We suggest the transformation of massive $z=2-3$ galaxies into 
$z\sim0$ galaxies will occur through a combination of dry major mergers, 
minor  mergers.  We will investigate in 
the relative importance and efficiency of these mechanisms in a future 
paper.

\acknowledgments
SJ, CJC, TW, MD, and RL  acknowledge support from  $HST$ grant GO-11082
from STScI, which is operated by AURA, Inc., for NASA, under NAS5-26555.
SJ and TW  also acknowledge support from the Norman Hackerman
Advanced Research Program (NHARP) ARP-03658-0234-2009,
National Aeronautics and Space Administration (NASA) LTSA grant
NAG5-13063, and NSF grant AST-0607748.
S.J. and T.W.  acknowledge support for this research by
the DFG cluster of excellence "Origin and Structure of
the Universe" (www.universe-cluster.de).
CJC acknowledges support from STFC and the Leverhulme Foundation.
We thank Knud Jahnke and Marco Barden for technical assistance with the
operation of FERENGI, and   Andreas Burkert,  Sadegh Khochfar,
T.~J. Cox,  Thorsten  Naab, and Ludwig Oser  for stimulating discussions.
We acknowledge the usage of the HyperLeda database (http://leda.univ-lyon1.fr).
Some/all of the data presented in this paper were obtained from the
Multimission Archive at the Space Telescope Science Institute (MAST). STScI is
operated by the Association of Universities for Research in Astronomy, Inc.,
under NASA contract NAS5-26555. Support for MAST for non-HST data is provided
by the NASA Office of Space Science via grant NAG5-7584 and by other grants
and contracts.
The Millennium Galaxy Catalogue consists of imaging data from the
Isaac Newton Telescope and spectroscopic data from the Anglo
Australian Telescope, the ANU 2.3m, the ESO New Technology Telescope,
the Telescopio Nazionale Galileo and the Gemini North Telescope. The
survey has been supported through grants from the Particle Physics and
Astronomy Research Council (UK) and the Australian Research Council
(AUS). The data and data products are publicly available from
http://www.eso.org/~jliske/mgc/ or on request from J. Liske or
S.P. Driver.

\clearpage
\begin{deluxetable}{ccccc}
\tablewidth{0pt}
\tablecolumns{5}
\tablecaption{Rest-Frame Optical S\'ersic Index $n$ and $r_e$ in Massive ($M_\star \geq 5\times 10^{10}$ $M_\odot$) Galaxies\label{tabmgc}}
\tablehead
{
 \colhead{$z$} & \colhead{ Morphology} & \colhead{$n\leq2$} & \colhead{$n>2$} & \colhead{$n>3$} }
v\startdata
\multicolumn{5}{c}{\vspace*{2 mm}\textbf{$M_\star \geq 5 \times 10^{10}$ $M_\odot$} }  \\
$z=2-3$ ($N=77$)   & All     &   \textbf{64.9 $\pm 5.44$\%}  &  35.1 $\pm 5.44$\% & 18.2 $\pm 4.40$\%\\
$z=1-2$ ($N=89$)   & All     &   \textbf{49.4 $\pm 5.30$\%}  &  50.6 $\pm 5.30$\% & 30.3 $\pm 4.87$\%\\
$z\sim0$ ($N=385$)    & All     &   \textbf{13.0 $\pm 1.71$\%}  &  87.0 $\pm 1.71$\% & 74.3 $\pm 2.23$\%\\
                   & E/S0    &  0.8 $\pm 0.45$\%  &  64.9 $\pm 2.43$\%& 58.7 $\pm 2.51$\%\\
                   & Sabc    &  10.4 $\pm 1.56$\% &  20.8 $\pm 2.07$\%& 14.8 $\pm 1.81$\%\\
                   & Sd/Irr  &  1.82 $\pm 0.68$\%  &  1.30 $\pm 0.58$\%& 0.78 $\pm 0.45$\% \vspace*{2 mm}\\
\hline
\multicolumn{5}{c}{\vspace*{2 mm}\textbf{$M_\star \geq 1 \times 10^{11}$ $M_\odot$} }  \\
$z=2-3$ ($N=41$)   & All     &   \textbf{58.5 $\pm 7.69$\%}   &  41.5 $\pm 7.69$\% & 17.1 $\pm 5.88$\%\\
$z=1-2$ ($N=41$)   & All     &   \textbf{34.1 $\pm 7.41$\%}   &  65.9 $\pm 7.41$\% & 43.9 $\pm 7.45$\%\\
$z\sim0$ ($N=115$)  & All     &   \textbf{10.4 $\pm 2.85$\%}   &  89.6 $\pm 2.85$\% & 80.9 $\pm 3.67$\%\\ 
                   & E/S0    &  1.7 $\pm 1.22$\%   &  72.2 $\pm 4.18$\% & 67.0 $\pm 4.39$\%\\
                   & Sabc    &  6.09 $\pm 2.23$\%   &  13.9 $\pm 3.23$\% & 12.2 $\pm 3.05$\%\\
                   & Sd/Irr  &  2.61 $\pm 1.49$\%   &  3.48 $\pm 1.71$\% & 1.74 $\pm 1.22$\% \vspace*{2 mm}\\
\hline
\vspace*{2 mm}$z$ & Morphology & $r_e\leq2$ kpc & $2<r_e\leq4$ kpc & $r_e>4$ kpc\\
\hline
\multicolumn{5}{c}{\vspace*{2 mm}\textbf{$M_\star \geq 5 \times 10^{10}$ $M_\odot$} }  \\
$z=2-3$ ($N=77$)   & All     &   \textbf{39.0 $\pm 5.56$\%}  &  42.9 $\pm 5.64$\% & 18.2 $\pm 4.40$\%\\
$z=1-2$ ($N=89$)   & All     &   \textbf{24.7 $\pm 4.57$\%}  &  48.3 $\pm 5.30$\% & 27.0 $\pm 4.70$\%\\
$z\sim0$ ($N=385$)    & All     &   \textbf{0.52 $\pm 0.37$\%}  &  1.8 $\pm 0.68$\% & 97.7 $\pm 0.77$\%\\
                   & E/S0    &  0.26 $\pm 0.26$\%  & 1.8 $\pm 0.68$\%  & 63.6 $\pm 2.45$\%\\
                   & Sabc    &  0.00 $\pm 0.00$\%  & 0.0 $\pm 0.00$\%  & 31.2 $\pm 2.36$\%\\
                   & Sd/Irr  &  0.26 $\pm 0.26$\%  & 0.00 $\pm 0.00$\%  & 2.86 $\pm 0.85$\% \vspace*{2 mm}\\
\hline
\multicolumn{5}{c}{\vspace*{2 mm}\textbf{$M_\star \geq 1 \times 10^{11}$ $M_\odot$} }  \\
$z=2-3$ ($N=41$)   & All     &  \textbf{39.0 $\pm 7.62$\%}  &  41.5 $\pm 7.69$\% & 19.5 $\pm 6.19$\%\\
$z=1-2$ ($N=41$)   & All     &  \textbf{22.0 $\pm 6.46$\%}  &  56.1 $\pm 7.75$\% & 22.0 $\pm 6.46$\%\\
$z\sim0$ ($N=115$)    & All     &  \textbf{ 0.87 $\pm 0.87$\%} & 1.74 $\pm 1.22$\%  & 97.39 $\pm 1.49$\%\\
                   & E/S0    &  0.00 $\pm 0.00$\%   & 1.7 $\pm 1.22$\%  & 72.2 $\pm 4.18$\% \\
                   & Sabc    &  0.00 $\pm 0.00$\%   & 0.00 $\pm 0.00$\%  & 20.0 $\pm 3.73$\% \\
                  & Sd/Irr  &   0.87 $\pm 3.73$\%   & 0.00 $\pm 0.00$\%  & 5.22 $\pm 2.07$\% \\
\enddata
\tablecomments{
Rows 1-12: For a given redshift (Column 1), morphology (Column 2), and stellar
mass range, Columns 3, 4, and 5 
list the fraction of galaxies in three separate bins of S\'ersic index $n$.  
Rows 13-24: Same as the above except that Columns 3, 4, and 5 reflect bins of
half-light radius $r_e$.
}
\end{deluxetable}

\clearpage
\begin{deluxetable}{ccc}
\tablewidth{0pt}
\tablecolumns{3}
\tablecaption{Fit of Massive Galaxies to $r_e/r_{e,z\sim0} = \alpha(1+z)^\beta$ Over $z=0-3$\label{beta}}
\tablehead
{
\colhead{Sample} & \colhead{$\alpha(\pm 1\sigma)$} & \colhead{$\beta(\pm 1\sigma)$} \\
\colhead{(1)} & \colhead{(2)} & \colhead{(3)} \\
}
\startdata
All $n$    &    1.15(0.30)  & -1.30(0.24) \\
$n\le2$    &    1.11(0.32)  & -1.30(0.29) \\
$n>2$      &    1.20(0.31)  & -1.52(0.26) \\
Non-AGN hosts with high SFR$_{\rm IR}$$^1$       &    1.15(0.33)  & -1.22(0.30) \\
Non-AGN hosts with low SFR$_{\rm IR} $$^2$        &    1.67(0.33)  & -1.67(0.28) \\
\enddata
\tablecomments{
$^1$ Non-AGN hosts with 24 $\mu$m flux above the $Spitzer$ 5$\sigma$ limit
(30 $\mu$Jy).
$^2$  Non-AGN hosts with 24 $\mu$m flux below the $Spitzer$ 5$\sigma$ limit
(30 $\mu$Jy).
}
\end{deluxetable}

\clearpage
\tiny
\begin{deluxetable}{ccccccc}
\tablewidth{0pt}
\tablecolumns{6}
\tablecaption{Fraction of Massive ($M_\star \geq 5\times 10^{10}$ $M_\odot$) Galaxies With 24 $\mu$m Detections\label{tabsfrprop}}
\tabletypesize{\scriptsize}
\tablehead
{
\colhead{$z$} & \colhead{ SFR$_{min}$} &  \colhead{Fraction with $f_{24\mu m} \geq 30 \mu$Jy} &     \colhead{Mean SFR (AGN $+$ non-AGN)} & \colhead{Mean SFR (Non-AGN)}\\
 \colhead{} &  \colhead{($M_\odot$ yr$^{-1}$)} &  \colhead{(\%)}&  \colhead{($M_\odot$ yr$^{-1}$)} & \colhead{($M_\odot$ yr$^{-1}$)} \\
\\
 \colhead{(1)} &  \colhead{(2) } &  \colhead{(3)} &  \colhead{(4)} & \colhead{(5)} \\
}
\startdata
$z=1-1.5$   & 4.29 &  43.6 $\pm 7.9$\%    & 63.8  $\pm$ 12.9 & 44.0 $\pm 7.3$       \\
$z=1.5-2$   & 12.4 &  48.0 $\pm 7.1$\%    & 222.8 $\pm$ 58.5 & 128.9  $\pm 37.6$     \\
$z=2-3$     & 47.2 &  42.9 $\pm 5.6$\%    & 1145.6 $\pm$ 274.5 & 418.8 $\pm 142.9$      \\
\enddata
\tablecomments{
Column 2 estimates the detection limit on SFR given the $5\sigma$
limit on $f_{24 \mu m}$ of 30 $\mu$Jy.  The expected SFR$_{\rm IR}$ at 30 $\mu$Jy is
determined by linear regression of the distribution of $f_{24 \mu m}$ versus SFR$_{\rm IR}$
in each redshift bin.
Column 3 lists the percentage of massive GNS galaxies
with $f_{24\mu m} > 30 \mu$Jy.  Column 4 shows the mean SFR among all galaxies having
$f_{24\mu m} > 30$ $\mu$Jy.  Column 5 shows the mean SFR among all galaxies without
any evidence for AGN activity (see \S\ref{agnsummary}). The error bars 
in Column 4 and Column 5 represent the standard error on the mean.
}
\end{deluxetable}

\clearpage
\tiny
\begin{deluxetable}{cccccccc}
\tablewidth{0pt}
\tablecolumns{8}
\tablecaption{Summary of AGN Detection and Properties\label{tabagnfrak}}
\tabletypesize{\scriptsize}
\tablehead
{
\colhead{$z$} & \colhead{Total Number} & \colhead{X-ray AGN} & \colhead{PLG} & \colhead{IR Excess AGN} & \colhead{AGN Fraction} & \colhead{Median $n$} & \colhead{Median $r_e$} \\
\colhead{} & \colhead{} & \colhead{} & \colhead{} & \colhead{} & \colhead{ } & \colhead{} & \colhead{(kpc)}       \\
\colhead{(1)} & \colhead{(2)} & \colhead{(3)} & \colhead{(4) } & \colhead{(5)} & \colhead{(6)}   & \colhead{(7)} & \colhead{(8)}   \\
}
\startdata
$z=1-1.5$ & 7  & 7  & 0 &  0  &  17.9 $\pm 6.1$\% & 2.12 & 4.48 \\
$z=1.5-2$ & 11 & 6  & 0 &  5  &  22.0 $\pm 5.9$\% & 1.85 & 3.73 \\
$z=2-3$   & 31 & 20 & 3 &  11 &  40.3 $\pm 8.8$\% & 1.42 & 2.83 \\
\enddata
\end{deluxetable}
\normalsize

\clearpage
\begin{deluxetable}{cccc}
\tablewidth{0pt}
\tablecolumns{4}
\tablecaption{Summary of Kolmogorov-Smirnov Test on Ellipticity\label{kstest}}
\tabletypesize{\scriptsize}
\tablehead
{
\colhead{Sample 1} & \colhead{Sample 2} & \colhead{Probability} & \colhead{KS Test D} \\
\colhead{} & \colhead{} & \colhead{(\%)} & \colhead{} \\
\colhead{(1)} & \colhead{(2)} & \colhead{(3)} & \colhead{(4) } \\
}
\startdata
\hline
$n\le2$ $z=2-3$    &     MGC E/S0      &  0                  &  0.489 \\
$n\le2$ $z=2-3$    &     MGC Spiral (Sabc + Sd/Irr)   &  4.78               &  0.221 \\
$n\le2$ $z=2-3$    &     MGC Sd/Irr     &  23.5             &  0.317 \\
\hline
$n>2$ $z=2-3$    &     MGC E/S0     &   34.3              &  0.184 \\
$n>2$ $z=2-3$    &     MGC Spiral (Sabc + Sd/Irr)    &    14.0             &  0.237 \\
$n>2$ $z=2-3$    &     MGC Sd/Irr    &   15.8              &  0.370 \\
\enddata
\tablecomments{
Columns 1 and 2 list the two samples for which ellipticity was compared in each KS test.
Column 3 lists the probability that Sample 1 and Sample 2 are drawn from the same distribution. 
Column 4 lists the Kolmogorov-Smirnov statistic specifying
the maximum separation between the cumulative ellipticity distribution functions 
for Sample 1 and Sample 2.
}
\end{deluxetable}

\clearpage
\begin{appendices}

\begin{center}
\textbf{\large Appendices}
\end{center}

\section{PSF Modeling}\label{decomppsf}
Knowledge of the PSF is important to assess data quality and for
deriving structural parameters. NIC3 is out of focus, so the PSF can deviate
from the theoretically expected one.
PSF convolution with GALFIT is commonly performed with a user-provided bright,
unsaturated star.
Not all of the GNS tiles contain suitably bright, unsaturated stars.
It is not advisable to adopt a set of PSF stars from a subset of
pointings because the NIC3 PSF depends on position within the NIC3 field
and is also subject to interpolation artifacts
introduced by \texttt{drizzle} that are dependent on the adopted dither pattern
(J. Krist, private communication).

As a result, the best available option for handling PSF convolution is to
make synthetic NIC3 PSFs with Tiny Tim (Krist 1995).  For each galaxy,
Tiny Tim PSFs were generated for all the galaxy's positions in the
individual, undrizzled exposures.  Telescope breathing was accounted for
with each PSF by refining the Tiny Tim parameters to match the Pupil Alignment
Mechanism (PAM) value recorded in the headers of the undrizzled frames.
Blank, zero-valued frames retaining the WCS information of the undrizzled
frames were made.  The synthetic PSFs were inserted into the blank frames
precisely where each galaxy would be in the individual frames.
The blank frames were drizzled together in the same way as the data with
a pixfrac of 0.7 and a final output platescale of $0\farcs1$/pixel.  This process
was repeated for all 166 massive ($M_\star \ge 5 \times 10^{10}$) galaxies in 
our sample.

This approach accounts both for variation in PSF with position on the
NIC3 field and for the dependence on the \texttt{drizzle} algorithm.  The
range of FWHM in the final drizzled synthetic PSFs is
$\sim0\farcs26-0\farcs36$,\footnote{The range in PSF FWHM comes from differing
positions in the NIC3 field and the PAM values used to create the synthetic
PSFs.}
with a mean value of $0\farcs3$.
The mean PSF diameter of the science images ($0\farcs3$) is 2.5 kpc at $z=2$,
under the adopted cosmology.

\section{Extra Tests on Systematic Effects}\label{appendix1}
\subsection{Tests on Robustness of Fits and Parameter Coupling}\label{decompcoupling}
How robust are the results that a dominant fraction  of the
massive galaxies at $z=2-3$  have  a low $n\leq2$ and that a large
fraction are ultra-compact? In
particular, how non-degenerate are the fits? Could some
of the galaxies with an $n\leq2$ have  similarly good fits with
higher $n$?

First, one should note that the errors quoted by GALFIT on 
the structural parameters cannot be used to assess the robustness
of the fits  because the errors quoted by GALFIT 
underestimate the true parameter errors 
(H{\"a}ussler et al. 2007), which 
are dominated by the systematics of galaxy structure, and in
particular, by parameter coupling and degeneracy.

The task of assessing the coupling between model parameters is 
complicated when models have a large number of free parameters. 
The single S\'ersic profiles 
fit to the NICMOS galaxy images have 7 free parameters
(centroid, luminosity, $r_e$, $n$, axis ratio, and position angle).
While GALFIT selects a best-fit by minimizing $\chi^2$ for a given 
set of input guesses,  it is not clear whether the  minimized $\chi^2$ 
is an absolute minimum or local minimum. 
Investigating the $\chi^2$ values for all combinations of fit parameters 
over the full multi-dimensional  parameter space is prohibitively time
consuming and  computationally expensive. Instead, we will 
adopt a simpler approach of  focusing on  strong coupling between 
$r_e$ and $n$, and exploring how  $\chi^2$ varies as these 
parameters are moved away from the initial solution picked by GALFIT.

One important point should be noted when using changes in 
$\chi^2$, or $\Delta\chi^2$, for fits to different models.
When errors are normally  distributed, 
the multi-dimensional  ellipsoids for a given $\Delta\chi^2$ contour 
can be associated with a statistical confidence level  (e.g., 
$\Delta\chi^2 \sim1$ corresponds to a 68\% confidence level).
However, since  the errors in the GALFIT models are not normally 
distributed,  but are instead dominated by the systematics of galaxy 
structure, this means that we cannot a priori assign a confidence level
to a given $\Delta\chi^2$. 
As outlined in the test below, we can still use the shape of  $\Delta\chi^2$  
as a function of $n$ or $r_e$ as a guide to the quality of fit in the
sense that sharp rises in  $\chi^2$ as $n$ is varied away from the
best-fit value are taken as indicative of poorer fits. 
But, we cannot a priori say  how much poorer the fits are in a statistical sense.  
This is a well-known and hard problem in structural fitting.
We will return to this point in section $\S$\ref{smonte}.

We carry out the test below for all galaxies in our sample
We  denote as $\chi^2_{\rm{min,0}}$,  the value of $\chi^2$ obtained
when  GALFIT fits the galaxy with  $n$ and $r_e$ as free parameters.
The associated best-fit parameters are $n_{\rm{min,0}}$ and $r_{e,\rm{min,0}}$.
We then fit single S\'ersic profiles with $n$ fixed at discrete values (0.5-10),
while allowing all other parameters to freely vary. The initial inputs
to these  fits were the same as those used to generate the model in which
$n$ is a free parameter. We let GALFIT find the best-fit for each of
these fixed $n$ models by minimizing  $\chi^2$, and we record for
each such best-fit the following quantities:  the fixed value of $n$, the 
best-fit value of $r_e$, and the associated minimum in $\chi^2$ called 
$\chi^2_{\rm{min}}$.  We then evaluate how the difference
$\chi^2_{\rm{min}}$ - $\chi^2_{\rm{min,0}}$
varies as a function of  $r_e$  and $n$, as we move to values
away from  $n_{\rm{min,0}}$ and $r_{e,\rm{min,0}}$.

\begin{figure}[]
\figurenum{B1}
\centering
\scalebox{1.00}{\includegraphics*{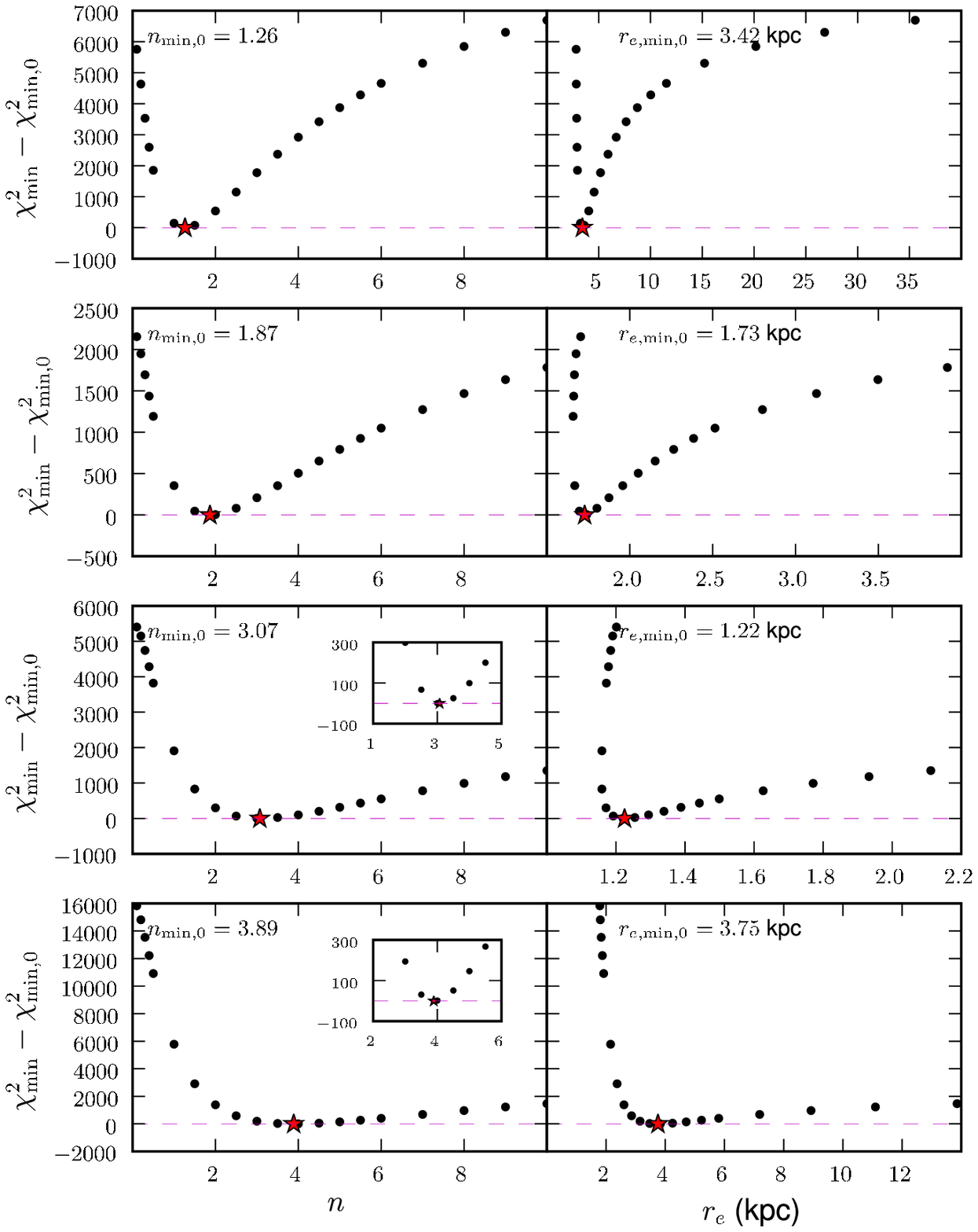}}
\caption{
For four representative galaxies with $n\sim1-4$,
the first and second columns show the
difference $\chi^2_{\rm{min}}-\chi^2_{\rm{min,0}}$
versus $n$ and $r_e$, respectively. $\chi^2_{\rm{min,0}}$ is the
minimum $\chi^2$ obtained when all parameters are freely fit, and $\chi^2_{\rm{min}}$
is the minimum $\chi^2$ when $n$ is held at discrete values (0.5-10).
The $r_e$ in the second
column are the best-fit results for a given $n$ and $\chi^2_{\rm{min}}$.
The red stars mark the best-fit $n_{\rm{min,0}}$ and
$r_{e,\rm{min,0}}$ corresponding to $\chi^2_{\rm{min,0}}$.
The insets in rows 3 and 4 of column 1 show a magnified view around
the minimum in  $\chi^2_{\rm{min}}-\chi^2_{\rm{min,0}}$.
Note that for galaxies with $n_{\rm{min,0}}<2$ (rows 1 and 2),
$\chi^2_{\rm{min}}$ - $\chi^2_{\rm{min,0}}$ rises sharply at
higher $n > n_{\rm{min,0}}$, thereby
making it unlikely  that a higher $n>2$ would provide
a similarly  good fit.
\label{chitest}}
\end{figure}

The test was carried out for all galaxies.
Figure~\ref{chitest} shows the results of the test for
four representative galaxies with $n\sim 1-4$.
The first  column of Figure \ref{chitest} shows how
($\chi^2_{\rm{min}}$-$\chi^2_{\rm{min,0}}$) changes when
$n$ is varied away from $n_{\rm{min,0}}$ at discrete values (0.5-10)
and GALFIT is allowed to vary all other parameters to get a
best-fit that yields  $\chi^2_{\rm{min}}$.
The second column shows the corresponding best-fit  $r_e$ for
that  $\chi^2_{\rm{min}}$.
Red stars in the plots denote $n_{\rm{min,0}}$ and $r_{e,\rm{min,0}}$,
which are associated with $\chi^2_{\rm{min,0}}$.
The shape of $\chi^2_{\rm{min}}$ - $\chi^2_{\rm{min,0}}$ is asymmetric 
for $n$ and $r_e$. The coupling between
$n$ and $r_e$ means ($\chi^2_{\rm{min}}$ - $\chi^2_{\rm{min,0}}$) 
varies  in a similar way with both $n$ and $r_e$.

\begin{figure}[]
\figurenum{B2}
\centering
\scalebox{1.00}{\includegraphics*{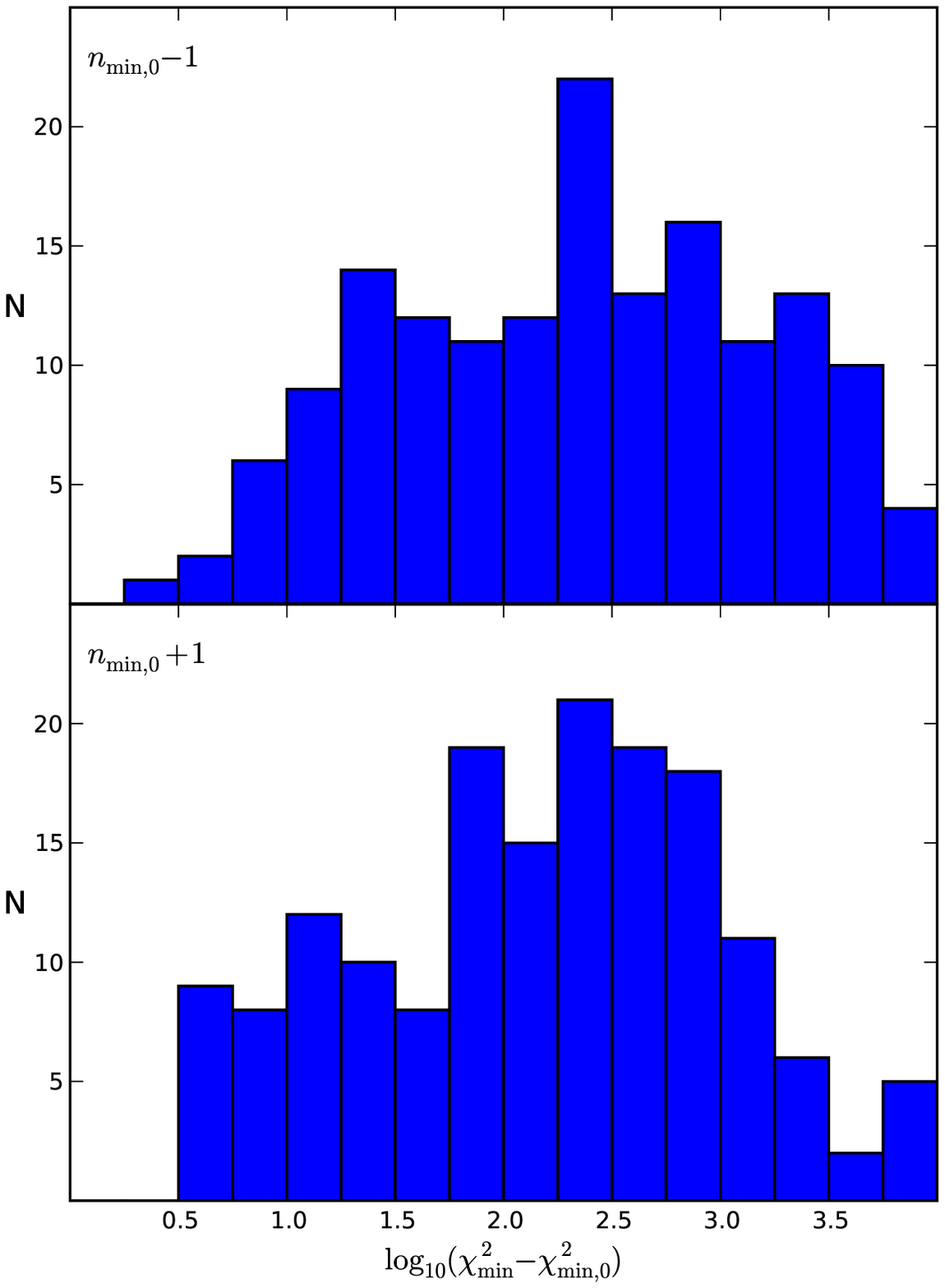}}
\caption{
The quantity $\chi^2_{\rm{min}}-\chi^2_{\rm{min,0}}$
from Figure \ref{chitest} is shown for all massive GNS
galaxies well fitted with a single S\'ersic profile.
The top panel evaluates  $\chi^2_{\rm{min}}-\chi^2_{\rm{min,0}}$ at 
$n_{\rm{min,0}}-1$, and the bottom panel evaluates  
$\chi^2_{\rm{min}}-\chi^2_{\rm{min,0}}$ at $n_{\rm{min,0}}+1$, where
$n_{\rm{min,0}}$ is the
best-fit S\'ersic index corresponding to $\chi^2_{\rm{min,0}}$.  
\label{chihist}}
\end{figure}

We can see that in  Figure~\ref{chitest}, the absolute minimum  
$\chi^2$ values occur at
the $n_{\rm{min,0}}$ and $r_{e,\rm{min,0}}$ values, which GALFIT
picked when it was allowed to freely fit the galaxies without fixing
$n$. Shifting  $n$ away from $n_{\rm{min,0}}$  (denoted by the red
stars) by $\pm1$ can  increase $\chi^2_{\rm{min}}$ by several 10s or
100s of $\chi^2$ units.  While only 4 representative galaxies are
shown in  Figure \ref{chitest}, we show results for the whole sample
in Figure \ref{chihist}. This figure illustrates that  the distribution of
($\chi^2_{\rm{min}}$ - $\chi^2_{\rm{min,0}}$) for  $n_{\rm{min,0}}-1$ 
(top panel) and $n_{\rm{min,0}}+1$ (bottom panel), and demonstrates for
the whole sample, $\chi^2_{\rm{min}}$ generally changes 
substantially when $n$ is shifted away from $n_{\rm{min,0}}$.
We draw two primary conclusions from Figure~\ref{chitest}: 
\begin{enumerate}
\item
For galaxies with $n_{\rm{min,0}}>2$ (rows 3 and 4),
$\chi^2_{\rm{min}}$ - $\chi^2_{\rm{min,0}}$
rises sharply  at lower $n < n_{\rm{min,0}}$, suggesting
that lower $n$ values are unlikely to yield a good fit
for such systems.
At $n > n_{\rm{min,0}}$, $\chi^2_{\rm{min}}$ - $\chi^2_{\rm{min,0}}$
rises less sharply,  but the rise is  still substantial as demonstrated
in by the high-magnification inset plots
in rows 3 and 4 of column 1.
\item
The most important point to take from Figure~\ref{chitest} is that
for galaxies with $n_{\rm{min,0}}<2$ (as in rows 1 and 2),
$\chi^2_{\rm{min}}$ - $\chi^2_{\rm{min,0}}$ rises rapidly at
higher $n > n_{\rm{min,0}}$, thereby
making it unlikely  that a higher $n>2$ would provide
a similarly  good fit. {\it Thus, we have a great degree of
confidence that we are not highly overestimating the number of
$n\leq2$ galaxies in the sample}.
\end{enumerate}

\subsection{Recovery of Parameters From Simulated Images}\label{smonte}

Section \S\ref{decompcoupling} tests how well parameters are
recovered in real galaxies, but we cannot a priori assign a confidence 
level to a given $\Delta\chi^2$
because the errors in the GALFIT models are not normally distributed.
However, we can  run an extra complementary test where we use
simulated {\it idealized} galaxies whose ($n$, $r_e$) are a priori known.
The drawback of using idealized galaxies as opposed to the real
galaxies fitted in \S\ref{decomp} is that the former lack the complexity of
real galaxies, since they are simply generated from GALFIT models
and exactly described by a functional form, such as a Sersic
model with a specified ($n$, $r_e$). However, the advantage is that
we do know  the ($n$, $r_e$) values  a priori and can therefore compare
these values to those obtained once these idealized galaxies
are inserted into frames with noise properties
corresponding to the NIC3 GNS images of our sample galaxies at $z=1-3$.

This test is performed by 
simulating 1000 galaxy images, each with a unique set of S\'ersic parameters:
surface brightness at the effective radius $\mu_e$, effective radius $r_e$, 
S\'ersic index $n$, axis ratio b/a, and position angle PA.
The parameters are chosen randomly from uniform distributions
spanning the parameter space of the observed galaxies.  The ranges in
$\mu_e$, $r_e$, $n$, b/a, and PA are 16 to 32 mag/arcsec$^2$, 
$0\farcs05$ to $1\farcs0$, 0.5 to 10, 0.3 to 1.0, and -90 to 90 degrees, 
respectively.  The chosen range in input $\mu_e$ mimics the effect of
surface brightness dimming, and the range in $r_e$ ensures the simulated
objects span the angular size of the real GNS galaxies.
The simulated galaxies were created with GALFIT and convolved with a 
drizzled PSF image.  They were set within a sky
background equivalent to the mean NIC3 sky background within GNS
(0.1 counts/sec).
Source noise, sky noise, and read noise (29 $e^-$) were added to the frames.

\begin{figure}[]
\figurenum{B3}
\centering
\scalebox{0.80}{\includegraphics*{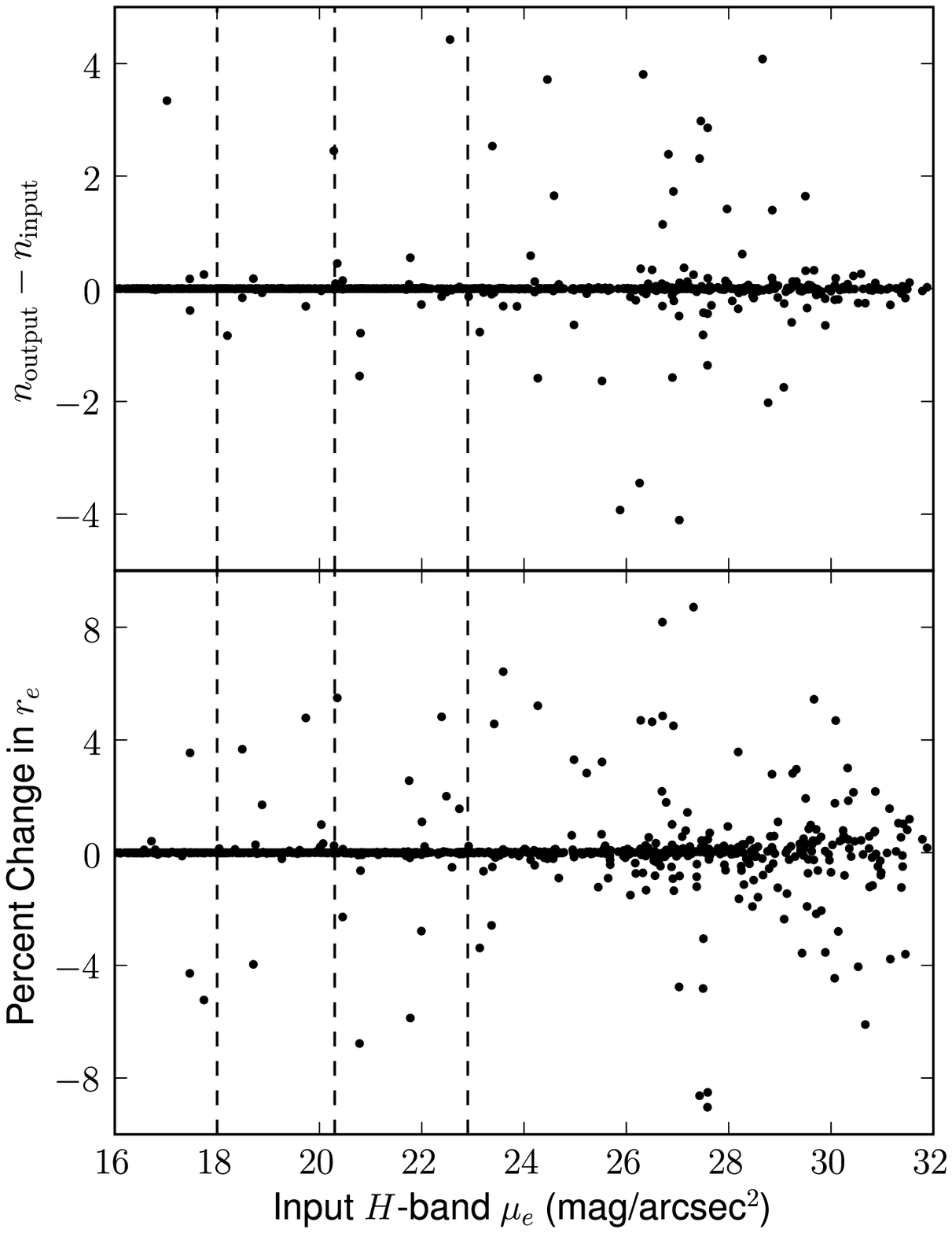}}
\caption{
For the simulations described in \S\ref{smonte}, 
the difference between input and output S\'ersic index $n$ and effective
radius $r_e$ are plotted against effective surface brightness $\mu_e$, the 
surface brightness at $r_e$.
The vertical lines correspond to the range in $\mu_e$  in the 
NIC3/F160W band for  the  massive galaxies at $z=1-3$ 
in our sample.
\label{plotmonte}}
\end{figure}

The simulated images were then re-fit with GALFIT to derive ($n$, $r_e$).
Initial guess parameters for ($\mu_e$, $r_e$, $n$, b/a, PA) were generated
randomly from uniform distributions spanning $\pm1.5$ mag/arcsec$^2$ in $\mu_e$,
$\pm0\farcs3$ in $r_e$,  $\pm2$ indices in $n$, 0.3 to 1 in b/a, and
-90 to 90 degrees in PA.  Figure~\ref{plotmonte} shows the recovery in
($n$, $r_e$)
plotted against surface brightness.  The dashed vertical lines represent the
minimum, median, and maximum $\mu_e$ for the observed massive galaxies. 
Figure~\ref{plotmonte} shows ($n$, $r_e$)
are well recovered across the full range in observed $\mu_e$. The recovery
as a function of $\mu_e$ severely degrades only at several mag/arcsec$^2$
fainter than observed $\mu_e$. 
In $\sim95\%$ of cases, $n$ and $r_e$ are recovered to within
10\% of their input values for the range of observed $\mu_e$ 
among the massive galaxies in our sample.

\subsection{Tests on MGC Fits}\label{smgcts}
The structural parameters for the massive galaxies  at $z\sim$~0,
are derived by Allen et al. (2006) by using the GIM2D code
(Simard et al. 2002)  to fit single S\'ersic  component
to the MGC $B$-band images.
We derived the structural parameters for the massive  galaxies  at
$z=1-3$, by using the GALFIT code (Peng et al. 2002) on the
NIC3/F610W images ($\S$\ref{decomp}).
One might  wonder whether  the dramatic shift in Figure~\ref{fnre}
of the $z=2-3$  galaxies toward lower ($n$, $r_e$) compared
to the the $z\sim$~0 MGC galaxies may be caused by systematic 
differences between the fitting techniques used by us versus
those by  Allen  et al. (2006). This would be the case only if the
fits  by  Allen  et al. (2006) give systematically higher 
($n$, $r_e$) than ours for the same galaxies. As we show below
this is not the case.

\begin{figure}[]
\figurenum{B4}
\centering
\scalebox{0.80}{\includegraphics*{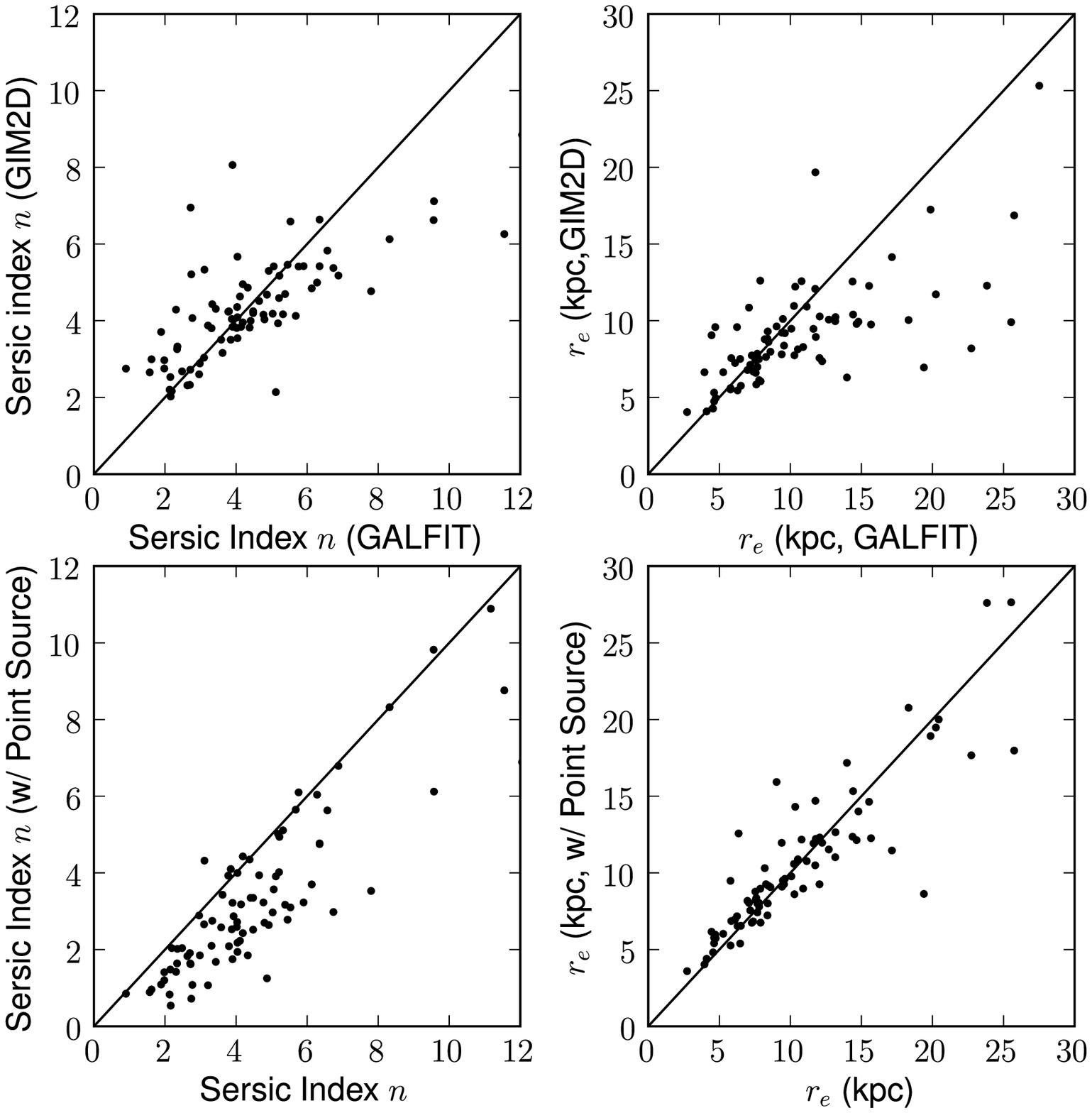}}
\caption{
Top row: We demonstrate for a subset of $z\sim 0$ galaxies in the MGC 
catalog that the GIM2D-based  ($n$, $r_e$) values  from Allen et al. 
(2006) are not biased to higher values compared to 
our GALFIT-based fits for the same galaxies. All fits are performed
on the $B$-band images from MGC.
Bottom row: We show the  effects of adding a point source in the GALFIT models
fitted to the $z\sim 0$ MGC galaxies.
The values obtained  using a model made of  a S\'ersic
component plus a point source are plotted along the y-axis,
while the x-axis shows the values obtained with a single S\'ersic
component.  The values of $r_e$ are not changed systematically.
The S\'ersic index is lowered by the addition of the point source,
but only 20\% of sources with $n>2$ in the single S\'ersic fit
have $n\leq2$ after including the point source.
\label{fmgc1}}
\end{figure}

In order to address this issue, we have applied  GALFIT 
to a subset of $B$-band MGC images and compared our 
resulting  structural parameters to the GIM2D-based results given 
in the MGC catalogue. The
comparison (top row of Figure~\ref{fmgc1}) shows that the GIM2D-based fits
of Allen  et al. (2006) are not biased to higher ($n$, $r_e$) 
compared to our GALFIT-based fits for the $z\sim$~0 MGC galaxies.
In fact, for large $r_e$, the GIM2D-based values may even be lower.
in many cases.

These results are consistent with extensive comparisons of
single component S\'ersic  fits  from GALFIT and GIM2D conducted
by H{\"a}ussler et al.(2007) on  both simulated and real galaxy
data. They concluded that both codes provide
reliable fits with little systematic error for galaxies with
effective surface brightnesses brighter than that of the sky,  
as long as one is not dealing with highly crowded fields.

Another possible source of difference between the  structural
parameters of the $z\sim$~0 and $z=2-3$ massive galaxies
might be the fact
that Allen et al. (2006) fitted the $z\sim$~0  massive galaxies
with only a single S\'ersic  component, and did not include
an extra point source component in galaxies with  evident
nuclear sources.  It seems unlikely that the much larger
fraction  of  higher ($n$, $r_e$) systems at $z\sim$~0  in
Figure~\ref{fnre} is mainly driven by this effect. To
illustrate this, we have fitted the  $z\sim$~0 MGC
galaxies in the top row of Figure~\ref{fmgc1} with a combination of
a single S\'ersic  component and a point source model using GALFIT.
The bottom row of Figure~\ref{fmgc1} shows the results. 
The values of $r_e$ are not 
changed systematically. The S\'ersic index is lowered by
the addition of the point source, but only 20\% of the sources with
$n>2$ in the single S\'ersic fit have $n\leq2$ after including the
point source. 
Since not all  $z\sim$~0 MGC galaxies in Figure~\ref{fnre}  will 
have nuclear sources, the fraction of sources impacted will be 
even less. We thus conclude that the presence of a point source 
in some of the $z\sim$~0 MGC galaxies and the inclusion of such 
a point source in the model fits, are not sufficient to shift 
the $z\sim$~0 MGC galaxies into 
the parameter space occupied by the $z=2-3$ massive 
galaxies in Figure~\ref{fnre}.
\end{appendices}

\end{document}